\documentclass[unnumsec,webpdf,contemporary,large]{oup-authoring-template}%

\graphicspath{{Fig/}}

\usepackage{xcolor}
\usepackage{soul}
\theoremstyle{thmstyleone}%
\theoremstyle{thmstyletwo}%
\theoremstyle{thmstylethree}%

\usepackage{comment}
\usepackage{booktabs}
\usepackage{longtable}
\providecommand{\rev}[1]{#1} 
\begin{document}

\journaltitle{Accepted to Interacting with Computers}
\copyrightyear{November 2026}
\pubyear{2026}

\firstpage{1}


\title[Human–AI Co-Ideation Across Socio-Technical Configurations]{Human–AI Co-Ideation Across Socio-Technical Configurations: An Interactional Analysis of Planar and Immersive Workspaces}

\author[1,$\ast$]{Orit Shaer\ORCID{0000-0002-0515-2957}}
\author[1]{Mary Calabro\ORCID{0009-0005-1223-866X}}
\author[1]{Zoe Mitchell\ORCID{0009-0007-3189-0332}}
\author[1]{Angelora Cooper\ORCID{0000-0003-0758-7740}}
\author[2]{Raffaella Sadun\ORCID{0000-0002-0987-8923}}
\author[3]{Andrew L. Kun\ORCID{0000-0001-9756-7748}}
\author[4]{Osnat Mokryn\ORCID{0000-0002-1241-9015}}

\address[1]{\orgdiv{Computer Science}, \orgname{Wellesley College}, \orgaddress{\street{106 Central st., Wellesley}, \postcode{02481}, \state{MA}, \country{USA}}}
\address[2]{\orgdiv{Business School}, \orgname{Harvard}, \orgaddress{\street{Boston}, 
\state{MA}, \country{USA}}}
\address[3]{\orgdiv{Electrical and Computer Engineering}, \orgname{University of New Hampshire}, \orgaddress{\street{Durham}, 
\state{NH}, \country{USA}}}
\address[4]{\orgdiv{Information Systems}, \orgname{University of Haifa}, \orgaddress{\street{Haifa},  \country{Israel}}}

\corresp[$\ast$]{Corresponding author. \href{email:oshaer@wellesley.edu}{oshaer@wellesley.edu}}


\abstract{
    Remote collaborative platforms increasingly integrate generative AI to support group ideation. These platforms are sociotechnical configurations in which spatial mediation, interaction modality, AI output format, and the mechanisms that make contributions visible to a group co-vary as properties of a deployed system. This paper examines how such configurations shape \rev{collaboration among participants and their} engagement with AI during creative work. We report a comparative study of two commercially available platforms that instantiate contrasting configurations: a planar shared canvas with text-based AI integration, and an immersive virtual reality workspace with concise AI output embedded in the shared space. \rev{Fourteen all-women triads (42 participants)} completed a structured ideation task consisting of human-only idea generation, AI-supported idea expansion, and group convergence. We analyzed collaboration patterns, \rev{selection of AI-generated ideas}, perceived idea quality, workload, presence, and the \rev{lexical} characteristics of generated ideas. The immersive configuration yielded greater spatial presence \rev{and introduced a participation paradox, in which spatial freedom enabled exploration and fragmented group coherence; it was also associated with more balanced role distribution during convergence, at the cost of} higher physical demand. \rev{In both configurations, AI multiplied the idea pool, and ideas groups authored with AI available were rated above participants' own. AI output format, about 21 words per idea on the planar canvas and 3 in the immersive workspace, tracked how participants evaluated AI-generated ideas and how often they selected them unaltered as finalists.} Neither configuration \rev{made AI contributions visible or attributable to the group:} participants invoked AI individually and asynchronously in both. We \rev{derive design implications for AI output format, group-visible AI invocation, and spatial support for convergence.}
}

\keywords{LLM, Brainwriting, human-AI Ideation}

\maketitle


\section{Introduction}
 
Remote collaborative ideation has become an increasingly important component of distributed teamwork, especially as remote and hybrid work arrangements have grown more common in the wake of the COVID-19 pandemic~\citep{barrero2023evolution, buckman2025measuring}. Research shows that remote teams, while effective in executing tasks, are less likely to engage in conceptual work or generate disruptive ideas~\citep{lin2023remote}. Challenges stem in part from constraints inherent to virtual collaboration environments including reduced spatial presence, limited access to non-verbal cues, and difficulty sustaining spontaneous engagement among participants~\citep{BrucksLevav2022, ansah2022need, Olson01092000, friedman2003attentional}. Women face additional challenges in these settings including greater tendencies to be interrupted, lower visibility of contributions, and differential reception of ideas, which may further inhibit participation in creative collaboration~\citep{Gupta2020OnlineMeetings, Yassien2021, schiller2011gender, HILL2025104107}.
 
A growing body of research suggests that virtual collaboration may not only limit interaction bandwidth but also alter patterns of attention and engagement~\citep{minet2024design,macchi2024virtual}. For example,~\cite{BrucksLevav2022} found that participants in virtual settings explored their environment less and recalled fewer unexpected details, reflecting shifts in attentional allocation that are directly related to ideation performance. Such findings highlight that collaborative technologies do not merely facilitate communication, but structure how participants attend, interact, and generate ideas.
 
At the same time, collaborative platforms are rapidly evolving. Contemporary tools range from planar shared canvases such as Miro~\citep{miro2024ai} to immersive virtual reality workspaces such as Noda~\citep{nodaNodaMind}, each offering different spatial, representational, and coordination affordances~\citep{arthur2023designthinking, dimenteers2023immersivecollab, fromm2020vr_brainstorming, gong2022virtual_brainstorming}. Increasingly, these systems also integrate generative AI technologies, including large language models, to augment idea generation and expansion during collaborative sessions~\citep{HeCanvasCHIWork24, miro2024ai, TaoDesingWeaver, PuIdeaSynth, ShaerBrainwriting24, wan2026diverse}. 
While prior research has often examined workspace dimensionality or AI assistance as isolated variables, less attention has been given to how integrated collaborative systems 
structure 
the joint activity of groups and AI. The relative lack of attention to integrated systems is the motivation for our work. 

\rev{The AI itself is a designed object within these systems, and how it is presented carries consequences for how groups engage with it. \cite{holzinger2022personas} adapt the personas method to AI systems and show that decisions about how an AI's role, competence, and limits are represented to users shape expectations, interpretation of output, and willingness to act on what it produces. \cite{seeber2020machines} argue that a machine occupying a team role alters the coordination and communication demands placed on the people around it, while \cite{evans2025collaborate} question whether the language of collaboration describes an asymmetric relationship in which one party sets the objectives. Empirically, \cite{wan2026diverse} find that assigning distinct personas to a generative model changes the diversity of the idea set a group produces with it. The two platforms we compare occupy the minimal end of this design space: in each, the AI is an unnamed function invoked on demand, with no stated role, no expressed competence, and no conversational turn. Persona is therefore not a variable we manipulate. What differs between the systems is the form in which the AI's contributions arrive in the shared workspace.}
 
\rev{Three areas of prior research inform this study. Each has typically examined one component of AI-supported group ideation. 
CSCW work on distributed collaboration established that geographic and technical arrangements shape what groups accomplish together~\citep{Olson01092000}, that persistent tensions separate social coordination needs from the technical mechanisms available to support them~\citep{Ackerman2000}, and that technologies and work practices are constituted through use~\citep{orlikowski2000using, Crowston19}. This tradition treats the system as a medium between people, and it developed before generative models began contributing content to the shared workspace. Research on immersive collaboration examines spatial and embodied properties, including immersion, co-presence, gesture, and avatar representation, and reports their effects on group creativity~\citep{fromm2020vr_brainstorming, gong2022virtual_brainstorming, CoralSpatial, IdeAvatar2020}; AI is largely absent from these designs. Finally, research on AI-supported group ideation varies the model, the prompt, or the interaction protocol on a fixed workspace, typically a single planar canvas or a text interface~\citep{HeCanvasCHIWork24, ShaerBrainwriting24, PuIdeaSynth}. Frameworks for human-AI co-creation describe the interaction itself, including turn-taking, contribution type, and communication~\citep{rezwana2023cofi}, and set out an agenda for machines acting as team members~\citep{seeber2020machines}, but they model the human-AI relationship without specifying the spatial and representational environment that hosts it.
}
 
\rev{The sociotechnical configuration lens we adopt takes the whole arrangement as the unit of analysis: spatial mediation, 
interaction modality, AI output format, and the awareness mechanisms that make contributions visible to the group. Deployed platforms bundle these properties, and the bundle is what a group encounters. Treating them as separable variables might conceal interactions between properties, for example between the display constraints of an immersive workspace and the verbosity a designer selects for its AI output. 
Therefore, we ask how configurations shape attention distribution, the negotiation of authority, and the interpretation and integration of AI contributions during AI-supported group ideation.
Our choice to focus on configurations provides ecological fidelity, but also carries an attributional cost, since a comparison of two complete configurations cannot isolate the contribution of any single property. We 
address this tradeoff in the 
Limitations section.}
 
This paper advances understanding of how sociotechnical configurations structure \rev{AI-supported group ideation}, through a comparative interactional analysis of two commercially available and widely adopted platforms that exemplify distinct configurations. One system provides a shared planar canvas with text-based AI integration, in which AI output appears as full-sentence nodes on a flat board. The other offers an immersive three-dimensional workspace with spatialized interaction and concise AI output spatially embedded in a virtual reality environment. \rev{Instead of isolating} dimensionality as a single causal factor, we examine how these integrated configurations shape collaboration patterns, collective uptake of AI suggestions, cognitive load, and the semantic characteristics of generated ideas.
 
 
\rev{
Our work 
distinguishes relations among people from relations between people and the AI system
. We use \emph{collaboration}, \emph{collaborative}, and \emph{team} for relations among the human participants only. We use \emph{co-ideation} for the joint activity of a group and an AI system during idea generation, and \emph{co-elaboration} for the act of a participant extending, combining, or generating new ideas from AI output. A \emph{collaborative platform} is a system that mediates collaboration among people; we do not describe any system as collaborating with its users.}
 
\rev{This usage follows \cite{evans2025collaborate}, who argue that collaboration in its full sense requires participants who hold autonomy over the ends of the joint activity and who can decline to take part. \cite{evans2025collaborate} propose using the term \emph{joint action} ``to describe what happens when humans work with machines.'' 
Others use \emph{collaboration} more inclusively for human-AI relations~\citep{seeber2020machines}. We adopt the restrictive usage for a reason internal to this study: in both configurations the AI was an on-demand function that could not decline a request, could not initiate a contribution, and held no turn in the group's exchange. As we report below, participants invoked it individually and asynchronously. 
Describing the AI as a team member would attribute to it a standing in the group's process that the configurations did not give it.}

The study focuses on all-women participant groups for both contextual and methodological reasons. Remote work has disproportionately affected women, offering flexibility while also introducing challenges such as increased interruptions and reduced visibility of contributions~\citep{Garcia_teams_22, peter2021gender_diversity, schiller2011gender, KOULOURI2017162}. Examining collaborative ideation within this population enables us to explore how platform configurations and AI integration interact with group dynamics in contexts where equitable participation is particularly salient. Methodologically, focusing on women allows us to reduce variability associated with gender-mixed communication patterns while generating insights relevant to \rev{a} growing body of work on inclusive design in technology-mediated collaboration~\citep{criado2019invisible, GuoGames, FoxFem, WestendorfCSCW}. Ecologically, this focus aligns with the growing participation of women in remote and hybrid knowledge work \citep{barrero2023evolution, owid_female_labor_supply, aksoy2025remote, buckman2025measuring}. Future work should extend this analysis to gender-mixed and male-identifying groups.
 
This paper makes the following contributions:
 

 \begin{enumerate}

\item \textbf{A sociotechnical configuration lens for co-ideation.}
Existing frameworks model the human-AI interaction independently of the
setting that hosts it \citep{rezwana2023cofi, seeber2020machines}, and
existing empirical work varies either workspace dimensionality or AI
properties with the other held fixed. The lens we introduce takes the
co-varying bundle of spatial mediation, interaction modality, AI output
format, and awareness mechanisms as the unit of analysis, and specifies
what a comparison of deployed configurations can and cannot attribute to
any single property.

\item \textbf{AI output format as a design variable in group ideation.}
In both configurations, the AI-supported phase enlarged the idea pool
(from 153 to 464 ideas on the planar canvas and from 155 to 796 in the
immersive workspace), and participants rated the ideas their group
authored with the AI available above the ideas they had proposed alone.
How groups treated the AI's own suggestions tracked the form in which
those suggestions arrived: about 21 words per idea on the planar canvas
and 3 in the immersive workspace. Participants rated the longer
suggestions as more insightful, and groups selected them unaltered as
finalists more often (10 of 21 finalists against 3 of 20). The two
platforms also run different underlying models; we weigh this
alternative account and state why the two cannot be fully separated in
this design.

\item \textbf{A shared deficit in group awareness of AI contributions.}
Neither configuration made AI contributions visible or attributable to
the group as a whole. Participants invoked the AI individually and
asynchronously, the prompt surface was a single node label with no
conversational channel, and co-elaboration was rare: 15 ideas across 14
groups, every one of which became a finalist. Most frameworks for
co-creative interaction model one creator and one system; this finding
specifies what changes when several people share one AI in a common
workspace.

\item \textbf{An interactional account of how the configurations
distributed participation.} The immersive configuration introduced a
participation paradox: the spatial freedom that let participants explore
and organize ideas also let them drift apart, fragmenting group
coherence. The comparison shows that dimensionality is one property
among several that shape collaboration. The immersive configuration was
also associated with more balanced role distribution during convergence
(five of seven groups against three of seven), with avatars oriented
toward one another and participants gathering near shared artifacts, at
the cost of higher physical demand.

\item \textbf{Design considerations for AI support in remote group
ideation.} From the second and third contributions: AI output whose
verbosity adapts to the workspace and to the phase of the task; shared,
synchronous AI invocation with attribution that remains legible across
the workspace; and a prompt surface the group can address jointly. From
the fourth: spatial support that preserves individual freedom during
divergence and draws participants together during convergence, with
orientation and proximity cues treated as designable awareness
mechanisms.

\end{enumerate}

In the remainder of the paper, we review related work on collaborative ideation, immersive collaboration, \rev{AI-supported creative systems, and the design of the AI support}. We then describe our comparative case methodology, followed by an analysis of interaction patterns, AI engagement, and idea characteristics across configurations. Finally, we discuss implications for the design of collaborative platforms that integrate AI support and outline directions for future research.

\section{Related Work}\label{related}
\subsection{Collaborative Ideation in Remote Settings}

Exploring collaboration in video-mediated environments has a long history in HCI research. Doug Engelbart's famous ``Mother of All Demos'' event~\citep{ju2007mouse}, held in 1968, introduced ideas such as the mouse and hypertext and demonstrated collaboration between remote participants working on shared files~\citep{engelbart1968research}. Since then, collaborative ideation in remote settings has been studied across multiple decades.

Olson and colleagues instructed small teams to write down requirements for a fictitious product, comparing face-to-face interaction with remote collaboration via video-plus-audio and audio-only channels~\citep{olson1995mix}. They concluded that teams connected via video could generate ideas indistinguishable in quality from face-to-face teams, though audio-only teams produced lower-quality work. This work established that remote collaboration can support idea generation under certain technological conditions.

More recent research has refined this picture. \cite{BrucksLevav2022} found that videoconferencing negatively affects divergent idea generation, although not convergent selection. 
They attribute this reduced performance to decreased variability in gaze direction, which narrows cognitive exploration. Prior research links reduced gaze variability to narrowed attentional focus and reduced creative performance~\citep{friedman2003attentional}. Interestingly, narrowed focus may benefit convergent tasks such as idea selection.

An important aspect of the Brucks and Levav study is the distinction between divergent and convergent phases of creativity~\citep{muller2011leaving}. Divergence involves generating multiple ideas, while convergence involves selecting promising options. This distinction provides a useful lens for examining how collaborative technologies structure different stages of ideation.

\cite{ansah2022need} operationalized divergent and convergent phases using two established creativity tests (the Alternate Uses Test~\citep{guilford1978alternate}, and the Remote Associates Task~\citep{mednick1968remote}), as well as real-world analog tasks. Their findings demonstrate how structured tasks and interaction conditions influence different phases of creativity.

\subsection{Structured Ideation and Visual Artifacts}

Structured approaches to generating and organizing ideas play a crucial role in creative processes. These include brainstorming~\citep{osborn1953applied}, brainwriting~\citep{Wilson2013}, Six Thinking Hats~\citep{Bono1999}, and concept and mind mapping~\citep{novak2008conceptmaps, buzan1996mindmap}. Collaborative ideation methods can lead to more creative solutions by exposing participants to diverse perspectives and supporting associative thinking~\citep{Herringinspired, Dowprototyping, Leeexamples, Siangliulue}.
Digital collaborative platforms extend structured ideation methods by embedding visual artifacts directly into shared workspaces. 

In this work we focus on one type of structured ideation: mind maps. Mind maps facilitate the externalization and spatial organization of ideas in a shared visual space~\citep{buzan1996mindmap}. They encourage associative thinking and help participants identify connections across ideas~\citep{novak2008conceptmaps, eppler2006conceptmaps}. Commercial platforms such as Miro~\citep{miro2024ai} and Xmind~\citep{xmind}, as well as immersive environments such as Noda~\citep{nodaNodaMind}, enable mind mapping by providing 
visual artifacts that represent 
linking, grouping, and visualizing ideas.

\subsection{Sociotechnical Systems, Distributed Cognition, and Embodiment}
Collaborative platforms can be understood as sociotechnical systems in which spatial affordances, representational structures, AI integration mechanisms, and coordination tools are interdependent. In CSCW and HCI, sociotechnical perspectives emphasize that technologies and social practices are intertwined in shaping collaborative work~\citep{orlikowski2000using, Crowston19}. Collaborative systems are not neutral conduits for communication. They are interdependent components of practice that structure what becomes visible, how attention is distributed, and how contributions are interpreted, negotiated, and integrated during group activity. Long-standing discussions of the social-technical gap further highlight that tensions often arise between social coordination needs and the technical mechanisms available to support them~\citep{Ackerman2000}. Understanding collaborative platforms as sociotechnical configurations foregrounds how integrated design choices shape collaborative ideation.

This perspective is grounded in theories of distributed cognition, which emphasize that cognitive processes extend beyond individuals to include interactions among people, artifacts, and environments ~\citep{Hutchins1995CognitionintheWild, Kirsh1995, Klein2006distributed}. From this view, collaborative ideation unfolds across participants and shared material structures, including spatial layouts, interface elements, and AI-generated artifacts. Representational tools such as mind maps, nodes, and spatial groupings do not merely record ideas but actively participate in structuring cognitive activity by externalizing thought, enabling comparison, and supporting iterative refinement.

Embodied cognition further suggests that thinking is shaped by bodily interaction with space~\citep{wilson2002six, shapiro2019embodied}. Research in immersive environments demonstrates that gestures, movement, and spatial positioning influence creative collaboration~\citep{gong2022virtual_brainstorming, gronbaek2024blended, IdeAvatar2020}. For example, the Blended Whiteboard integrates physical whiteboards with an expandable virtual canvas and embodied avatars to support flexible collaborative configurations~\citep{gronbaek2024blended}. Such work illustrates how embodied interaction becomes one component of a broader sociotechnical configuration that shapes how groups generate, explore, and converge on ideas.

The sociotechnical tradition in CSCW developed around systems that mediate between people. A generative model in the shared workspace changes what the framework must account for: the system now contributes content that groups evaluate, attribute, and negotiate over, and the presentation of that content is a design decision on a par with spatial layout.

Together, these perspectives suggest that collaborative ideation emerges from the dynamic interplay among participants, artifacts, spatial arrangements, and AI systems. Analyzing collaborative platforms as sociotechnical configurations provides a framework for examining how design choices redistribute attention, effort, authority, and uptake of contributions within AI-supported group creativity.

\subsection{Immersive Collaboration Platforms}

Immersive technologies provide affordances such as spatial co-presence, embodied interaction, and shared artifacts~\citep{CoralSpatial, gong2022virtual_brainstorming}. Fromm et al.~\citep{fromm2020vr_brainstorming} demonstrated that immersion and co-presence can reduce negative group effects during brainstorming and facilitate freer idea exchange. Gong et al.~\citep{gong2022virtual_brainstorming} emphasized that immersion and symbolic gestures encourage divergent thinking and inclusivity. Ide et al.~\citep{IdeAvatar2020} showed that symbolic gestures can improve communication clarity and increase engagement.

At the same time, immersive environments introduce trade-offs. Prolonged immersion may lead to cognitive fatigue~\citep{fromm2020vr_brainstorming}. Symbolic gestures may increase cognitive load~\citep{IdeAvatar2020}. Accessibility and usability concerns also limit adoption~\citep{CoralSpatial}. These findings suggest that immersive environments reshape collaboration through integrated spatial and technical properties rather than dimensionality alone.

Recent telepresence research extends this design space with avatars and
environments reconstructed from captured imagery at near-photographic
fidelity: URAvatar ~\citep{uravatar24} produces head avatars that can be
relit to match the shared scene, and GaussianNexus ~\citep{Huang25}
reconstructs a room in real time so remote participants share a rendering
of the same physical space.
These systems expand the range of representational fidelity available to immersive collaboration. Commercially available platforms sit well below this frontier, and they vary widely in representational fidelity and awareness mechanisms. This variation reinforces the importance of analyzing collaborative technologies as sociotechnical configurations in which spatial mediation, representation, and interaction affordances jointly shape group dynamics. Across this literature, the collaborating parties are human; generative AI is largely absent from the systems studied, and its integration into immersive workspaces remains an open design question that the following sections take up.

\subsection{AI Support for Group Ideation}

Research indicates that LLM-powered frameworks can enhance ideation by supporting both divergent and convergent thinking. Tools such as Scideator~\citep{radensky2024scideator}, IdeaSynth~\citep{PuIdeaSynth}, and AI-Augmented Brainwriting~\citep{ShaerBrainwriting24} enable iterative idea generation and refinement. Integrated LLM features can surface conceptual links and dynamically organize content~\citep{ishizaka2024sparkit, HeCanvasCHIWork24}.

AI integration reshapes not only idea production but also group dynamics. ~\cite{dell2025cybernetic} argue that AI alters performance patterns, expertise signaling, and collaboration norms in knowledge work. At the same time, generative models encode biases (Bender et al., 2021), raise individual creativity while reducing the diversity of what a population collectively produces (Doshi and Hauser, 2024), and can reduce user agency (Shaer et al., 2024). An open challenge is therefore integration in which AI enlarges the group's idea space without displacing the contributions participants would otherwise make, and without homogenizing what different groups produce.

\subsection{Designing the AI's Role and Presentation}
 
\rev{The work above characterizes the AI by what it produces. A second line of work characterizes it by how it is presented to the people working with it. \cite{holzinger2022personas} 
ground this in the properties that distinguish human-AI interfaces from conventional ones: the system holds a degree of autonomy, its outputs are non-deterministic, and its internal process is opaque to the user. Their open toolbox supplies guidelines, templates, and visualization materials for constructing AI personas, with a worked example from medical AI. The method is a design instrument, and the claim it supports is that representation decisions operate before any single output is evaluated: a user's reading of an AI contribution depends on the role, competence, and limits the system has been given.}
 
\rev{A substantial vocabulary already exists for the roles an AI can occupy in creative work. \cite{lubart2005partners} named four ways computers partner in the creative process, as nanny, pen-pal, coach, and colleague. \cite{guzdial2019friend} derived roles empirically from creators working with an AI game level editor, friend, collaborator, student, and manager, and found that the role a creator perceived shaped their experience of the same underlying system. \cite{guo2025pen} report the archetypes creative writers themselves apply to AI in their practice, including muse, ghostwriter, and collaborator. At the team scale, \cite{mccomb2023focus} organize AI participation in design teams along two dimensions, whether the AI serves as tool or guide and whether it reacts or intervenes proactively, yielding classes of tools, partners, analytics, and coaches.}
 
\rev{A parallel line describes how the co-creative exchange itself is structured. \cite{kantosalo2016modes} distinguish alternating from task-divided co-creativity. The COFI framework of \cite{rezwana2023cofi} maps this space in the most detail, organizing co-creative systems by interaction between the collaborating parties (participation style, task distribution, timing of initiative, and the communication channels available in each direction) and by interaction with the shared product; coding 92 systems against it, they find most systems communicate nothing to the user beyond the shared product itself. \cite{koch2019mayai} demonstrate cooperative ideation in which the system learns which suggestions to offer next from the designer's responses. \cite{zhang2025agency} synthesize two decades of this work through the lens of agency, mapping agency patterns and control mechanisms across 134 papers from top-tier HCI venues. General guidance for human-AI interaction addresses when a system should take initiative and how it should behave~\citep{horvitz1999mixed, amershi2019guidelines}, though such guidance is calibrated to task completion and error recovery, while creative settings can treat ambiguity and open interpretation as design resources~\citep{gaver2003ambiguity, sengers2006staying}.}
 
\rev{Whether an AI can occupy a team role at all is contested. \cite{seeber2020machines} derive a research agenda from a survey of 65 collaboration researchers, organized into design areas spanning the machine artifact, the collaboration, and the institution, and framed as dualities in which each anticipated benefit carries a corresponding harm. \cite{evans2025collaborate} argue the opposing position introduced: 
in their analysis, current systems hold at most autonomy over means, never over ends, which makes them designed artifacts within a joint action, not collaborators within it. Empirically, presentation choices have measurable group-level consequences: \cite{wan2026diverse} find that ideating with several differently configured AI personas preserves the diversity of the resulting idea set, where a uniform configuration reduces it.}
 
\rev{Two properties of this literature situate our study. First, its vocabularies presuppose an AI whose role is expressed to users through naming, framing, or conversational conduct; the platforms we examine express none of these, which concentrates the AI's entire presentation in one channel, the form its output takes in the shared workspace. Second, nearly all of this work operates at the scale of one creator and one system, from Lubart's roles through COFI's dyad to the writers \cite{guo2025pen} interviewed; \cite{mccomb2023focus} and the machines-as-teammates agenda are team-level exceptions, and neither examines how role and presentation operate when several people share a single AI in a common workspace. That is the arrangement in deployed group platforms, and it is the setting our comparison addresses.}
\\

Each of these bodies of literature supplies one piece of the setting our study examines: how remote groups ideate, how structure and visual artifacts scaffold that work, how sociotechnical arrangements and embodiment shape it, what immersion changes, what AI support contributes, and how the AI itself can be designed. Our study focuses on their conjunction, where spatial mediation, AI output format, and awareness mechanisms vary together as properties of deployed platforms.

\section{Methods}
We conducted a between-subjects comparative study of two widely adopted commercial collaborative platforms that represent distinct sociotechnical configurations for AI-supported ideation. Each configuration integrates spatial mediation, representational structure, interaction modality, and AI output style in different ways. Studying mature, real-world systems enables examination of how contemporary configurations of spatial design, interaction techniques, and AI integration shape group interaction, collective engagement with AI suggestions, cognitive load, and collaborative outcomes~\citep{olson1995mix, BrucksLevav2022}.

We addressed the following research questions:
\begin{itemize}
    \item \textbf{RQ1:} \textit{How do distinct sociotechnical configurations of collaboration shape interaction patterns, attention distribution, and perceived collaboration during AI-supported ideation?}\\

    \item \textbf{RQ2:} \textit{How do different collaborative platforms shape the collective evaluation, negotiation, and integration of AI-generated ideas across divergent and convergent phases?}\\

\end{itemize}

\subsection{Operational Definitions}

Table~\ref{tab:definitions} fixes the terms used throughout the paper, following the usage committed in the Introduction.

\begin{table}[t]
\centering
\caption[Operational definitions]{Operational definitions of collaboration-related terms as used in this paper.}
\label{tab:definitions}
\begin{tabular}{@{}p{0.28\linewidth}p{0.64\linewidth}@{}}
\toprule
Term & Denotes \\
\midrule
group, triad & the three human participants in a session \\
collaboration, team & relations among the human participants \\\\
co-ideation & the joint activity of a group and the AI system during idea generation \\\\
co-elaboration & the act of a participant extending, combining, or generating new ideas from AI output \\\\
group-authored (ideas) & ideas authored or edited by participants during the AI-supported phase \\\\
collaborative platform & a system that mediates collaboration among people \\\\
\bottomrule
\end{tabular}%

\end{table}

In the idea corpus (Table~\ref{tab:idea_dataset_curation}), the source of each idea records authorship and phase: participant-authored during the human-only phase, AI-generated during the AI-supported phase, or group-authored during the AI-supported phase. In the post-task questionnaire (Tables~\ref{tab:miro_anova} to~\ref{tab:idea_quality}), participants rated three mutually exclusive sets of ideas: the ideas they personally proposed during the human-only phase (Subtask 1); the ideas produced by the platform's AI feature (Subtask 2); and the ideas their group authored during Subtask 2, whether elaborating on the pooled human and AI ideas or composing new ideas while working with the AI. This third rated set corresponds to the group-authored corpus category. It contains no unmodified AI output and no ideas from the human-only phase, so the three rated sets are mutually exclusive.

\subsection{Ideation Technique}

We employed \textit{mind mapping} as the shared ideation structure. It externalizes ideas into persistent, spatially organized artifacts and supports non-linear, associative thinking ~\citep{buzan1996mindmap, novak2008conceptmaps}. 
Mind mapping facilitates parallel contributions, flexible reorganization, and clear visual representation of idea structure, which aligns well with our goals of examining how configurations shape group processes of idea generation, modification, negotiation, and AI integration.

Our study compares two commercially available platforms, both of which support synchronous mind mapping. The two platforms represent distinct sociotechnical configurations, which we describe next.

\subsection{Sociotechnical Configurations}
We conducted a between-subjects experiment in which our participants completed a three-phase ideation task in one of the following two distinct setups:
\subsubsection {Configuration 1: Planar Shared Canvas with Text-Based AI Integration (Miro)} 
\begin{figure*}[htbp]
\centering\includegraphics[width=1\textwidth]{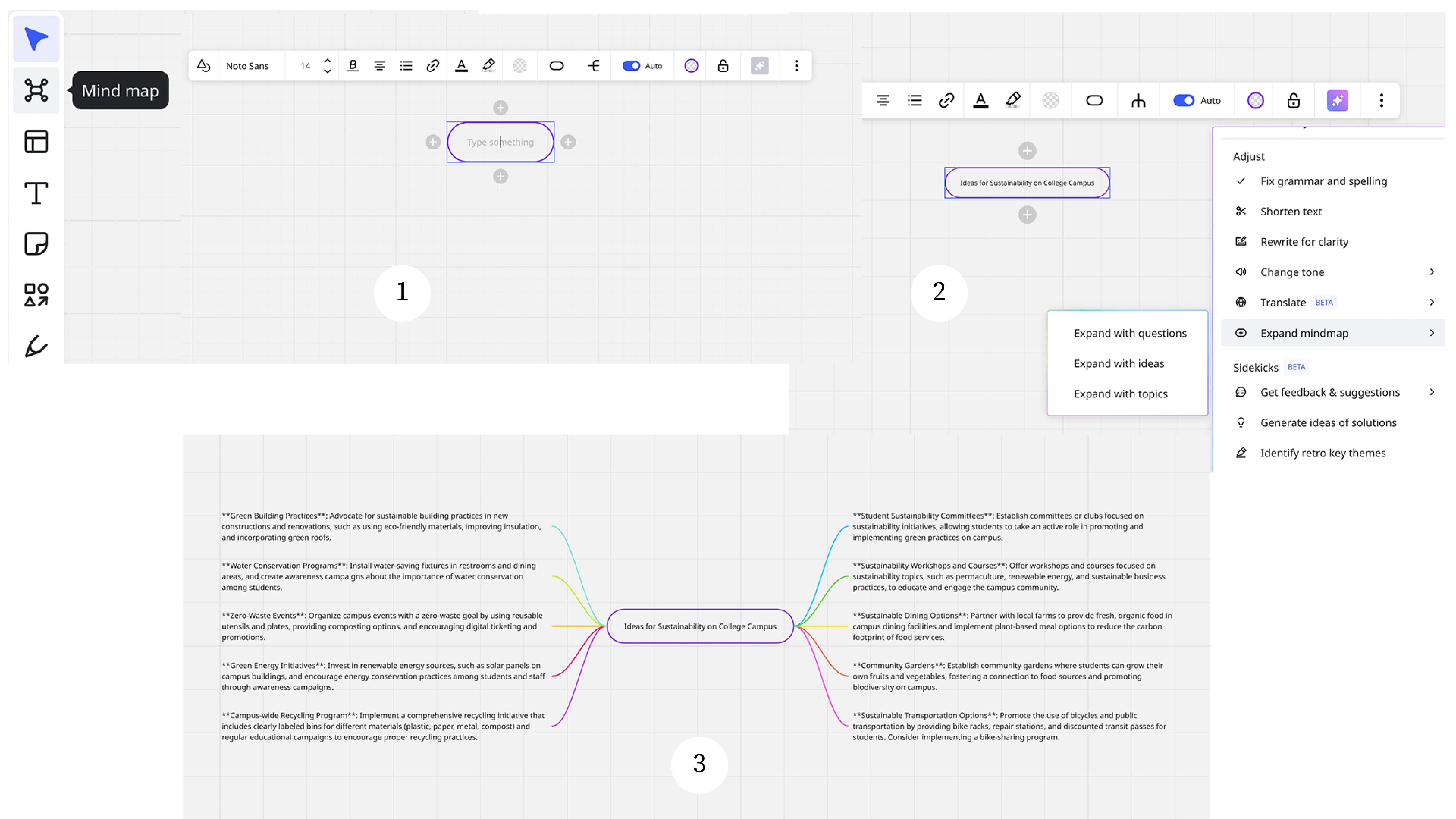}
    \caption{A planar Miro mind map created with AI using the ``Expand mindmap'' tool. (1) The user creates a new mind map and labels it with the desired AI prompt. (2) User selects the AI feature and selects ``Expand With Ideas''. (3) Once selected, Miro generates a mind map of ideas based on the label of the user-created node.}
    \label{fig:mindmap-in-miro}
\end{figure*}

Miro is a digital whiteboard that supports infinite canvas mind mapping with real-time multi-user editing. AI support was integrated as a text-based expansion feature that generated detailed, full-sentence suggestions organized automatically into mind map nodes \citep{miro2024ai}. For AI support, Miro uses the LLM GPT-4o-mini (see Figure \ref{fig:mindmap-in-miro}).

Participants worked on desktop computers in separate areas of the lab and communicated via open audio. Interaction occurred through mouse and keyboard input, and AI-generated ideas appeared as full-sentence text nodes on the flat canvas. 

This configuration emphasized detailed AI output read on a two-dimensional surface, and screen-based interaction.

\subsubsection {Configuration 2: Immersive VR Workspace with Spatialized AI Integration (Noda)}

Noda is a three-dimensional immersive VR workspace to create and manipulate mind map nodes. Users were represented as avatars within a shared virtual environment and interacted with artifacts through embodied gestures and spatial positioning.

\begin{figure*}[htbp]
    \centering
    \includegraphics[width=1\textwidth]{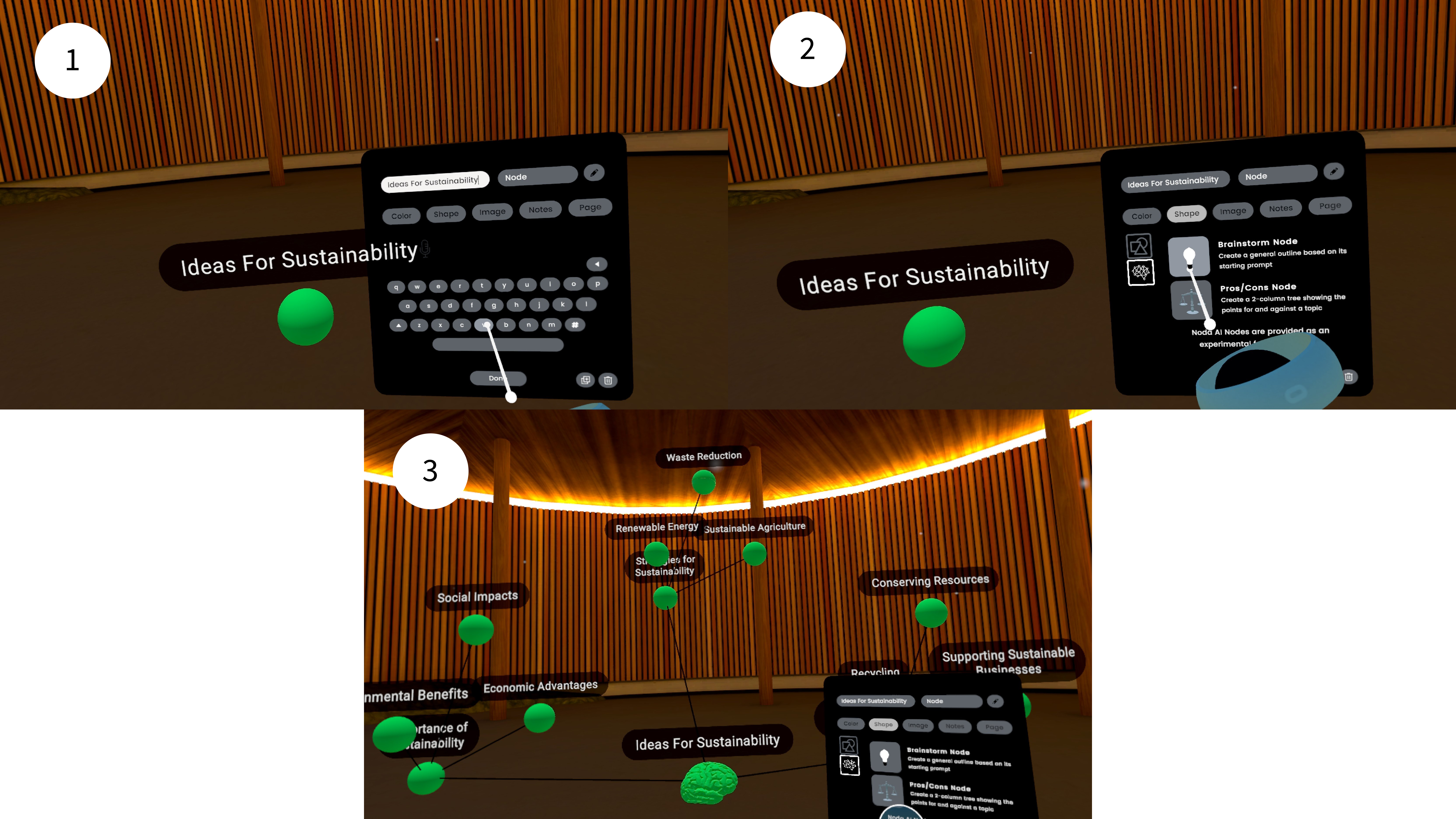}
    \caption{An immersive environment Noda mind map created with AI using the ``brainstorm'' node feature. (1) The user creates a new node and labels it with the desired AI prompt. (2) User switches the shape of the node to ``brainstorm'' to activate the AI feature. (3) Once the shape has changed, Noda generates a mind map based on the label of the brainstorm node.}
    \label{fig:brainstorm-in-Noda}
\end{figure*}

AI support was embedded as a node-based brainstorming feature that generated concise idea suggestions inserted directly into the spatial mind map. AI output in this configuration was brief and spatially embedded within the immersive environment. Noda's AI feature is powered by the GPT-3.5 Turbo model. Figure~\ref{fig:brainstorm-in-Noda} illustrates the steps taken to use the AI ``brainstorm'' feature. 
Participants used the Meta Quest 2 VR headset while seated or standing in separate areas of the lab and communicated via voice within the virtual environment. In both platforms, nodes remained in the workspace for the session; they differ in how they are read. Noda labels float in three-dimensional space, and their legibility depends on a participant’s distance from and orientation toward them (See Figure \ref{fig:noda-arrangement}).

This configuration emphasized embodied interaction through handheld controllers and avatar movement, placement of nodes in three-dimensional space, avatar presence, and concise AI contributions inserted into the shared space.

\subsubsection{AI Features as Deployed}
\rev{Both AI features are closed commercial products, with limited documentation. In each platform, participants used the feature's default behavior: the only input available to users is the text label of the node being expanded, which serves as the prompt (Figures~1 and~2 show the invocation flows). Neither platform exposes system prompts, generation parameters such as temperature, or model configuration, and neither vendor documents them; we could not control or record these settings, and differences in them cannot be ruled out as contributors to the observed differences in output. The constrained prompt surface is itself a property of the configurations: both features expose a single invocation gesture and no conversational channel, a constraint we return to in the design implications.}

\rev{The two models also differ in benchmark capability; OpenAI reports GPT-4o-mini outperforming GPT-3.5 Turbo on MMLU (82\% against 70\%)~\citep{openai2024gpt4omini}. Model capability is therefore one of the differences between the configurations, alongside output format.}
\\\\
\rev{The two configurations differ on multiple properties including spatial mediation, AI model, the length and format of AI output (Miro's full-sentence suggestions averaging 21 words against Noda's keyword-style suggestions averaging 3), in interface design and interaction modality (mouse and keyboard at a desktop against controller-based interaction in a headset), in how collaborators are represented (cursors against avatars), in text-entry method (keyboard typing against pointer-based typing on a virtual keyboard, one character at a time), in voice channel (a separate conferencing call against voice within the virtual environment), and in training procedure (7.5 minutes of video plus 3 minutes of practice against 11.5 minutes of mixed instruction). We treat this bundle as constitutive of the comparison: the study contrasts two complete sociotechnical configurations as deployed, and no observed difference can be attributed to any single property. The Discussion weighs alternative accounts for the principal findings, and the Limitations state the boundaries this design places on causal claims.}

\subsection{Study Task: Ideation using Mind Mapping}
The study task was designed to support both divergent and convergent phases of collaborative ideation within a realistic scenario. Participants assumed the role of a philanthropic committee tasked with allocating \$100,000 to increase sustainability on campus. The task consisted of three structured stages, allowing examination of how each configuration shaped idea generation, AI engagement, and collective decision-making. This task was adapted from the framework outlined in ~\cite{ansah2022need} and comprised the following subtasks:

\begin{enumerate}
    \item \textbf{Subtask 1: Human-only divergent thinking (8 minutes):} Participants first engaged in individual ideation for three minutes, during which each participant created an initial mind map within the shared workspace. This was followed by five minutes of group expansion and discussion, during which participants merged, reorganized, and extended ideas collectively. Figures \ref{fig:miro_board_noAI}
 and \ref{fig:noda_noAI} show the planar (Miro) and immersive (Noda) workspaces after subtask 1.
 
\begin{figure*}[htbp]
    \centering
\includegraphics[width=1\textwidth]{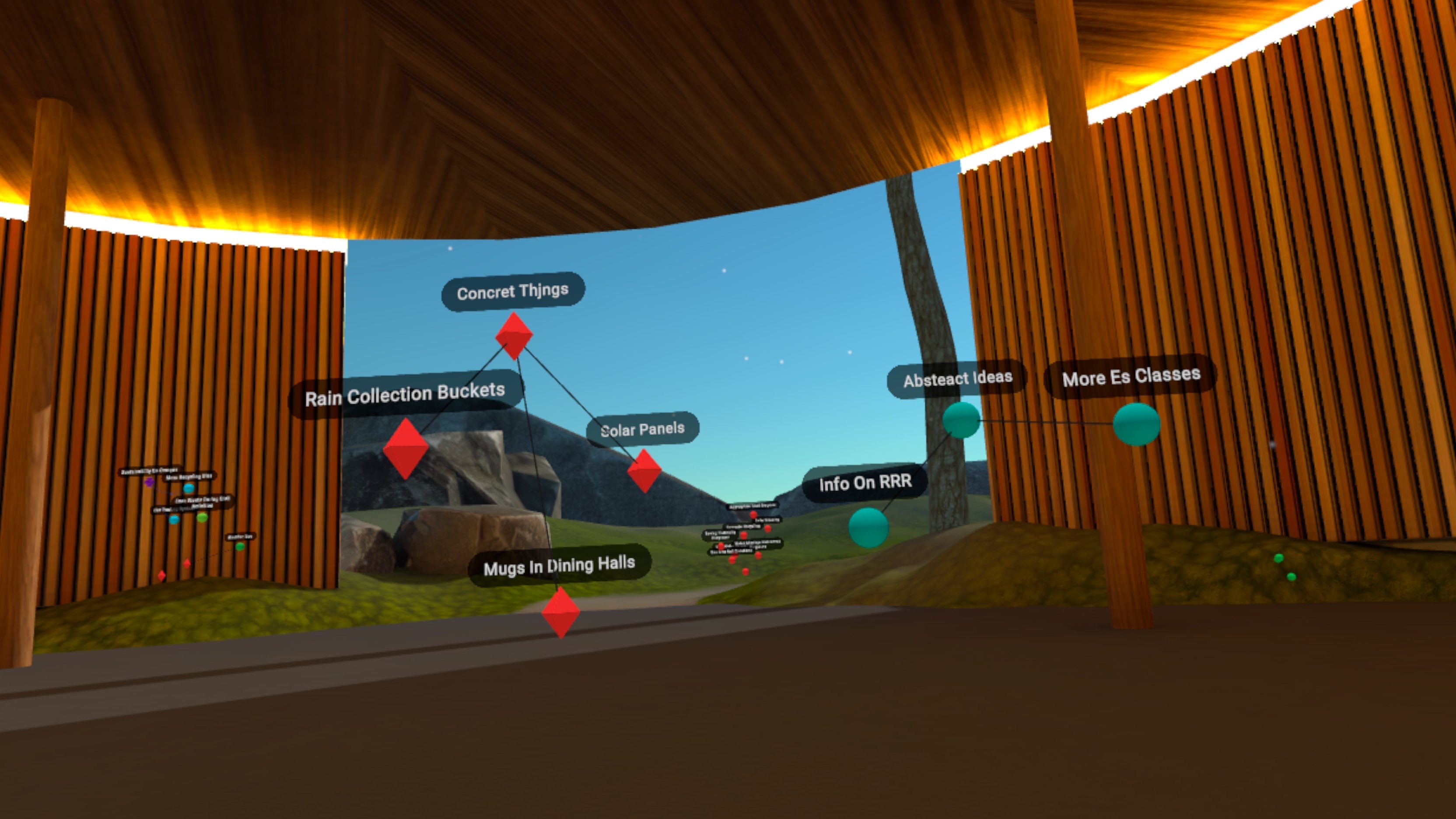}
    \caption{Immersive environment (Noda) map after the completion of subtask 1 (generating a list of potential philanthropic causes to fund), before a group has ideated with AI.}
    \label{fig:noda_noAI}
\end{figure*}

\begin{figure*}[htbp]
    \centering\includegraphics[width=1\textwidth]{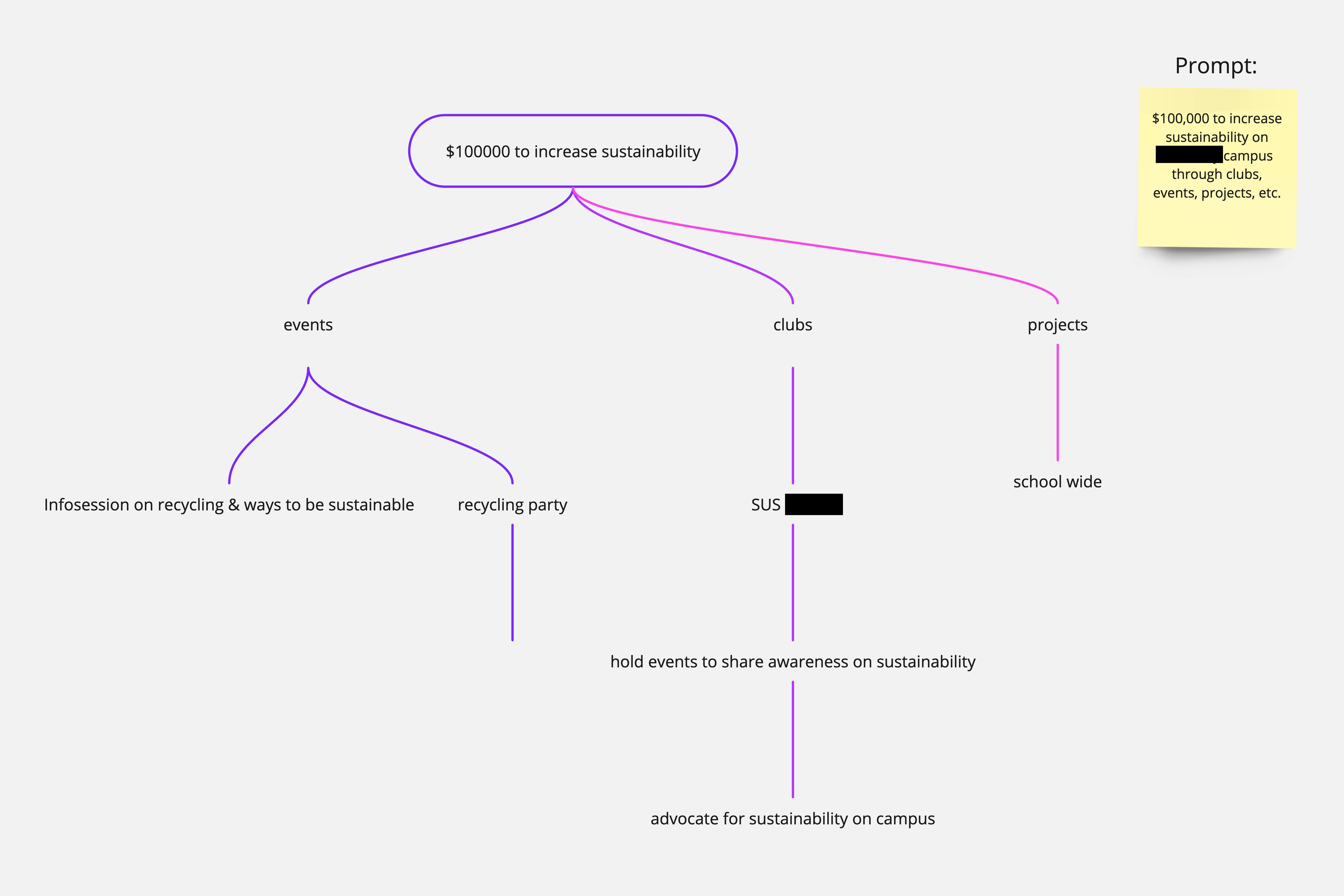}
    \caption{Planar Board (Miro) after the completion of subtask 1 (generating a list of potential philanthropic causes to fund), before a group has ideated with AI. (We blacked out some content to maintain anonymity.)}
    \label{fig:miro_board_noAI}
\end{figure*}

    \item \textbf{Subtask 2: Configuration-Mediated Human-AI Divergence (7 minutes):}  
    Groups invoked the platform's AI feature, which inserted its suggestions directly into the shared mind map through the configuration-specific mechanism (Figures~1 and~2). Participants then reviewed the pooled human and AI ideas as a group, repositioning, editing, and elaborating them and composing new ideas of their own. Figures \ref{fig:miro_board_AI} and \ref{fig:noda} show both workspaces at the end of this subtask.  
    
    \item \textbf{Subtask 3: Human-only idea selection (convergence) (10 minutes):} In the final stage, groups selected three ideas to fund: one high-cost idea (\$60,000), one medium-cost idea (\$30,000), and one low-cost idea (\$10,000). Participants reached consensus through discussion and marked selected ideas within the shared workspace. This phase encouraged the group to engage in convergent thinking, requiring consensus and decision-making. Participants completed this task by discussing ideas without additional AI tools, and were instructed to mark their chosen ideas by labeling or highlighting them.
\end{enumerate}

Groups completed the above task by creating mind maps. An example mind map is shown in Figure \ref{fig:miro_board_noAI}. In this example, a team that used the planar Miro platform created a mind map with a top node denoting the idea ``\$100000 to increase sustainability''. This mind map was created for \textit{Subtask 1}, where the team worked without AI support. The team identified three possible avenues for this, which they recorded as branches of the mind map: ``events,'' ``clubs,'' and ``projects''. These ideas are further developed in sub-branches; e.g., the team recorded the idea that ``projects'' should be ``school-wide'' efforts.

Figure~\ref{fig:miro_board_AI} shows an example of a mind map that a team created in the planar platform  (Miro, Configuration 1) for \textit{Subtask 2}: Collaborative human-AI divergent thinking. Here the team was supported by AI. 
AI-generated content starts with two stars (**). Figure~\ref{fig:miro_board_AI} demonstrates the fact that Miro's AI produced lengthy and detailed suggestions. 
This stands in contrast to the brief notes that the human team members put down.

Figure~\ref{fig:noda} shows the VR workspace (Noda, Configuration 2) after \textit{Subtask 2}, which participants completed using AI.

\begin{figure*}[htbp]
    \centering
    \includegraphics[width=1\textwidth]{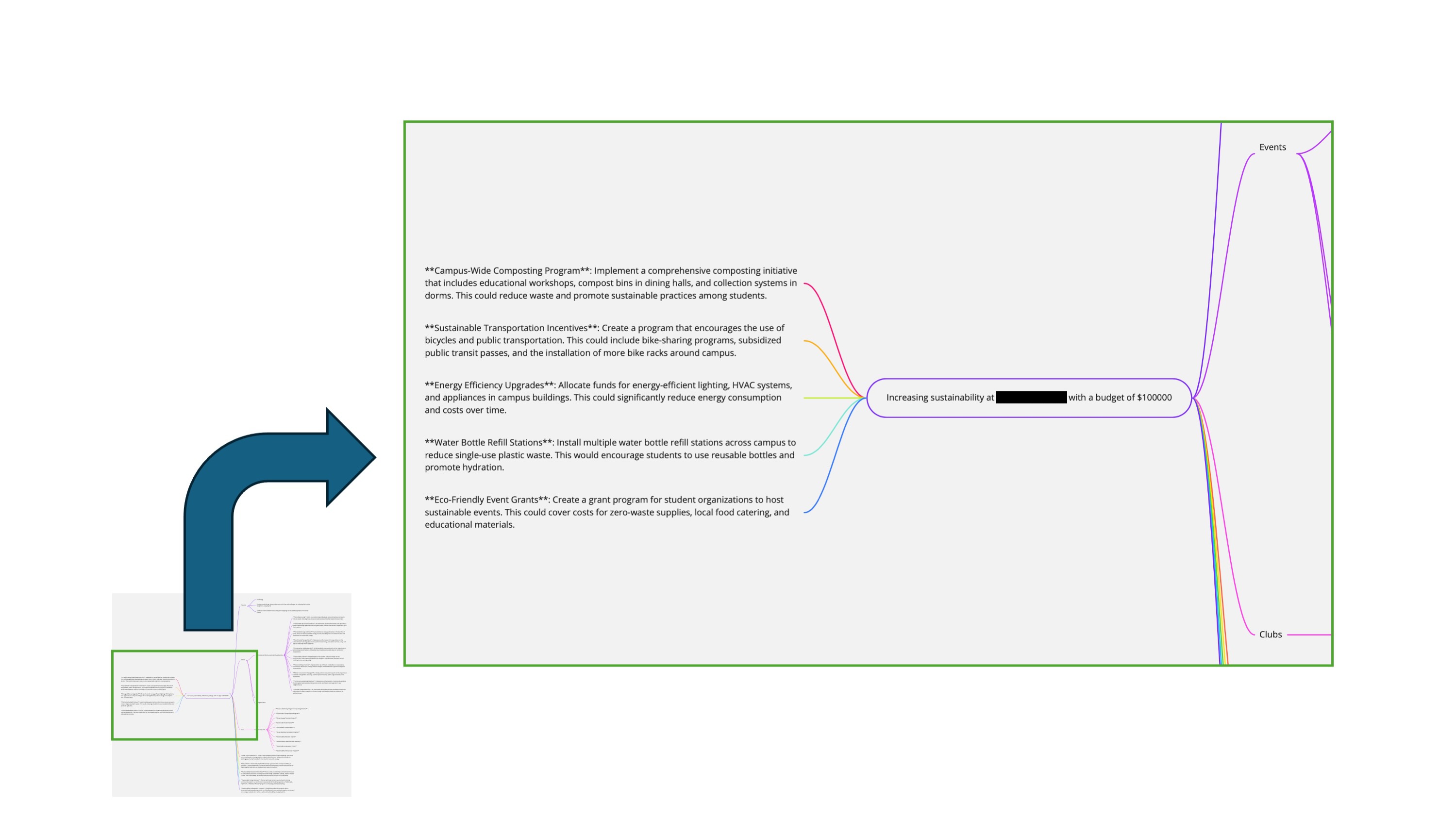}
    \caption{Planar  Board (Miro) after a group has ideated with AI. Ideas that were generated by AI have a double asterisk (**) around them. }
    \label{fig:miro_board_AI}
\end{figure*}

\begin{figure*}[htbp]
    \centering
    \includegraphics[width=1\textwidth]{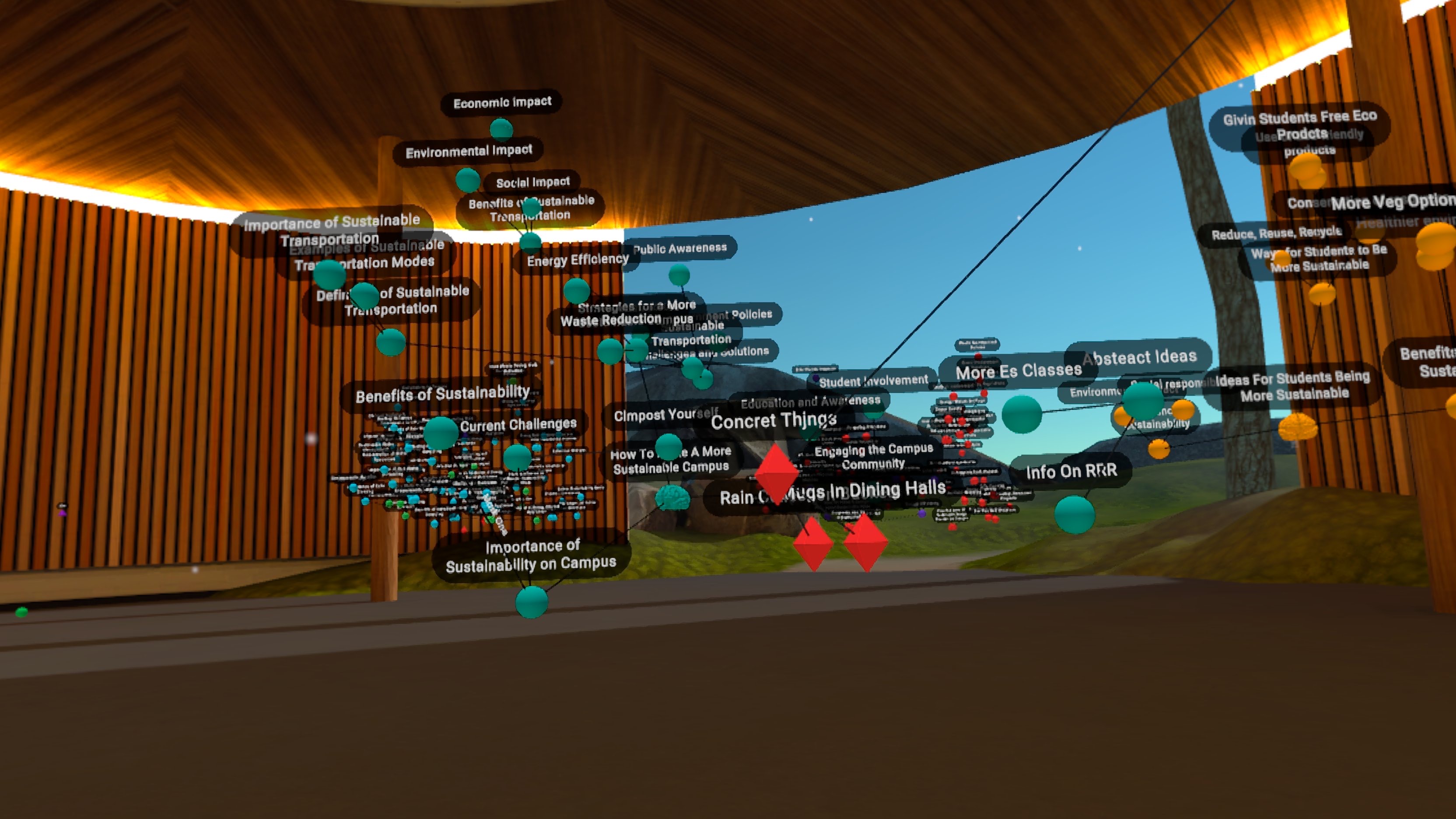}
    \caption{Immersive platform (Noda) map after a group has ideated with AI in subtask 2. In this image, AI generated ideas are marked by having a blue-colored node.}
    \label{fig:noda}
\end{figure*}

\subsection{Study Design}
We conducted a comparative study (between-subjects) involving 42 participants. Participants were randomly assigned to groups of three, forming a total of 14 groups. These groups were evenly distributed across the two configurations, with seven groups in each condition. Randomization ensured an unbiased allocation of the groups to each configuration. The study protocol and procedures were reviewed and approved by the institutional ethics committee.

\subsection{Procedure}
At the beginning of the study, the researchers briefed participants on the purpose and procedures of the study. Next, participants signed an informed consent form. The rest of the procedure was as follows for the two configurations:

\textbf{Platform Configuration 1: Planar Shared Canvas with Text-Based AI Integration (Miro).} Each participant was seated at a computer in a separate area of the research lab and provided with noise-canceling headphones. All participants, along with one researcher in another room, joined a Zoom meeting. A second researcher remained in the room to observe and help with any technical difficulties. Next, the researcher on Zoom shared links to two tutorial videos for using Miro, which totaled 7.5 minutes. The participants then accessed a sample Miro board and were given three minutes to explore its features using a practice prompt. Once all participants completed the training, the researcher on Zoom provided a link to the shared Miro board. The participants collaborated on the study task within Miro, using Zoom for voice communication. The session was recorded via Zoom. The Miro board was saved for later analysis.

\textbf{Platform Configuration 2: Immersive VR Workspace with Spatialized AI Integration (Noda).}
Each participant was positioned in a separate area of the research lab. Researchers assisted them on how to wear and adjust a Meta Quest 2 VR headset to ensure comfort and proper fit. Participants were also given brief instructions about the handheld controls and navigation buttons. Next, participants underwent a 5-minute general VR training on how to wear the headset and use the controllers. This was followed by a 5-minute guided in-app tour of the Noda application. This guided tour was provided by the app itself and allowed participants to practice how to create mind maps, edit nodes, and move around the virtual space. A 1.5-minute video detailing how to use the AI feature in Noda was shown to participants in addition to the guided tour. Once training was completed, one researcher, also equipped with a VR headset, created a shared Noda meeting room. Participants joined the meeting room in VR and collaborated on the study task within the immersive 3-D environment. A second researcher sat in the lab without a headset to track time and read out task instructions. The session was recorded from within the 3-D environment for further analysis.
\\

Upon completing the task, in either platform, each participant completed a post-task questionnaire assessing their experiences and perceptions. We collected the following information: platform used; spatial presence, using the MEC-SPQ (Vorderer et al., 2004); perceived workload, using the NASA Task Load Index (NASA-TLX), which measures workload across six dimensions with responses recorded on a 1–7 scale (Hart and Staveland, 1988); perceptions of the collaborative process; and ratings of idea quality. The MEC-SPQ was developed for, and validated in, immersive virtual environments; its items apply unevenly to a planar shared canvas, and cross-platform presence comparisons should be read with this asymmetry of instrument fit in mind, a point we return to in the Limitations. For idea quality, participants rated three sets of ideas on relevance, innovation, and insightfulness, following the idea-evaluation framework of Shaer et al. (2024). Each item was presented with the stem ``Reflecting on your idea generation session, to what extent do you agree with the following statement?'' and answered on a five-point agreement scale (strongly disagree to strongly agree). The statements referred to (a) the ideas the participant personally proposed during Subtask 1  ``The ideas I contributed were..."); (b) the ideas generated by the AI assistant during Subtask 2 (``The ideas generated by AI were..."); and (c) the ideas their group authored during Subtask 2 (``Overall, the ideas created by the team were..."). \rev{Throughout the paper we refer to this third rated set as \emph{group-authored} ideas; the questionnaire item itself used the word ``team.''} Each statement was completed, per dimension, with ``relevant to the problem", ``innovative—reflected original and creative ways to think of the problem", or ``insightful—reflected deep and perceptive ways to think of the problem". The third set comprises elaborations on the pooled human and AI ideas and new ideas composed while working with the AI; it contains no unmodified AI output and no Subtask 1 ideas, so the three sets are mutually exclusive. The complete questionnaire is included in the Appendix, 
Table~\ref{tab:questionnaire}.

\subsection{Participants}
We report on data from 42 undergraduate students, who were recruited 
by email (no male participants) to participate in the study in groups of three (14 groups). Participants received a \$30 gift card.

We chose to focus on all-women groups to minimize gender-related variability and allow a more controlled examination of how the configuration influences group ideation. This decision aligns with the rising participation of women in the remote workforce~\citep{owid_female_labor_supply,barrero2023evolution, buckman2025measuring, aksoy2025remote}. 

\subsection{Data Analysis}
We followed a mixed-methods workflow that combined statistical analysis, computational text analytics, and qualitative techniques. 

\subsubsection{Data Cleaning and Preparation:}
We conducted 19 study sessions with 52 participants. 
Five sessions were removed due to corrupted video recordings, no-show participants, or a combination of these two factors. Only sessions that included three participants were used, resulting in seven Platform Configuration 1 (Miro) and seven Platform Configuration 2 (Noda) sessions for a total of 14 study sessions with 14 x 3 = 42 participants. 

We curated a corpus of ideas generated during the study. The final corpus comprised 1,260 distinct ideas exported from the Miro boards and Noda room logs. Each entry in the dataset was encoded with the following fields: idea id, group id, configuration, source (Human/AI/Group-authored), text of the idea, and word count of idea text.
Table \ref{tab:idea_dataset_curation} provides an overview of this corpus.

\begin{table*}[]
\centering
\begin{tabular}{|p{2cm}|p{7cm}|p{7cm}|p{1cm}|}
\hline
\multicolumn{1}{c}{Source} &
  \multicolumn{1}{c}{\textbf{Origin \& Timing}} &
  \multicolumn{1}{c}{\textbf{Curation Workflow}} &
  \multicolumn{1}{c}{\textbf{n}} \\ \hline
Human &
  Entered by individual participants during Subtask 1 (human-only divergence); snapshot frozen at the end of Subtask 1. &
  Deduplicated identical notes; prompt-label nodes removed; word count recorded. &
  308 \\ \hline
AI &
  Inserted by the platform's AI feature during Subtask 2 (AI-supported divergence) when a participant invoked it ("Expand with ideas" in Miro; the "brainstorm" node in Noda); snapshot frozen at the end of Subtask 2. &
 Formatting artifacts stripped (for example, Miro's leading asterisks); word count recorded. &
  937 \\ \hline
  Group-authored &
Distinct from both Human and AI categories. Authored or edited by group members during Subtask 2 after reviewing the pooled human and AI ideas; snapshot frozen at the end of Subtask 2. May draw on AI output, but contains no unmodified AI text. &
Word-count recorded. & 15 \\ \hline
\end{tabular}
\caption{Idea corpus by source: origin, curation, and counts of distinct ideas exported from the Miro boards and Noda room logs (N = 1,260).}
\label{tab:idea_dataset_curation}
\end{table*}

\subsubsection{Statistical Analysis} 

\rev{Statistical analyses were conducted in R (version 4.3.3) using base functions only; the analysis scripts, their full output, and the underlying data are provided as supplementary material. The quantitative variables analyzed include session time, the number of ideas generated, idea length, and post-task questionnaire responses (see Appendix – Supplementary Tables). 
}
 
\rev{All questionnaire measures are single Likert items. Shapiro-Wilk tests (\texttt{shapiro.test}) 
indicated departures from normality for most measures in at least one platform. We report parametric tests where the submission did, and nonparametric checks agreed with every reported significance decision.}
 
\rev{To compare perceptions of the three rated idea sets within each platform (own, AI-generated, group-authored), we used one-way repeated-measures ANOVA, fitted as a multivariate linear model on the three responses per participant (\texttt{lm}, \texttt{anova.mlm}). Sphericity was assessed with Mauchly's test (\texttt{mauchly.test}); it indicated violations for relevance ($W = 0.530$, $p = .002$) and innovation ($W = 0.668$, $p = .022$) in the Noda configuration, where Greenhouse-Geisser corrections were applied ($\varepsilon = 0.680$ and $0.751$) and corrected $p$ values remained below $.001$. No violations were indicated in the Miro condition. Pairwise comparisons are paired t-tests (\texttt{t.test}, df~=~20) with Bonferroni correction over the three comparisons per dimension; unlike comparisons based on a pooled error term, these do not depend on the sphericity assumption the Noda-configuration data violate. The method is stated in the table notes.}
 
\rev{To compare perceptions of the same rated set across platforms, we used independent-samples t-tests with pooled variance (\texttt{t.test} with \texttt{var.equal = TRUE}, df~=~40) and report Cohen's $d$ computed with the pooled standard deviation. Welch tests and Mann-Whitney tests were run as sensitivity checks and agreed with all reported significance decisions except AI-generated innovation, where the t-test gave p = .050 and the Mann-Whitney test p = .046; we report this comparison as marginal. Comparisons of individual workload, presence, and collaboration items between platforms used Mann-Whitney tests (\texttt{wilcox.test}); reported $U$ statistics take the Miro group first, and all $p$ values are two-tailed. Item-level results for these measures are reported as medians and interquartile ranges (Tables~A.1 to~A.3), matching the rank-based tests; measures analyzed with ANOVA and t-tests are reported as means and standard deviations. Session time was compared at the group level ($n = 7$ per platform) with the same tests. The significance tests for term frequencies are described under Computational Text Analytics.}
 
\rev{Not all tests include all 42 participants. Four participants omitted presence items: Miro Participant 32 answered only the first presence item; Miro Participant 1 omitted four items; Miro Participant 20 omitted two items; and Noda Participant 51 omitted the item ``I had a sense of acting in the virtual space, rather than operating something from outside.'' Analyses use all available responses per item, and per-item Ns are reported in the supplementary tables. Workload, collaboration, and idea-quality items had no missing responses ($N = 42$).}
\subsubsection{Qualitative Analysis}

\rev{We conducted a thematic analysis~\citep{braun2012thematic} of the open-text post-task responses, using a codebook developed collaboratively by two researchers. The analysis proceeded in five stages. First, both researchers independently familiarized themselves with all responses, attending to explicit content and latent meaning, and generated initial codes for each question. Second, they compared their codes and consolidated them into a shared codebook per question, resolving disagreements by returning to the original data, merging overlapping codes, and eliminating redundant ones. Third, each researcher independently applied the consolidated codebook to the full dataset. Fourth, the researchers collaboratively identified themes, defined as patterns of shared meaning relevant to the research questions, and refined them iteratively until both reached agreement. Fifth, inter-coder reliability of the independent coding pass was assessed as percentage agreement, ranging from 85.6\% to 99.0\% across the six open-text questions (challenges and positive experiences of team interaction, AI contribution and hindrance to idea generation and to teamwork, idea-source preference, and platform helpfulness; Supplementary Tables ~B.1 to B.6 give the full wording). All codebooks are included in the Appendix.}
 
\subsection{Video Analysis}
\label{ref:Svideo}
\rev{Two researchers independently analyzed the video recordings of all 14 sessions (screen and audio recordings for the Miro platform; recordings captured within the virtual environment for the immersive platform), coding each subtask for spatial arrangement, movement, and collaboration patterns. The codebook is provided in Table \ref{tab:codebook-video}.}
 
\rev{The spatial codes were operationalized per platform. Participants were coded as working at short distance if positioned next to one another: in the Noda platform, an avatar needing only to turn in place to see the other avatars; in the Miro platform, mind maps and cursors appearing together in the viewport. They were coded as at long distance when reaching another participant's work required teleporting (Noda) or when cursors and maps fell outside the viewport (Miro). Orientation codes (facing toward or away from other participants) apply only to the Noda platform, where avatars have a facing direction; the planar Miro platform offers no analogous cue. Collaboration patterns in Subtask~3 were coded from role distribution: groups in which one or two participants acted as navigators (providing verbal or written suggestions) while the others acted as drivers (controlling the platform, implementing ideas, or leading the discussion) were coded as unequal collaboration, and groups in which all three participants alternated between these roles, contributing both written and verbal input, were coded as equal collaboration.}
 
\rev{These operationalizations capture analogous but not identical constructs across the two platforms. In particular, spatial separation in the Miro platform partly reflects the canvas area required for text, so cross-platform comparisons of distance codes are reported descriptively. The equal and unequal collaboration classifications rest on the role-distribution codes, which do not depend on spatial readings.}
 
\rev{Inter-coder reliability, assessed as percentage agreement, was 100\%, 89.64\%, and 85.87\% for Subtasks 1, 2, and 3, respectively.}

\subsubsection{Computational Text Analytics} 
\label{sec:da-lvs}
 We used a divergence-based method for identifying distinctive and underrepresented terms, implemented via Learning via Surprisability (LvS)~\citep{mokryn2025interpretable,mokryn2021domain}. LvS identifies surprising terms in each category by comparing the term distribution in that category with a reference distribution derived from the dataset, using a decomposition of Jensen--Shannon divergence. This enables the identification of both overrepresented terms and underrepresented or absent terms. Importantly, underrepresented or absent terms are not simply rare overall: they are terms that are relatively common in the broader dataset but appear much less frequently, or not at all, in a specific category~\citep{benshoshan2026making}. In our framework, these absences are informative because they reveal ideas that would normally be expected but are missing from a particular category. Here, we refer to idea-level appearance as one occurrence, regardless of the number of times the term is mentioned in that idea. 

\paragraph{LvS Data Preparation } For each idea we normalized the text (NFKC), removed leading cost markers
of the form ``(high)'', ``(medium)'', or ``(low)'', lowercased, tokenized,
dropped possessive suffixes, removed the 318 English stop words in the
scikit-learn list, and reduced the remaining tokens to stems with the
Porter stemmer. Each idea contributes at most one count per stem,
regardless of how often the stem occurs in it. From these idea-level
counts we formed a normalized presence distribution for each category and
a reference distribution equal to the unweighted mean of the two category
distributions.

\paragraph{LvS Operation } 
Following \cite{mokryn2025interpretable} and \cite{mokryn2021domain}, each stem's signed contribution to the
Jensen--Shannon divergence between a category and the reference was
computed with the reference implementation
(\url{https://github.com/ScanLab-ossi/LvS-for-Datasets}). A stem entered a
category's signature if its absolute contribution was at least 1\% of that
category's total absolute divergence; stems meeting the threshold for
either category were assigned to the category with the larger signed
contribution. Tables 8 and 9 print stems as their most frequent surface
form; \emph{organ} covers organize and organization, and \emph{import}
covers importance.

\paragraph{Significance} 
Significance was assessed with Pearson's chi-square statistic (one degree of
freedom, no continuity correction) when both groups had adequately populated
counts, and with a two-sided Fisher exact test otherwise. Fisher's test was
used whenever a term appeared in fewer than five ideas in either group or
whenever any expected cell count fell below five; chi-square was used only
when neither condition applied. All reported $p$ values are two-tailed.
Resulting $p$ values were corrected with the Benjamini--Hochberg
false-discovery-rate procedure, with the correction family limited to the
terms that passed the LvS filter for that comparison, since the filter had
already narrowed the candidate set before testing began. Odds ratios for
Fisher tests are the conditional maximum-likelihood estimates returned by
\texttt{fisher.test}. Signature extraction was implemented in Python~3
(pandas, NLTK, scikit-learn) using the reference LvS implementation;
significance tests used base R (version 4.3.3: \texttt{chisq.test},
\texttt{fisher.test}, \texttt{p.adjust}). All scripts and their output are
included in the supplementary bundle.

\section{Results}

\subsection{RQ1: Shaping interaction patterns, attention distribution, and perceived collaboration}

To answer RQ1, we examined: (a) collaboration strategies; (b) participants’ perceived workload (NASA-TLX) and sense of presence; and (c) participants’ perceptions of the collaborative process. 

\subsubsection{Collaborative Strategies}
Here we describe the findings from the video analysis. See Appendix Table~\ref{tab:codebook-video} for the video analysis code book.

During subtask 1, which involved individual work, participants had no restrictions regarding movement or the placement of their maps on either platform. In this subtask, the majority of groups (12/14) were observed working at long distance from one another, as defined in the Video Analysis section. 

In Miro, participants tended to build their maps away from each other on different areas of the shared whiteboard. In Noda, participants usually worked in separate corners of the virtual room. Only in two groups (N1 and N7) did participants work within a short distance of each other. 
In addition, a notable behavior was observed in all but one Noda group (N3) participants’ virtual avatars \textit{faced away} from each other.  Since Noda mind maps are designed to appear as if hanging in mid-air, it is sensible for users to turn away from their colleagues so as to have a relatively clear background for their own work.  

Movement during subtask 1 in both platforms was minimal. In Miro, six of the seven groups found a space on the board and did not adjust or shift their mind map during the task. 
Within the Noda platform, similar patterns were exhibited, where in six of seven groups, participants stayed in the spot they had spawned in and did not move, other than to turn their heads. 
\\

\begin{figure*}[t]
  \centering
  \includegraphics[width=\textwidth]{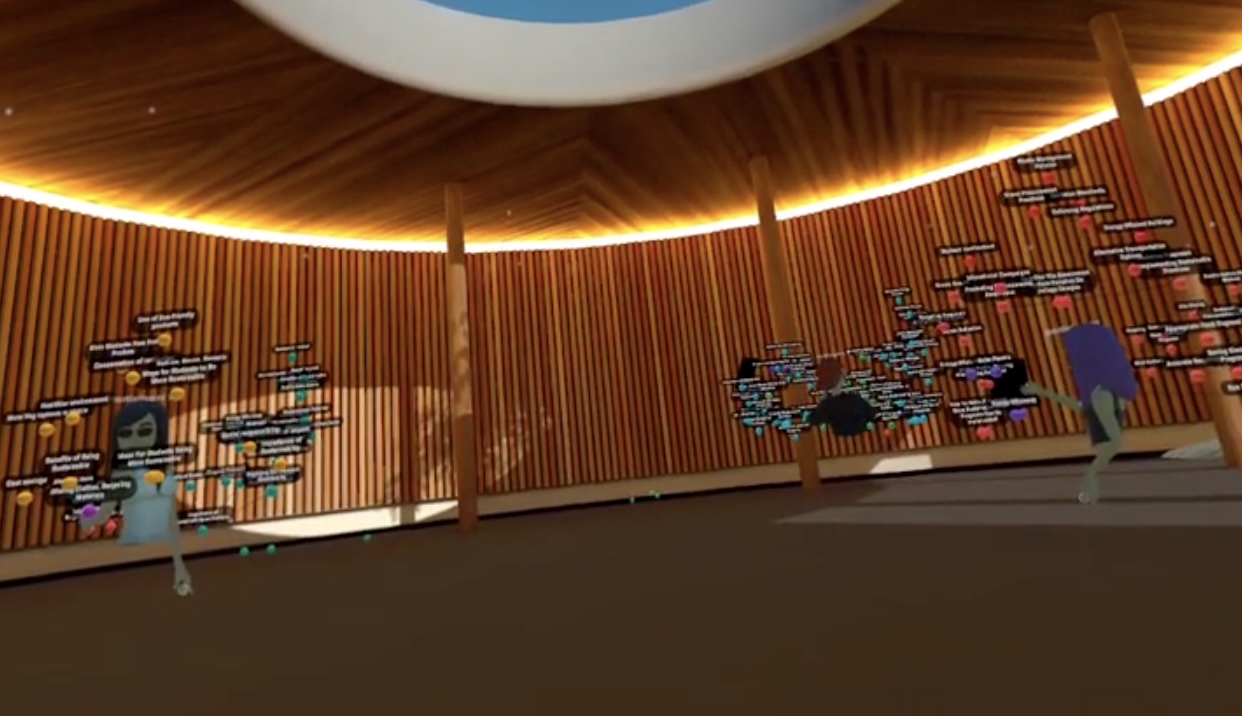}
  \caption{The immersive Noda workspace at the end of Subtask 2, captured from the viewpoint of the researcher who hosted the shared room. All three participants are in frame, each at a separate idea cluster: one stands at the left cluster near the wall, facing into the room; a second stands inside the large central cluster with its back to the viewer, partly hidden by its own nodes; the third, in the purple shirt at right, is turned toward the right-hand cluster. The central and right participants are within a few steps of each other, while the left participant is at the far side of the room; even the two nearer participants are oriented toward their own clusters, not toward each other. The three clusters, expanded by AI output, occupy distinct regions of the room, and no participant faces another. This is the long-distance, not-facing arrangement coded in the video analysis (Table~\ref{tab:codebook-video}). Node labels are not intended to be legible at this scale.}
  \label{fig:noda-arrangement}
\end{figure*}

In \textit{subtask 2}, which introduced AI, participants were again not given instructions on how to interact with each other. Similar to subtask 1, a majority of participants, regardless of condition, worked individually. 
Among the Noda groups, 4/7 still had participants facing away from each other during the task.
Figure~\ref{fig:noda-arrangement} shows this arrangement in one group at the end of the subtask.

In the Miro platform, the addition of the AI did not lead to positional changes in most groups, and 6 of the 7 groups continued developing their individual mind maps from Subtask 1. However, one Miro group (M1) moved to a central area of the board to collaboratively create a new map using AI-generated ideas. A similar pattern was observed among the Noda groups. Six out of seven groups maintained work on their original maps from subtask 1, while in one group (N7), a participant moved to a new location within close proximity to another participant to create a new AI-augmented mind map.

As participants familiarized themselves with the Miro or Noda platforms during subtask 2, more participants would visit or move closer to the maps of other participants. This increased movement to check the mind maps of other participants did not seem to be collaborative, as it involved a participant staying near another participant's mind map for a few seconds before returning to their own mind map. 
\\

In \textit{Subtask 3}, participants were instructed to work collaboratively. Although all groups demonstrated some level of collaboration, they differed in how evenly responsibilities were distributed across the three group members. Groups were coded as showing equal or unequal collaboration from the distribution of navigator and driver roles, as defined in the Video Analysis section.



In the Miro platform, 3 of the 7 groups demonstrated equal collaboration during \textit{Subtask 3}, with participants sharing both navigation and driving responsibilities. The remaining 4 groups showed unequal collaboration, with one or two participants taking the driver role. 

In the Noda platform, 5 of the 7 groups demonstrated equal collaboration, with all participants contributing ideas verbally and taking turns modifying the mind map. In these five groups, participants also tended to face one another when marking their final three choices. During discussion, participants generally faced the current speaker, except when the speaker briefly turned away to consult their notes. The two groups that did not exhibit equal navigation and driving (N3 and N4) worked at greater distances from one another, and in both groups the shared map was modified primarily by a single participant.

Compared with \textit{Subtask 2}, more participants in both platforms moved around during \textit{Subtask 3}. In the Miro platform, 4 of the 7 groups showed substantial movement, with participants creating entirely new mind maps for the three selected ideas. The remaining three groups also showed some movement, primarily while marking the selected ideas with sticky notes. In the Noda platform, 5 of the 7 groups demonstrated substantial movement. In these five groups (which were also the ones coded for collaborating equally), participants also tended to remain closer to one another. Two movement patterns were observed: participants gathering in the center of the room and participants moving between one another’s mind maps.

\subsubsection{Workload (NASA-TLX) and Presence}


We measured \textit{workload} using the NASA-TLX. 
The only statistically significant difference between the two platforms was observed for \textit{Physical Demand} (\textit{How physically demanding was the task}). A Mann--Whitney U test indicated that participants in the Miro platform (median = 1.0) reported significantly less physical demand than participants in the Noda platform (median = 1.9, U = 126.0, p = 0.015). The full results are presented in Table \ref{tab:tlx-appendix}.


Participant perceptions of \emph{presence} differed between the two platforms. \rev{Participants in the immersive Noda platform reported significantly higher ratings than those in the planar Miro platform on four statements: ``In the computer generated world I had a sense of `being there.' '' (Mdn = 6 vs.\ 4, $p < .001$), one of the largest differences in the questionnaire, ``Somehow I felt that the virtual world surrounded me'' ($p < .001$), ``I had a sense of acting in the virtual space, rather than operating something from outside'' ($p = .010$), and ``I felt present in the virtual space'' ($p = .003$). Participants in the planar Miro platform rated two statements significantly higher: ``I felt like I was just perceiving pictures'' ($p < .001$), an item on which higher agreement indicates lower presence, and ``The virtual world seemed more realistic than the real world'' ($p = .002$). Two further items differed significantly in directions that likely reflect how the MEC-SPQ items read on a planar display: participants in Noda reported greater awareness of the real world ($p = .017$), and one of the two ``How real did the virtual world seem to you?'' items was rated higher in Miro ($p = .039$). We return to this asymmetry of instrument fit in the Limitations. Full item-level results, with per-item sample sizes and the remaining six nonsignificant items, are reported in Table~\ref{tab:presence-appendix} in the Appendix.

\subsubsection{Participant Perceptions of Collaboration}
Participants in the immersive Noda platform reported slightly higher levels of comfort with the collaborative process than those in the planar Miro platform, although most differences were not statistically significant. One exception was comfort with feedback (\textit{Did you feel comfortable providing feedback to your team members and receiving feedback from them?}). Participants in the Noda platform (median = 5) reported greater comfort in giving and receiving feedback than participants in the Miro platform (median = 4), this difference was statistically significant (Mann--Whitney $U = 148$, $p = 0.047$). Full results are reported in Table~\ref{tab:collab-appendix} in the Appendix.


Participants were asked two open-ended questions about the challenges and positive experiences they encountered while interacting with their team (see Questions Q19 and Q21 in Table~\ref{tab:questionnaire} in the Appendix). The main challenges they reported were a perceived sense of disconnection from teammates, technical issues, and feelings of unfamiliarity with one another or with the platform. Six participants reported no negative experiences of team interaction (4 in the Miro platform and 2 in the Noda platform). Approximately one quarter of participants (11/42; 5 in Miro and 6 in Noda) described the lack of face-to-face communication as challenging. 
In the Miro platform, participants could see only one another’s cursors, whereas in the Noda platform, each participant was represented by a generic,explicitly unexpressive avatar with no resemblance to their real-world appearance. One participant in the Noda platform noted that it was ``odd interacting without being able to perceive the team's body language or facial expressions''. Another form of disconnection was a perceived sense of virtual distance, which was reported only by participants in the 
Noda platform (5/42). These participants felt that the virtual environment negatively affected their spatial awareness. As one participant explained, ``it just felt like we were very disconnected from each other as we were very far away in the virtual space''.

In contrast, participants reported positive experiences related to compatibility and respectful discourse with their teammates. More than half of the participants (24/42) described their group interactions as compatible and respectful. Of these participants, 14 were in the Miro platform and 10 were in the Noda platform. With respect to verbal interaction, 12 of the 42 participants attributed their positive experience to a general sense of agreement among teammates (7 in Miro and 5 in Noda).

Another positive experience that several participants noted was related to the tasks and the platform used. Some participants (8/42) found that the tasks gave them new insight and perspectives (4 in Miro, and 4 in Noda). Some participants in the Noda platform (4/21) mentioned positively the ability to spatially view each other's ideas as they were brainstorming. One such participant recalls, ``It was cool to see the brainstorm nodes like from a different perspective as the other person was working on theirs''.

\subsection{RQ2: collective evaluation, negotiation, and integration of AI-generated ideas across phases}

\subsubsection{Contribution of the AI during the divergent phase of ideation across platforms}
The divergent phase of ideation consisted of two subtasks. In the first (\textit{Subtask 1}), participants generated human ideas only. In the second (\textit{Subtask 2}), they engaged in AI-supported idea generation by prompting the large language model assistant for additional ideas. Participants then reviewed the AI-generated ideas and added new group-authored ideas.
\begin{table}[t]
    \centering
    \caption{Divergence phase of the ideation process: number of Ideas by Collaborative Platform and Source Type}
    \begin{tabular}{lcccc}
        \toprule
        \textbf{Platform} & \textbf{Human Ideas} & \textbf{AI Ideas} & \textbf{Group-authored Ideas} & \textbf{Total} \\
        \midrule
        Miro & 153 & 304 & 7 & 464 \\
        Noda  & 155 & 633 & 8 & 796 \\
        \bottomrule
    \end{tabular}
    \label{tab:ideas_count}
\end{table}
Table \ref{tab:ideas_count} shows the idea counts for Subtask 1 (human-only ideas) and Subtask 2 (AI ideas and group-authored ideas). In both platforms, the AI contributed more ideas than the participants. However, this imbalance was much greater in the Noda platform than in the Miro platform, resulting in a substantially higher total number of ideas in Noda.

\begin{table*}[ht]
    \centering
    \caption{Idea‐length (in number of words) statistics and cross-platform effect sizes}
    \begin{tabular}{lccc}
        \toprule
        \textbf{Type} &
        \textbf{Miro: Mean (SD)} &
        \textbf{Noda: Mean (SD)} &
        \textbf{Cohen’s $d$ (Noda vs Miro)} \\
        \midrule
        AI ideas    & 21.41 (6.37) & 2.88 (1.70) & $-4.77$ \\
        Human ideas & 5.03 (5.16)  & 3.50 (2.08) & $-0.39$ \\
        \bottomrule
    \end{tabular}
    \label{tab:sentencelen}
\end{table*}
Table~\ref{tab:sentencelen} compares the average length of ideas produced in Miro and Noda platforms for both AI-generated and human-generated ideas. The difference for AI-generated ideas is particularly striking: AI ideas in Miro averaged 21.4 words ($SD = 6.4$), whereas AI ideas in Noda averaged only 2.9 words ($SD = 1.7$). The cross-platform effect size, measured by Cohen's $d$, was extremely large ($d = -4.77$), indicating very little overlap between the two distributions. 

Human-generated ideas showed a much smaller cross-platform difference. Participants in the Miro platform produced ideas averaging 5.0 words ($SD = 5.2$), compared with 3.5 words in the Noda platform ($SD = 2.1$), corresponding to a small effect size as measured by Cohen's $d$ ($d = -0.39$). Thus, the length of human-generated ideas was broadly comparable across the two platforms, whereas the length of AI-generated ideas differed dramatically. This pattern suggests that platform design may moderate the AI's contribution more strongly than it shapes human idea expression.


\subsubsection{Participant Perceptions of Quality of Ideas and Ideation Process across platforms} 
\rev{The question this section addresses is whether ideation with the AI available improved on human-only ideation, as judged by participants, and how that judgment differed across configurations.} To assess idea quality, we asked participants to rate, on relevance, innovation, and insightfulness, (a) the ideas they personally proposed during the human-only phase; (b) the set of ideas generated by the AI feature; and (c) the set of ideas their group authored during the AI-supported phase. These rating scales draw on the idea-evaluation framework of \citet{ShaerBrainwriting24}}, which operationalizes idea quality along the dimensions of relevance (how directly an idea addresses the stated problem), innovation (the degree to which an idea is novel), and insightfulness (the extent to which an idea reflects deep conceptual understanding).

We first conduct an in-depth analysis of how participants \textit{within} each collaborative platform evaluated their own ideas, the AI ideas, and group authored ideas.

\begin{table*}[htbp]
\centering
\caption[Idea quality perceptions, Miro platform]{\rev{Within-platform comparison of idea quality perceptions in the Miro platform (repeated-measures ANOVA, $N = 21$). Participants rated three mutually exclusive sets of ideas on five-point agreement scales: their own ideas from the human-only phase (Subtask 1), the ideas generated by the platform's AI feature (Subtask 2), and the ideas their group authored during the AI-supported phase (Subtask 2); see Operational Definitions, Table~1. Pairwise comparisons are Bonferroni-corrected paired t-tests (df = 20).}}
\label{tab:miro_anova}
\begin{tabular}{@{}llcccccll@{}}
\toprule
Quality Dimension & Idea Source & Mean & SD & F & df & p & $\eta^2_p$ & Pairwise Comparisons \\
\midrule
\multirow{3}{*}{\textbf{Relevance}} & Human & 4.33 & 0.73 & \multirow{3}{*}{2.30} & \multirow{3}{*}{2,40} & \multirow{3}{*}{.113} & \multirow{3}{*}{0.103} & \multirow{3}{*}{No significant differences} \\
 & AI & 4.24 & 0.77 & & & & & \\
 & Group-authored (Team) & 4.62 & 0.50 & & & & & \\
\midrule
\multirow{3}{*}{\textbf{innovation}} & Human & 3.38 & 1.02 & \multirow{3}{*}{4.02} & \multirow{3}{*}{2,40} & \multirow{3}{*}{.026*} & \multirow{3}{*}{0.167} & \multirow{3}{*}{\begin{tabular}[c]{@{}l@{}} Group-authored $>$ AI (p $<$ .05);\\ 
\end{tabular}} \\
 & AI & 3.38 & 1.20 & & & & & \\
 & Group-authored (Team) & 3.95 & 0.86 & & & & & \\
\midrule
\multirow{3}{*}{\textbf{Insightfulness}} & Human & 3.43 & 0.98 & \multirow{3}{*}{4.00} & \multirow{3}{*}{2,40} & \multirow{3}{*}{.026*} & \multirow{3}{*}{0.167} & \multirow{3}{*}{\begin{tabular}[c]{@{}l@{}}Group-authored $>$ Human (p $<$ .01)\\ \end{tabular}} \\
 & AI & 3.76 & 1.14 & & & & & \\
 & Group-authored (Team) & 4.10 & 0.83 & & & & & \\
\bottomrule
\end{tabular}
\begin{tablenotes}
\small
\item {\textit{Note:} *** $p < .001$, ** $p < .01$, * $p < .05$. $\eta^2_p$ = partial eta squared. Mauchly's test indicated no sphericity violations; uncorrected tests are reported.}
\end{tablenotes}
\end{table*}
Table~\ref{tab:miro_anova} presents the results of a repeated-measures ANOVA examining Miro participants' evaluations of the quality of the generated ideas.
Overall, participants found most ideas relevant, regardless of their source. However, they considered the group-authored ideas more innovative than AI-generated ideas, and more insightful than (their own) human ideas. There was no statistically significant difference in the quality of human vs. AI-generated ideas.

\begin{table*}[htbp]
\centering
\caption[Idea quality perceptions, Noda platform]{\rev{Within-platform comparison of idea quality perceptions in the Noda platform (repeated-measures ANOVA, $N = 21$). Participants rated three mutually exclusive sets of ideas on five-point agreement scales: their own ideas from the human-only phase (Subtask 1), the ideas generated by the platform's AI feature (Subtask 2), and the ideas their group authored during the AI-supported phase (Subtask 2); see Operational Definitions, Table~1. Pairwise comparisons are Bonferroni-corrected paired t-tests (df = 20).}}
\label{tab:noda_anova}
\begin{tabular}{@{}llcccccll@{}}
\toprule
Quality Dimension & Idea Source & Mean & SD & F & df & p & $\eta^2_p$ & Pairwise Comparisons \\
\midrule
\multirow{3}{*}{\textbf{Relevance}} & Human & 4.62 & 0.50 & \multirow{3}{*}{15.08} & \multirow{3}{*}{2,40} & \multirow{3}{*}{$<$.001***} & \multirow{3}{*}{0.430} & \multirow{3}{*}{\begin{tabular}[c]{@{}l@{}}Human $>$ AI (p $<$ .05);\\Group-authored $>$ Human (p $<$ .05);\\Group-authored $>$ AI (p $<$ .001) \end{tabular}} \\
 & AI & 3.95 & 0.92 & & & & & \\
& Group-authored (Team) & 4.95 & 0.22 & & & & & \\
\midrule
\multirow{3}{*}{\textbf{innovation}} & Human & 3.52 & 0.93 & \multirow{3}{*}{23.00} & \multirow{3}{*}{2,40} & \multirow{3}{*}{$<$.001***} & \multirow{3}{*}{0.535} & \multirow{3}{*}{\begin{tabular}[c]{@{}l@{}}Human $>$ AI (p $<$ .05);\\ Group-authored $>$ Human (p $<$ .01);\\ Group-authored $>$ AI (p $<$ .001) \end{tabular}} \\
  & AI & 2.62 & 1.24 & & & & & \\
  & Group-authored (Team) & 4.29 & 0.56 & & & & & \\
\midrule
\multirow{3}{*}{\textbf{Insightfulness}} & Human & 3.71 & 1.01 & \multirow{3}{*}{23.94} & \multirow{3}{*}{2,40} & \multirow{3}{*}{$<$.001***} & \multirow{3}{*}{0.545} & \multirow{3}{*}{\begin{tabular}[c]{@{}l@{}}Human $>$ AI (p $<$.01);\\ Group-authored $>$ AI (p $<$ .001);\\ \end{tabular}} \\
 & AI & 2.57 & 1.29 & & & & & \\
 & Group-authored (Team) & 4.24 & 0.62 & & & & & \\
\bottomrule
\end{tabular}
\begin{tablenotes}
\small
\item {\textit{Note:} *** $p < .001$, ** $p < .01$, * $p < .05$. $\eta^2_p$ = partial eta squared. Mauchly's test indicated sphericity violations for relevance and innovation; Greenhouse-Geisser corrections were applied ($\varepsilon = 0.680$ and $0.751$) and corrected $p$ values remained below .001.}
\end{tablenotes}
\end{table*}
In the Noda platform (Table~\ref{tab:noda_anova}), AI-generated ideas were rated significantly lower than both participants' own ideas and group-authored ideas on all three dimensions.

Specifically, for relevance, both group-authored (Team) and human-generated ideas were rated significantly higher than AI-generated ideas, with group-authored ideas receiving the highest ratings. For innovation, participants rated group-authored ideas significantly higher than both human-generated and AI-generated ideas, and human-generated ideas were also rated significantly higher than AI-generated ideas. For insightfulness: both group-authored ideas and human-generated ideas received higher ratings than AI-generated ideas, but group-authored ideas were not rated higher than human-generated ideas. 


We continue to evaluate how perceptions of the same type of ideas varied \textit{across} collaborative platforms.  Table~\ref{tab:idea_quality} directly compares how ideas from the same source (human, AI, or group-authored) were perceived across Miro and Noda platforms.

\begin{table*}[htbp]
\centering
\caption{Participant Perceptions of Idea Quality by Platform (Independent-Samples t-Test Results)}
\label{tab:idea_quality}
\begin{tabular}{llcccccccc}
\hline
\textbf{Idea Quality Measure} & \textbf{Platform} & \textbf{N} & \textbf{Mean} & \textbf{SD} & \textbf{t} & \textbf{df} & \textbf{p} & \textbf{95\% CI} & \textbf{Cohen's d} \\
\hline
\multicolumn{10}{l}{\textbf{Human-generated ideas}} \\
\hline
Relevance & Miro & 21 & 4.33 & 0.73 & -1.48 & 40 & 0.146 & [-0.68, 0.10] & -0.46 \\
 & Noda & 21 & 4.62 & 0.50 & & & & & \\
\hline
innovation & Miro & 21 & 3.38 & 1.02 & -0.47 & 40 & 0.638 & [-0.75, 0.47] & -0.15 \\
 & Noda & 21 & 3.52 & 0.93 & & & & & \\
\hline
Insightfulness & Miro & 21 & 3.43 & 0.98 & -0.93 & 40 & 0.357 & [-0.91, 0.33] & -0.29 \\
 & Noda & 21 & 3.71 & 1.01 & & & & & \\
\hline
\multicolumn{10}{l}{\textbf{AI-generated ideas}} \\
\hline
Relevance & Miro & 21 & 4.24 & 0.77 & 1.09 & 40 & 0.281 & [-0.24, 0.82] & 0.34 \\
 & Noda & 21 & 3.95 & 0.92 & & & & & \\
\hline
innovation & Miro & 21 & 3.38 & 1.20 & 2.02 & 40 & 0.050 & [-0.00, 1.53] & 0.62 \\
 & Noda & 21 & 2.62 & 1.24 & & & & & \\
\hline
Insightfulness & Miro & 21 & 3.76 & 1.14 & 3.18 & 40 & 0.003** & [0.43, 1.95] & 0.98 \\
 & Noda & 21 & 2.57 & 1.29 & & & & & \\
\hline
\multicolumn{10}{l}{\textbf{Group-authored (Team) ideas}} \\
\hline
Relevance & Miro & 21 & 4.62 & 0.50 & -2.81 & 40 & 0.008** & [-0.57, -0.09] & -0.87 \\
 & Noda & 21 & 4.95 & 0.22 & & & & & \\
\hline
innovation & Miro & 21 & 3.95 & 0.86 & -1.48 & 40 & 0.146 & [-0.79, 0.12] & -0.46 \\
 & Noda & 21 & 4.29 & 0.56 & & & & & \\
\hline
Insightfulness & Miro & 21 & 4.10 & 0.83 & -0.63 & 40 & 0.532 & [-0.60, 0.32] & -0.19 \\
 & Noda & 21 & 4.24 & 0.63 & & & & & \\
\hline
\end{tabular}
\begin{flushleft}
\textit{Note}: *p $<$ .05, **p $<$ .01. Cohen's d interpretation: small effect (0.2), medium effect (0.5), large effect (0.8). Positive Cohen's $d$ values indicate higher ratings in the Miro platform, whereas negative values indicate higher ratings in the Noda platform. Pooled-variance t-tests (df = 40). The three rated sets are mutually exclusive (see Operational Definitions). For AI-generated innovation, p = .050 at the boundary.
\end{flushleft}
\end{table*}
As shown in Table~\ref{tab:idea_quality}, group-authored ideas were rated as significantly more relevant in the Noda platform than in the Miro platform, with a large effect size ($d = -0.87$). AI-generated ideas were rated as more insightful in the Miro platform (p = .003, d = 0.98). The difference in innovation was of medium size but marginal (p = .050, d = 0.62); the rank-based check reached significance (Mann-Whitney p = .046, Table A.4), the pooled-variance t-test did not. No significant cross-platform differences were observed for human-generated ideas.

\rev{Read against the human-only baseline, the AI-supported phase improved on it in both configurations. Participants rated the ideas their group authored with the AI available above the ideas they had proposed alone: on insightfulness in Miro (Table~\ref{tab:miro_anova}) and on relevance and innovation in Noda (Table~\ref{tab:noda_anova}). All 15 group-authored ideas were selected as finalists (Table~\ref{tab:origin}). The AI-supported phase also multiplied the idea pool, from 153 to 464 ideas in Miro and from 155 to 796 in Noda (Table~\ref{tab:ideas_count}). Participants in both configurations agreed that AI led the group to consider new ideas (Mdn = 5 in Miro, 4 in Noda) and raised the quantity of ideas (Mdn = 5 and 4); agreement that it raised quality was lower and did not differ between configurations (Mdn = 4 in both; Table~\ref{tab:quality-process-appendix}, Q22\_10 to Q22\_12). The design includes no human-only control for workload, so whether the improvement came at lower effort cannot be assessed from these data.}

\subsubsection{Semantic Differences Between Human- and AI-Generated Ideas Across Platforms}
\label{sec:lvs1}
To examine whether human- and AI-generated ideas differed in their lexical framing, we compared the idea-level prevalence of terms across both platforms. 
The pooled corpus comprised 937 AI-generated ideas and 308 human-generated ideas. We used LvS as an exploratory procedure to identify terms that contributed disproportionately to either source's lexical signature. 
Six stems accounted for at least 1\% of the divergence budget of the source they favored. All six also showed significant idea-level prevalence differences after Benjamini--Hochberg correction within this LvS-selected family (Table~\ref{tab:table8}).
 
\begin{table}[h]
\centering
\caption{LvS-selected words associated with AI- and human-generated ideas. Counts represent the number of ideas containing each word at least once. Words meeting the LvS threshold were subsequently tested for differences in idea-level prevalence. $N_{\text{AI}}=937$ and $N_{\text{Human}}=308$.}
\label{tab:table8}
\begin{tabular}{lrrlrc}
\toprule
Term & AI ideas & Human ideas & Test & Statistic & Sig. \\
\midrule
promote   & 68 & 0  & Fisher     & $OR=\infty$   & *** \\
practice  & 84 & 2  & Fisher     & $OR=15.051$   & *** \\
organ    & 64 & 1  & Fisher     & $OR=22.482$   & *** \\
host     & 29 & 0  & Fisher     & $OR=\infty$   & *** \\
sus       & 0  & 6  & Fisher     & $OR=0.000$    & *** \\
hall     & 19 & 16 & chi-square & $\chi^2=8.510$ & ** \\
\bottomrule
\end{tabular}
\par\vspace{2pt}
\parbox{0.9\linewidth}{\footnotesize Significance from Benjamini--Hochberg corrected $q$-values: * $q<0.05$, ** $q<0.01$, *** $q<0.001$.}
\end{table}
Table~\ref{tab:table8} shows the results. 
Four terms were favored in AI-generated ideas: \textit{promote}, \textit{practice}, \textit{organ}, and \textit{host}. The first three appeared in 68, 84, and 64 AI-generated ideas, respectively, but in no more than two human-generated ideas. The word \textit{host} appeared in 29 AI-generated ideas and no human-generated ideas. Collectively, these words are consistent with a more programmatic and action-oriented style in which ideas are framed around promoting practices, organizing initiatives, and hosting activities. 

Two terms were favored in human-generated ideas. The informal shorthand \textit{sus} (referring to the institute's student organization led sustainability task force) appeared in six human-generated ideas and no AI-generated ideas. The term \textit{hall} appeared in 16 of 308 human-generated ideas (5.2\%) and 19 of 937 AI-generated ideas (2.0\%), making it approximately 2.6 times as prevalent in human ideas. In context, \textit{hall} may reflect references to concrete campus locations. Together, these terms are consistent with human ideas being expressed in a more abbreviated and locally situated manner, while AI ideas more often adopted the language of organized programs and interventions.

\subsubsection{Semantic Differences Between Ideas Across Collaborative Platforms}
To evaluate the semantic differences between ideas across the two platforms, Miro and Noda, we repeated the analysis. 

 Pooling AI- and human-generated ideas within each platform (group-authored ideas excluded
throughout: 788 Noda ideas, 457 Miro ideas), fourteen terms clear the 1\% LvS threshold in
either platform's signature. Only five remain significant after Benjamini--Hochberg
correction within this 14-term family.
 
\begin{table}[h]
\centering
\caption{LvS-selected terms associated with Noda and Miro ideas. Counts represent the number of ideas containing each word at least once. Words meeting the LvS threshold were subsequently tested for differences in idea-level prevalence. Group-authored ideas were excluded; $N_{\text{Noda}}=788$ and $N_{\text{Miro}}=457$.}
\label{tab:table9}
\begin{tabular}{lrrlrc}
\toprule
Term & Noda ideas & Miro ideas & Test & Statistic & Sig. \\
\midrule
economy   & 11 & 0  & Fisher     & $OR=\infty$    & * \\
organ    & 3  & 62 & Fisher     & $OR=0.024$     & *** \\
encourage & 2  & 46 & Fisher     & $OR=0.023$     & *** \\
install   & 0  & 28 & Fisher     & $OR=0.000$     & *** \\
focus    & 0  & 21 & Fisher     & $OR=0.000$     & *** \\
\addlinespace
college   & 19 & 3  & Fisher     & $OR=3.736$     & \\
lake     & 7  & 0  & Fisher     & $OR=\infty$    & \\
example   & 7  & 0  & Fisher     & $OR=\infty$    & \\
limit    & 7  & 0  & Fisher     & $OR=\infty$    & \\
impact   & 30 & 11 & chi-square & $\chi^2=1.780$ & \\
donate    & 15 & 4  & Fisher     & $OR=2.196$     & \\
import   & 27 & 11 & chi-square & $\chi^2=1.016$ & \\
benefit  & 29 & 12 & chi-square & $\chi^2=1.010$ & \\
reduct   & 26 & 13 & chi-square & $\chi^2=0.197$ & \\
\bottomrule
\end{tabular}
\par\vspace{2pt}
\parbox{0.9\linewidth}{\footnotesize Significance from Benjamini--Hochberg corrected $q$-values: * $q<0.05$, ** $q<0.01$, *** $q<0.001$; blank = not significant.}
\end{table}
Table~\ref{tab:table9} depicts the results. After Benjamini--Hochberg correction, all four Miro-favored terms: \textit{organ}, \textit{encourage}, \textit{install}, and \textit{focus} remained significant ($q<.001$), whereas only one Noda-favored term, \textit{economy}, remained significant ($11/788$ Noda ideas versus $0/457$ Miro ideas; $q<.05$). None of the other nine LvS-selected Noda candidates survived correction.
The lexical comparison revealed a clearer platform-associated pattern for Miro than for Noda. Miro ideas disproportionately contained the terms organ, encourage, install, and focus. In context, these terms may reflect a more procedural and implementation-oriented style, in which ideas were expressed as organized programs, behavioral recommendations, concrete interventions, and explicitly framed priorities.
\subsubsection{Ideation Outcomes} 
All fourteen groups completed the experimental task; one Noda group (N4) did not record a medium-cost selection. At the end of each session (\textit{Subtask 3}), each group selected three projects to fund while meeting the requirements specified in the task instructions. Table~\ref{tab:ideas_count} presents the total number of ideas produced in each platform without AI (\textit{Subtask 1}), with AI, and authored as a group  (\textit{Subtask 2}) in each platform. The average time required to complete all three subtasks was approximately 27 minutes in both platforms. 

Table~\ref{tab:finalists} in the  Appendix presents the ideas selected by each group in \textit{Subtask 3}, during which participants chose three finalist ideas from the full set of generated ideas: one high-cost idea, one medium-cost idea, and one low-cost idea. 

Table~\ref{tab:origin} presents a summary of the outcomes of \textit{Subtask 3}. The source of each selected idea was categorized as human (a Subtask 1 participant-entered node), AI (an unmodified AI-generated node), or group-authored (a node the group composed during Subtask 2; see Operational Definitions). The table reports the distribution of these categories across the two platforms. In Miro, participants selected 10 AI-generated ideas and 4 human-generated ideas as finalists, whereas in Noda, participants selected 3 AI-generated ideas and 9 human-generated ideas as finalists. All 15 group-authored ideas in the corpus (Table 3) were selected as finalists, 7 of 7 in Miro and 8 of 8 in Noda.


\begin{table}[h]
    \centering
    \caption{\textbf{\small Origin of Selected Ideas - results of subtask 3}}
    \begin{tabular}{c|r|ccc}
    \hline
        && \textbf{AI} & \textbf{Human} & \textbf{Group-authored} \\ \hline
        \multirow{3}{*}{Miro} & High Cost  & 3 & 1 & 3 \\ 
         &Medium Cost & 3 & 2 & 2 \\ 
         &Low Cost & 4 & 1 & 2 \\ \hline
         \multicolumn{2}{c|}{\textbf{Miro Total}}   & \textbf{10} & \textbf{4} & \textbf{7}\\ \hline
        \multirow{3}{*}{Noda}&  High Cost  & 0 & 2 & 5 \\ 
        &Medium Cost & 3 & 2 & 1 \\ 
        &Low Cost & 0 & 5 & 2 \\ \hline
        \multicolumn{2}{c|}{\textbf{Noda Total}}   & \textbf{3} & \textbf{9} & \textbf{8}\\ \hline
    \end{tabular}
    \label{tab:origin}

\end{table}

\subsubsection{Group Awareness of AI Contributions}
\rev{In both configurations, the AI was invoked by individual participants on their own nodes. The video analysis records no instance in which a group invoked the feature jointly or discussed an invocation before it was made; in \textit{Subtask 2}, six of seven groups in each configuration continued on the individual maps they had started in \textit{Subtask 1}, and the one group in each that moved (M1 and N7) did so to build a new map from AI output. Participants' open-text responses to Q25 describe ``It seemed we mainly used AI features beforehand that we shared with each other, but did not use while collaborating,'' (P40, Noda), and ``I think because it was so easy, there wasn't a huge reason to collaborate when generating ideas,'' (P1, Miro). Where several participants invoked the feature at once, some participants reported that the result was hard to read, ``When we would all click the AI tool, a whole bunch of branches would appear making it hard to read,'' (P15, Miro). Full codes are given in Table~\ref{tab:codebook-q25}.}

\section{Discussion}
This study examines how different collaborative platforms affect two aspects of human–AI ideation: first, how they structure interaction, attention, and effort within groups (RQ1), and second, how they shape the evaluation and integration of AI-generated ideas (RQ2).
Across both research questions, the findings show that differences between platforms are driven by how spatial layout, interaction modality, and AI output format structure group interaction and AI use.

The Noda platform provides spatial co-presence, embodied interaction, and concise, spatially embedded AI suggestions. The Miro platform provides 
text-based interaction, and longer, more elaborated AI output.
In Noda, participants collaborate more evenly and report a stronger sense of presence, but also higher physical demand. The AI produces a larger number of short, spatially embedded suggestions. These suggestions are less often selected unaltered as finalists (3 of 20 in Noda, 10 of 21 in Miro; Table 10) and are rated lower 
than both human-generated and group-authored ideas. In Miro, interaction is more text-based. The AI produces longer suggestions, which are more frequently selected as finalists and are rated as more insightful.

These differences might reflect how AI outputs are presented and processed within the workspace. In Noda, suggestions are brief and distributed in space, which requires participants to locate, interpret, and compare them. In Miro, suggestions are longer and sit as full sentences on a flat canvas, which makes them easier to read, evaluate, and discuss. As a result, the same AI function contributes differently to group decisions across configurations.

In the following, we discuss each of these findings in depth, and derive design implications for sociotechnical collaborative platforms that aim to support both effective human-human collaboration and productive human-AI co-ideation.
\subsection{RQ1: Shaping Interaction Patterns, Attention Distribution, and Perceived Collaboration 
}
As expected, the Noda platform yielded significantly higher spatial presence and a stronger sense of immersion than the Miro platform, aligning with theories of distributed and embodied cognition that link richer spatial environments to associative thinking and creative engagement \citep{Hutchins1995CognitionintheWild, kirsh2010thinking, Klein2006distributed}. The Noda configuration also imposed greater physical demands, while mental demand and effort did not differ significantly (Table \ref{tab:tlx-appendix}). This is a more limited cost than the cognitive fatigue reported for complex virtual environments \citep{fromm2020vr_brainstorming, IdeAvatar2020}.
Beyond confirming these expected presence and workload differences, our findings offer three contributions that extend current understanding.

First, we identify a participation paradox in the immersive configuration: the spatial freedom that let participants explore and organize ideas also let them drift apart, reducing collaborative awareness and group coherence. Although participants reported greater immersion and greater comfort giving and receiving feedback in Noda, a distinct subset reported a sense of disconnection attributable to large virtual distances between avatars. This finding builds on \cite{kutak}'s call for ``careful spatial design'' in immersive systems, but sharpens the claim: spatial distance is not merely a usability concern but a mechanism that could directly counteracts the social presence and participation benefits that immersive environments are designed to produce. 

Second, by examining commercially available platforms as integrated sociotechnical configurations, we show that the observed collaboration patterns are not a function of two-dimensional against three-dimensional space alone. 
Miro's cursor-based awareness, sentence-length text nodes, and screen-bounded shared canvas constitute an integrated sociotechnical configuration that structures attention and coordination in specific ways, just as Noda’s avatar embodiment and three-dimensional placement of nodes constitute a distinct one. This framing provides an ecologically valid account of how contemporary collaborative tools, as they are actually designed and deployed, relate to how groups experience ideation and how evenly they participate.

Third, the Noda configuration was associated with more balanced role distribution during convergence: five of seven Noda groups shared navigation and driving roles, against three of seven in Miro. Previous work on VR brainstorming has focused primarily on idea quantity and creative outcomes \citep{fromm2020vr_brainstorming, gong2022virtual_brainstorming}. 
Our video analysis adds role distribution as a further dimension. In the five balanced Noda groups, the pattern co-occurred with specific embodied behaviors: avatars facing one another during discussion, and participants gathering at the center of the room or moving between one another’s maps during convergence. These are observed coordination behaviors, recorded independently of self-report, and they suggest that spatial co-presence, even through low-fidelity avatar representations, can accompany more balanced group participation. Our coding cannot order the two: groups that had settled into balanced participation may have oriented toward one another as a consequence, orientation may have supported balance, or both may reflect group dynamics we did not measure. We therefore read facing as a behavioral marker of balanced collaboration. The design implication, making orientation cues available, does not depend on the direction of the effect.
This extends \cite{gong2022virtual_brainstorming}'s account that symbolic gestures and proximity cues in VR can support inclusivity, by grounding the claim in behavioral data from a structured ideation task using a commercially available platform.

Taken together, these findings indicate that immersive environments similar to Noda could offer meaningful advantages for presence and balanced participation, but might impose cognitive and communicative overhead that current platform designs have not fully resolved. Notably, neither platform adequately supported group awareness of AI contributions: participants invoked AI individually and asynchronously, and there were no mechanisms to make AI-generated content immediately visible or attributable to all group members simultaneously — a limitation that shaped how groups evaluated and negotiated AI suggestions across both configurations. The challenge for design is therefore not simply to maximize immersion, but to structure spatial configurations, awareness mechanisms, and AI integration styles that together translate embodied engagement into productive, equitable collaboration — a point we return to in the design implications section.

\subsection{RQ2: Mediating Human-AI Co-Ideation}
\subsubsection{Evaluation of AI-generated ideas}
Participants' evaluations of AI-generated ideas differed across the two configurations. In Miro, participants rated AI-generated ideas as more insightful, and marginally more innovative, than participants in Noda did (Table~\ref{tab:idea_quality}). In Noda (Table~\ref{tab:noda_anova}), AI-generated ideas were rated significantly lower than both human-generated and group-authored ideas on relevance, innovation, and insightfulness, with large effect sizes. In Miro (Table~\ref{tab:miro_anova}), participants rated group-authored ideas highest but showed smaller differences between AI-generated and human-generated ideas. \rev{Selection followed the same pattern (Table~\ref{tab:origin}): 10 of the 21 finalist ideas in Miro were unaltered AI-generated nodes, against 3 of the 20 finalists in Noda. Selection of an unaltered AI node is the only measure of uptake the workspace export supports; group-authored ideas may draw on AI output without leaving a recoverable trace, so uptake through co-elaboration is not measured.}

\rev{The comparison that matters for a platform intended to augment group ideation is between what groups produced with the AI available and what they produced without it, and on that comparison both configurations came out ahead: group-authored ideas were rated above participants' own human-only ideas, every group-authored idea was selected as a finalist, and the idea pool grew threefold in Miro and fivefold in Noda. Two qualifications apply. The within-participant comparison is between rated sets, not matched ideas, so it shows that the AI-supported phase yielded ideas participants valued more, not that any particular idea improved. And \textit{Subtask 3} required selection by design, so the refinement observed in the convergent phase reflects the structure of the workflow as much as the configuration. The configurations differ in what participants selected, with unaltered AI nodes in Miro and human-authored nodes in Noda. The rest of this section addresses this difference.}

\rev{The finding we consider most consequential for design is that these differences track the form in which AI output arrived in the workspace. Miro's AI produced ideas averaging 21 words; Noda's averaged 3 (Table~\ref{tab:sentencelen}). This \textit{sevenfold} difference in verbosity is a design choice made by each platform's developers in response to its display constraints, and it co-occurred with large differences in perceived innovation and insightfulness and in the rate at which AI-generated ideas were selected as finalists. AI output format therefore functions as a first-class design variable in collaborative ideation platforms, on a par with spatial layout and awareness mechanisms. Work on user-centered constraints on LLM output has shown that length and structure shape how individuals integrate generated content into their work \citep{liu2024structured}; our data indicate that the same choices shape how groups evaluate, negotiate, and adopt that content collectively.}

\rev{The principal alternative account is model capability.} The Miro AI feature was powered by OpenAI GPT-4o-mini, whereas the Noda AI feature was powered by OpenAI GPT-3.5 Turbo, \rev{and the two models differ on benchmarks \citep{openai2024gpt4omini}. Two observations weigh against capability as the sole explanation.} For \textit{relevance}, the most surface-level of the three judgments, there was no significant cross-platform difference (Table~\ref{tab:idea_quality}), suggesting both models drew on comparable domain knowledge. Where the platforms diverged was on \textit{innovation} and \textit{insightfulness}, the dimensions most sensitive to depth, elaboration, and persuasiveness. Further, the Noda AI produced substantially more ideas than Miro's (633 against 304), \rev{so GPT-3.5 Turbo was not failing generatively; it was presenting its output in a format constrained by interface design. On our reading, presentation explains the pattern better than capability alone. The two accounts are nevertheless bundled in this design and cannot be fully separated (see Sociotechnical Configurations and Limitations).}

\rev{Two mechanisms are consistent with the presentation account.} Prior research suggests that more detailed textual output generated by large language models can be perceived as more persuasive, thereby increasing acceptance in collaborative decision-making contexts \citep{breum2024persuasive}. The Miro AI may also have adhered more closely than the Noda AI to Grice's cooperative principle, which holds that contributions to conversation should provide an appropriate amount of information and do so clearly \citep{grice1975logic}. \rev{At three words per idea, the Noda AI's contributions may have been too brief to be informative or credible to participants weighing them against their own.}

\rev{A further account concerns the participants and not the AI.} Brucks and Levav \citep{BrucksLevav2022} found that face-to-face collaboration supported better divergent thinking than remote collaboration, \rev{and attributed the difference to variability in gaze direction.} Participants in Miro collaborated through a screen-based interface, which likely concentrated visual attention on a shared display, whereas participants in Noda interacted at least part of the time in a more face-to-face arrangement within the virtual environment. \rev{If participants in Noda experienced greater gaze variability and stronger divergent thinking during \textit{Subtask 1}, they may have generated stronger ideas of their own and found the AI's contributions less insightful by comparison.} We did not record gaze direction in either platform, so this interpretation remains speculative. The questionnaire offers only partial support for this account: group-authored ideas were rated as more relevant in Noda than in Miro, but ratings of participants' own ideas did not differ across platforms (Table~\ref{tab:idea_quality}), so any strengthening of human ideation in Noda was not strong enough to register there.

\subsubsection{Group Awareness of AI Contributions}
\rev{Neither configuration made the AI's contributions legible to the group. In both, the AI was an on-demand function invoked by one participant on one node: the invocation was not announced to the others, its output appeared at the invoking participant's location, and the only marker of AI authorship was a leading asterisk in Miro or a node color in Noda (Figures~\ref{fig:miro_board_AI} and~\ref{fig:noda-arrangement}).} 

\rev{Both configurations accept a single node label as the prompt for AI contributions and expose no conversational channel, so a group cannot steer the model, ask it to revise, or tell it which of its suggestions were taken up. The corpus shows how little co-elaboration followed: 15 ideas across the 14 groups were authored during the AI-supported phase (7 in Miro, 8 in Noda; Table~\ref{tab:idea_dataset_curation}), and all 15 became finalists (Table~\ref{tab:origin}). Groups valued what they built from AI output when they built it, but neither configuration made producing such an idea an easy or a group-visible move.}

\rev{This finding locates a gap in the frameworks reviewed in Related Work. Role vocabularies and the COFI framework model one creator and one system \citep{rezwana2023cofi}; the team-level work of \citet{mccomb2023focus} and \citet{seeber2020machines} addresses the AI's position in a team without specifying how its output is surfaced, attributed, and steered when several people share it in one workspace. The two deployed platforms answer that question by default, with individual invocation and no attribution, and the result in both was an AI that enlarged each participant's idea space without entering the group's exchange. We discuss the design consequences, shared invocation, attribution, and a group-steerable prompt surface, under Implications for Design.}
\subsection{Focusing on Women Participants}
Focusing on women participants allowed us to minimize the variability related to gender dynamics and provided clearer insights into specific ideation processes for women. Previous studies indicate that gender composition significantly affects group interactions, communication styles, and creativity in collaborative settings \citep{schiller2011gender,  HILL2025104107, criado2019invisible}. Women often encounter different challenges in remote collaborative ideation, including participation barriers in environments mediated by digital media \citep{Gupta2020OnlineMeetings, Marchant2021ZoomCalls, bradley2021negativity,Yassien2021}, biases in idea evaluation \citep{Proudfoot2015GenderBias}, and differences in the use of AI \citep{humlum2025unequal}. 

Our findings contribute to the growing body of work that emphasizes the need for inclusive design practices that advocate and account for gender-specific collaboration dynamics, particularly in remote and AI-supported contexts \citep{DelcourtCHI22, FoxFem, criado2019invisible}. Future research should extend these insights by investigating mixed-gender and other demographic configurations to further inform inclusive collaborative technology design.

\subsection{Implications for Design}
Our findings offer insights for the design of sociotechnical collaborative platforms for online group ideation, concerning the presentation of LLM output, the legibility of AI contributions to the group, spatial design, and awareness among participants.

\subsubsection{Integrating LLMs for Effective Ideation Support}

\rev{If AI output format shapes collective evaluation and adoption, as the RQ2 findings indicate, then platforms should treat verbosity, structure, and spatial embedding as design parameters open to users and not as fixed properties of the feature.} This points to a need for \textit{platform-aware, context-adaptive AI output design}. Concretely, we recommend that platforms support adaptive AI verbosity: users and groups should be able to request brief keyword-style suggestions when they want to maintain ideation momentum, or detailed elaborations when they want to evaluate and compare specific ideas. This is particularly critical in immersive environments, where display constraints make longer text difficult to read and navigate, but where brief AI contributions may be insufficiently informative to be persuasive or credible. In these settings, voice-based AI delivery may serve as a viable alternative modality for richer AI contributions, though this requires attention to voice design principles to avoid reinforcing gender biases \citep{HwangCHI24, SuminCHI25, Zhang2025AIVoice}.

Beyond verbosity, our findings also suggest that AI integration should be designed to \rev{reinforce human creative contribution instead of bypassing it}. Across both platforms, participants rated the ideas their group authored during the AI-supported phase more highly than the AI-generated ideas alone. Our corpus also indicates how little co-elaboration occurred: 15 ideas across all fourteen groups were authored during that phase (origin not separable; Table \ref{tab:idea_dataset_curation}), and neither configuration made producing one an easy or a group-visible move. This supports a design principle of scaffolding human-AI co-elaboration: instead of presenting AI suggestions as finished ideas to be accepted or rejected, platforms should make it easy for groups to build on, rephrase, and combine AI contributions with their own ideas. Design principles for generative AI applications \citep{weitz2024designpriciples} suggest a distinction between ``design for exploration'' and ``design for optimization''\rev{, a distinction that} maps well onto the divergent and convergent phases of collaborative ideation. Future platforms should support phase-specific AI behavior: more expansive and varied during divergence, more focused and elaborative during convergence, with mechanisms for groups to signal which mode they are in.

\subsubsection{Making AI Contributions Legible to the Group}

\rev{The deficit reported under Group Awareness of AI Contributions calls for a class of awareness mechanism directed at the AI's role in the workspace, distinct from the mechanisms that make participants aware of one another. In both platforms, participants invoked AI individually and asynchronously, producing uneven awareness of AI-generated content, and the only markers of AI authorship, a leading asterisk in Miro and a node color in Noda, were not legible across the workspace. Both platforms also constrain the prompt surface to a single node label, with no conversational channel, a design choice that bounds how groups can steer the generative model. Three mechanisms follow. First, shared, synchronous AI invocation that announces itself to all members, so that an invocation is a group event and its output arrives where the group can see it. Second, attribution markers that persist and remain legible at a distance, so that the provenance of a node is visible during convergence as well as at the moment of generation. Third, a prompt surface the group can address jointly, so that steering the model, asking it to revise, and telling it which suggestions were taken up become group acts and not private ones. This design space remains largely unaddressed in current tools.}

\subsubsection{Spatial Design for Immersive Collaboration}

Our findings point to a specific design tension that goes beyond the general call for ``careful spatial design'' in immersive environments such as Noda \citep{kutak}: the same three-dimensional affordances that enable rich, embodied engagement with ideas also allow participants to drift apart spatially in ways that reduce group coherence. This tension has direct implications for how immersive ideation platforms should be structured.

\rev{Designers should treat spatial layout as an active mechanism} for regulating the balance between individual exploration and group convergence\rev{, and not as a neutral backdrop for collaboration}. Concretely, this suggests designing for \textit{phase-sensitive spatial dynamics}: during divergent ideation, participants may benefit from spatial freedom to create and explore their own idea clusters; during convergence, the platform should actively support or encourage proximity, perhaps through subtle spatial cues, avatar pull-toward gestures, or shared focal artifacts that draw participants into a common area. The Blended Whiteboard \citep{gronbaek2024blended} provides an example of reconfigurable collaborative space. Future platforms should extend this logic to structured ideation workflows by aligning spatial affordances with task phases.

Additionally, our findings indicate that even minimal avatar representations are relevant to participation equity: the Noda groups whose avatars faced one another were the groups coded as showing balanced collaboration. This behavioral finding suggests that avatar orientation, which is currently largely incidental in commercial VR platforms, should be treated as a designable awareness mechanism. Platforms should support orientation cues, directional audio, and spatial proximity indicators that make other participants' attention and engagement legible without requiring explicit communication.

\subsubsection{Enhancing Awareness and Communication Among Participants}

Both platforms lacked robust support for non-verbal communication, but the mechanisms of awareness differed in illuminating ways: in the Miro configuration, awareness was mediated through cursor position and artifact co-location; in the Noda configuration, it emerged through avatar orientation and spatial proximity. This contrast reveals that workspace awareness is \rev{a property to be designed in alignment with the interactional affordances of each configuration, and not a single feature to be added}.

For planar platforms, features such as live cursor tracking \citep{Gutwin04, Yang23}, collaborative highlighting, and contribution visualization \citep{Gutwin04, Ollesch2021} can make participants' activities legible in real time. Our findings support extending these mechanisms to make \textit{convergence activity} particularly visible, specifically, who is moving toward, selecting, or annotating ideas during decision phases\rev{, since \textit{Subtask 3} was the phase in which groups came together and in which unequal role distribution appeared}. Prior work has shown that shared awareness of contribution patterns can support more equitable participation \citep{Rocha}, a finding our data extends to AI-augmented group ideation. For immersive platforms, avatar expressiveness \citep{uravatar24}\rev{,} high-fidelity environment\rev{s} \citep{Mueller22}\rev{,} and telepresence \citep{Huang25} are promising directions.

\subsection{Limitations and Future Work}

This study has several limitations related to the sample, task design, and platform differences, which may affect the generalizability of the findings. Below we outline these limitations and highlight directions for future work.

\textit{Sample}. The study involved 42 student participants, all of whom were women. While this relatively homogeneous sample reduced variability in the collected data, the results may not generalize to mixed-gender groups or to professional teams with domain expertise. Future studies should examine more diverse participant populations, including mixed-gender teams and groups with varying levels of expertise, to better understand how group composition influences collaborative ideation in sociotechnical platforms.

\textit{Task}. The experimental task was designed to simulate short-term creative work with distinct phases for divergent and convergent thinking. Although this structure allowed us to evaluate immediate ideation outcomes, it may not reflect the dynamics of longer-term collaborative projects, in which users gradually adapt to platform affordances and develop shared working practices. Because creativity, collaboration patterns, and reliance on AI can evolve over time, longitudinal studies are needed to understand how sustained collaboration, group learning, and AI use develop across repeated or extended sessions.

\textit{Platform differences}. The study compared two commercially
available platforms, Miro and Noda, as deployed. They differ in the
properties listed under Sociotechnical Configurations, including the
underlying model, the length and format of AI output, interaction
modality, and text entry. This design has the ecological validity of
studying what groups actually use, at the cost that no observed
difference can be attributed to any single property of a configuration.
Text entry illustrates the problem. In Noda, participants typed one
character at a time on a virtual keyboard with a pointer; in Miro they
used a physical keyboard. This cost falls on human-authored text only,
and it offers a second account of two Noda-side observations: the shorter
human ideas (3.5 against 5.0 words, Table~4) and the small number of
group-authored ideas composed during the AI-supported phase (8 of the 796
Noda ideas). Where entering a sentence is slow, participants may keep
ideas terse and elaborate verbally instead of in the workspace, which the
workspace export does not capture. Novelty is a further candidate: participants were not screened for prior VR experience, and unfamiliarity with an immersive workspace may have raised engagement and exploration in Noda independently of its design. Future
work should hold the platform constant and vary one property at a time,
for example the verbosity of AI output within a single workspace, to
separate the contributions of spatial design, interaction modality,
novelty, and AI integration.

\rev{\textit{Measures.} The MEC-SPQ was developed for and validated in immersive virtual environments (Vorderer et al., 2004), and its items apply unevenly to a planar shared canvas. Two items differed across platforms in directions consistent with this asymmetry of instrument fit (Table A.1), and cross-platform presence comparisons should be read accordingly. In addition, the idea-quality items referred to ideas created by ``the team''; the intended referent, stated in Methods, was the set of ideas the group authored during the AI-supported phase, and we cannot rule out that individual participants read the word as referring to their fellow participants more generally.}

\section{Conclusion}

This study provides empirical insight into how sociotechnical collaborative platforms and LLM integration jointly shape remote collaborative ideation. By comparing two different sociotechnical configurations, we show that platform configuration \rev{shaped} the quantity and form of generated ideas\rev{, as well as} collaboration patterns, spatial presence, physical effort, and the ways groups relate to AI contributions. AI support \rev{multiplied} idea production in both platforms\rev{, and the ideas groups authored with AI available were rated above participants' own. Participants nevertheless} evaluated\rev{ and} selected AI-generated ideas differently\rev{ across configurations, a difference that tracked how AI output was designed to appear in each configuration, about 21 words per idea in Miro and 3 in Noda, more closely than it tracked model capability, though the two cannot be fully separated in this design.}

More specifically, the Noda configuration introduced a participation paradox in which spatial freedom supported exploration and fragmented group coherence, and was associated with more balanced role distribution, with five of seven groups sharing roles \rev{and moving toward a shared map} during convergence.

\rev{In both configurations, AI output entered the workspace through individual, asynchronous invocation, with no mechanism that made it visible or attributable to the group and no prompt surface the group could address jointly. Together with the role of AI output format in shaping evaluation and selection, this defines a design space for AI in group ideation: shared invocation, group-visible attribution, and output whose form adapts to the workspace and to the phase of the task.}

By focusing on women participants, this study also contributes to ongoing efforts to increase gender-inclusive design research in the context of remote and hybrid knowledge work with AI support. This matters in light of the growing participation of women in hybrid and remote knowledge work, where equitable participation is both particularly salient and increasingly consequential. Future research should extend this work by examining more diverse participant populations, studying collaboration longitudinally, and conducting more controlled comparisons that isolate the effects of spatial design, AI output format, and awareness mechanisms. Such work can help refine design principles for sociotechnical platforms that balance embodiment, workload, and effective human-AI co-ideation.

\section{Conflicts of interest}
The authors declare that they have no competing interests.


\section*{Data availability}
The data and code underlying the reported results are available at
\url{https://tinyurl.com/28adpshk}.
The deposit contains a single workbook (\texttt{noda\_miro\_statistical\_data.xlsx})
holding the idea corpus (1,260 ideas by group, configuration, and source),
the finalist ideas and session times per group, and the post-task
questionnaire responses (presence, workload, collaboration, idea-quality,
and open-text items for all 42 participants), together with the R scripts
for the idea-quality analyses (Tables~5 to~7), the post-task Mann-Whitney
tests (Tables~A.1 to~A.4), and the term-prevalence tests (Tables~8 and~9),
and the Python scripts for the LvS signature extraction.

\bibliographystyle{oup-abbrvnat}
\bibliography{reference}

@article{holzinger2022personas,
  author  = {Holzinger, Andreas and Kargl, Michaela and Kipperer, Bettina
             and Regitnig, Peter and Plass, Markus and M{\"u}ller, Heimo},
  title   = {Personas for Artificial Intelligence {(AI)} an Open Source Toolbox},
  journal = {IEEE Access},
  volume  = {10},
  pages   = {23732--23747},
  year    = {2022},
  doi     = {10.1109/ACCESS.2022.3154776},
  url     = {https://doi.org/10.1109/ACCESS.2022.3154776},
  issn    = {2169-3536},
  publisher = {IEEE}
}

@article{wan2026diverse,
  author  = {Wan, Yun and Kalman, Yoram M.},
  title   = {Diverse {AI} personas can mitigate the homogenization effect in
             human-{AI} collaborative ideation},
  journal = {Computers in Human Behavior: Artificial Humans},
  volume  = {8},
  pages   = {100289},
  year    = {2026},
  month   = may,
  doi     = {10.1016/j.chbah.2026.100289},
  url     = {https://doi.org/10.1016/j.chbah.2026.100289},
  issn    = {2949-8821},
  publisher = {Elsevier}
}

@article{evans2025collaborate,
  author  = {Evans, Katie D. and Robbins, Scott A. and Bryson, Joanna J.},
  title   = {Do We Collaborate With What We Design?},
  journal = {Topics in Cognitive Science},
  volume  = {17},
  number  = {2},
  pages   = {392--411},
  year    = {2025},
  doi     = {10.1111/tops.12682},
  url     = {https://doi.org/10.1111/tops.12682},
  issn    = {1756-8757},
  publisher = {Wiley}
}

@article{seeber2020machines,
  author  = {Seeber, Isabella and Bittner, Eva and Briggs, Robert O. and
             de Vreede, Triparna and de Vreede, Gert-Jan and Elkins, Aaron
             and Maier, Ronald and Merz, Alexander B. and
             Oeste-Rei{\ss}, Sarah and Randrup, Nils and Schwabe, Gerhard
             and S{\"o}llner, Matthias},
  title   = {Machines as teammates: A research agenda on {AI} in team collaboration},
  journal = {Information \& Management},
  volume  = {57},
  number  = {2},
  pages   = {103174},
  year    = {2020},
  month   = mar,
  doi     = {10.1016/j.im.2019.103174},
  url     = {https://doi.org/10.1016/j.im.2019.103174},
  issn    = {0378-7206},
  publisher = {Elsevier}
}

@article{rezwana2023cofi,
  author    = {Rezwana, Jeba and Maher, Mary Lou},
  title     = {Designing Creative {AI} Partners with {COFI}: A Framework for
               Modeling Interaction in Human-{AI} Co-Creative Systems},
  journal   = {ACM Transactions on Computer-Human Interaction},
  volume    = {30},
  number    = {5},
  numpages  = {28},
  pages     = {1--28},
  year      = {2023},
  month     = oct,
  doi       = {10.1145/3519026},
  url       = {https://dl.acm.org/doi/10.1145/3519026},
  issn      = {1073-0516},
  publisher = {Association for Computing Machinery}
}

@article{mokryn2021domain,
  title={Domain-based Latent Personal Analysis and its use for impersonation detection in social media},
  author={Mokryn, Osnat and Ben-Shoshan, Hagit},
  journal={User Modeling and User-Adapted Interaction},
  volume={31},
  number={4},
  pages={785--828},
  year={2021},
  doi={10.1007/s11257-021-09295-7},
  url={https://doi.org/10.1007/s11257-021-09295-7},
  publisher={Springer}
}

@inproceedings{Huang25,
  author = {Huang, Xincheng and Frehlich, Dieter and Xia, Ziyi and Gholami, Peyman and Xiao, Robert},
  title = {GaussianNexus: Room-Scale Real-Time AR/VR Telepresence with Gaussian Splatting},
  year = {2025},
  isbn = {9798400720376},
  publisher = {Association for Computing Machinery},
  address = {New York, NY, USA},
  url = {https://doi.org/10.1145/3746059.3747693},
  doi = {10.1145/3746059.3747693},
  booktitle = {Proceedings of the 38th Annual ACM Symposium on User Interface Software and Technology},
  articleno = {189},
  numpages = {18},
  series = {UIST '25}
}

@article{Mueller22,
  author = {M\"{u}ller, Thomas and Evans, Alex and Schied, Christoph and Keller, Alexander},
  title = {Instant neural graphics primitives with a multiresolution hash encoding},
  year = {2022},
  issue_date = {July 2022},
  publisher = {Association for Computing Machinery},
  address = {New York, NY, USA},
  volume = {41},
  number = {4},
  issn = {0730-0301},
  url = {https://doi.org/10.1145/3528223.3530127},
  doi = {10.1145/3528223.3530127},
  journal = {ACM Trans. Graph.},
  month = jul,
  articleno = {102},
  numpages = {15}
}

@inproceedings{benshoshan2026making,
  author = {Ben Shoshan, Hagit and Lanir, Joel and Goldstein, Pavel and Mokryn, Osnat},
  title = {Making Absence Visible: The Roles of Reference and Prompting in Recognizing Missing Information},
  year = {2026},
  isbn = {9798400719844},
  publisher = {Association for Computing Machinery},
  address = {New York, NY, USA},
  url = {https://doi.org/10.1145/3742413.3789150},
  doi = {10.1145/3742413.3789150},
  booktitle = {Proceedings of the 31st International Conference on Intelligent User Interfaces},
  pages = {1100--1111},
  numpages = {12},
  series = {IUI '26}
}

@article{orlikowski2000using,
  author    = {Orlikowski, Wanda J.},
  title     = {Using Technology and Constituting Structures: A Practice Lens for Studying Technology in Organizations},
  journal   = {Organization Science},
  volume    = {11},
  number    = {4},
  pages     = {404--428},
  year      = {2000},
  publisher = {INFORMS},
  doi       = {10.1287/orsc.11.4.404.14600},
  url       = {https://doi.org/10.1287/orsc.11.4.404.14600}
}

@inproceedings{uravatar24,
  author = {Li, Junxuan and Cao, Chen and Schwartz, Gabriel and Khirodkar, Rawal and Richardt, Christian and Simon, Tomas and Sheikh, Yaser and Saito, Shunsuke},
  title = {URAvatar: Universal Relightable Gaussian Codec Avatars},
  year = {2024},
  isbn = {9798400711312},
  publisher = {Association for Computing Machinery},
  address = {New York, NY, USA},
  url = {https://doi.org/10.1145/3680528.3687653},
  doi = {10.1145/3680528.3687653},
  booktitle = {SIGGRAPH Asia 2024 Conference Papers},
  articleno = {128},
  numpages = {11},
  location = {Tokyo, Japan},
  series = {SA '24}
}

@article{Ackerman2000,
  author = {Ackerman, Mark S.},
  title = {The intellectual challenge of CSCW: the gap between social requirements and technical feasibility},
  year = {2000},
  issue_date = {September 2000},
  publisher = {L. Erlbaum Associates Inc.},
  address = {USA},
  volume = {15},
  number = {2},
  issn = {0737-0024},
  url = {https://doi.org/10.1207/S15327051HCI1523_5},
  doi = {10.1207/S15327051HCI1523_5},
  journal = {Hum.-Comput. Interact.},
  month = sep,
  pages = {179--203},
  numpages = {25}
}

@article{Crowston19,
  author = {Crowston, Kevin and Saltz, Jeff S. and Rezgui, Amira and Hegde, Yatish and You, Sangseok},
  title = {Socio-technical Affordances for Stigmergic Coordination Implemented in MIDST, a Tool for Data-Science Teams},
  year = {2019},
  issue_date = {November 2019},
  publisher = {Association for Computing Machinery},
  address = {New York, NY, USA},
  volume = {3},
  number = {CSCW},
  url = {https://doi.org/10.1145/3359219},
  doi = {10.1145/3359219},
  journal = {Proc. ACM Hum.-Comput. Interact.},
  month = nov,
  articleno = {117},
  numpages = {25}
}

@article{humlum2025unequal,
  author    = {Humlum, Anders and Vestergaard, Erik},
  title     = {The unequal adoption of ChatGPT exacerbates existing inequalities among workers},
  journal   = {Proceedings of the National Academy of Sciences of the United States of America},
  volume    = {122},
  number    = {1},
  pages     = {e2414972121},
  year      = {2025},
  doi       = {10.1073/pnas.2414972121},
  url       = {https://doi.org/10.1073/pnas.2414972121}
}

@inproceedings{GuoGames,
  author = {Freeman, Guo},
  title = {Making Games as Collaborative Social Experiences: Exploring An Online Gaming Community},
  year = {2016},
  isbn = {9781450339506},
  publisher = {Association for Computing Machinery},
  address = {New York, NY, USA},
  url = {https://doi.org/10.1145/2818052.2869076},
  doi = {10.1145/2818052.2869076},
  booktitle = {Proceedings of the 19th ACM Conference on Computer Supported Cooperative Work and Social Computing Companion},
  pages = {265--268},
  numpages = {4},
  location = {San Francisco, California, USA},
  series = {CSCW '16 Companion}
}

@book{criado2019invisible,
  author    = {Criado-Perez, Caroline},
  title     = {Invisible Women: Data Bias in a World Designed for Men},
  year      = {2019},
  publisher = {Abrams Press},
  address   = {New York, NY},
  isbn      = {978-1-4197-3207-7}
}

@article{wilson2002six,
  author = {Wilson, Margaret},
  title = {Six views of embodied cognition},
  journal = {Psychonomic Bulletin \& Review},
  volume = {9},
  number = {4},
  pages = {625--636},
  year = {2002},
  doi = {10.3758/BF03196322},
  url = {https://doi.org/10.3758/BF03196322}
}

@book{shapiro2019embodied,
  author = {Shapiro, Lawrence},
  title = {Embodied Cognition},
  edition = {2},
  publisher = {Routledge},
  year = {2019},
  address = {New York, NY},
  isbn = {9781138746985}
}

@incollection{Klein2006distributed,
  author = {Klein, Gary and Feltovich, Paul J. and Bradshaw, Jeffrey M. and Woods, David D.},
  title = {Common ground and coordination in joint activity},
  booktitle = {Organizational Simulation},
  editor = {Rouse, William B. and Boff, Kenneth R.},
  pages = {139--184},
  publisher = {Wiley},
  year = {2005},
  doi = {10.1002/0471739448.ch6},
  url = {https://doi.org/10.1002/0471739448.ch6}
}

@article{Kirsh1995,
  author = {Kirsh, David},
  title = {The intelligent use of space},
  journal = {Artificial Intelligence},
  volume = {73},
  number = {1-2},
  pages = {31--68},
  year = {1995},
  doi = {10.1016/0004-3702(94)00017-U},
  url = {https://doi.org/10.1016/0004-3702(94)00017-U}
}

@book{Hutchins1995CognitionintheWild,
  author = {Hutchins, Edwin},
  title = {Cognition in the Wild},
  publisher = {MIT Press},
  year = {1995},
  address = {Cambridge, MA},
  isbn = {9780262082310}
}

@article{WestendorfCSCW,
  author = {Westendorf, Lauren and Shaer, Orit and Varsanyi, Petra and van der Meulen, Hidde and Kun, Andrew L.},
  title = {Understanding Collaborative Decision Making Around a Large-Scale Interactive Tabletop},
  year = {2017},
  issue_date = {November 2017},
  publisher = {Association for Computing Machinery},
  address = {New York, NY, USA},
  volume = {1},
  number = {CSCW},
  url = {https://doi.org/10.1145/3134745},
  doi = {10.1145/3134745},
  journal = {Proc. ACM Hum.-Comput. Interact.},
  month = dec,
  articleno = {110},
  numpages = {21}
}

@inproceedings{FoxFem,
  author = {Fox, Sarah and Ulgado, Rachel Rose and Rosner, Daniela},
  title = {Hacking Culture, Not Devices: Access and Recognition in Feminist Hackerspaces},
  year = {2015},
  isbn = {9781450329224},
  publisher = {Association for Computing Machinery},
  address = {New York, NY, USA},
  url = {https://doi.org/10.1145/2675133.2675223},
  doi = {10.1145/2675133.2675223},
  booktitle = {Proceedings of the 18th ACM Conference on Computer Supported Cooperative Work \& Social Computing},
  pages = {56--68},
  numpages = {13},
  location = {Vancouver, BC, Canada},
  series = {CSCW '15}
}

@inproceedings{Yassien2021,
  author = {Yassien, Amal and Makled, ElHassan B. and Elagroudy, Passant and Sadek, Nouran and Abdennadher, Slim},
  title = {Give-Me-A-Hand: The Effect of Partner's Gender on Collaboration Quality in Virtual Reality},
  year = {2021},
  isbn = {9781450380959},
  publisher = {Association for Computing Machinery},
  address = {New York, NY, USA},
  url = {https://doi.org/10.1145/3411763.3451601},
  doi = {10.1145/3411763.3451601},
  booktitle = {Extended Abstracts of the 2021 CHI Conference on Human Factors in Computing Systems},
  articleno = {392},
  numpages = {6},
  location = {Yokohama, Japan},
  series = {CHI EA '21}
}

@article{KOULOURI2017162,
  title = {The influence of visual feedback and gender dynamics on performance, perception and communication strategies in CSCW},
  journal = {International Journal of Human-Computer Studies},
  volume = {97},
  pages = {162--181},
  year = {2017},
  issn = {1071-5819},
  doi = {10.1016/j.ijhcs.2016.09.003},
  url = {https://www.sciencedirect.com/science/article/pii/S1071581916301033},
  author = {Theodora Koulouri and Stanislao Lauria and Robert D. Macredie}
}

@inproceedings{schiller2011gender,
  title        = {Gender Differences in Virtual Collaboration on a Creative Design Task},
  author       = {Schiller, Shu and Nah, Fiona and Mennecke, Brian and Siau, Keng},
  booktitle    = {Proceedings of the 32nd International Conference on Information Systems (ICIS '11)},
  year         = {2011},
  pages        = {1--12},
  publisher    = {AIS},
  series       = {ICIS 2011 Proceedings},
  url          = {https://aisel.aisnet.org/icis2011/proceedings/hci/12}
}

@inproceedings{Garcia_teams_22,
  author = {Garcia, Rita and Liao, Chieh-Ju and Pearce, Ariane and Treude, Christoph},
  title = {Gender Influence on Communication Initiated within Student Teams},
  year = {2022},
  isbn = {9781450390705},
  publisher = {Association for Computing Machinery},
  address = {New York, NY, USA},
  url = {https://doi.org/10.1145/3478431.3499279},
  doi = {10.1145/3478431.3499279},
  booktitle = {Proceedings of the 53rd ACM Technical Symposium on Computer Science Education - Volume 1},
  pages = {432--438},
  numpages = {7},
  location = {Providence, RI, USA},
  series = {SIGCSE 2022}
}

@article{minet2024design,
  title={Design thinking in physical and virtual environments: Conceptual foundations, qualitative analysis, and practical implications},
  author={Minet, Alice and Wentzel, Daniel and Raff, Stefan and Garbas, Janina},
  journal={Technological Forecasting and Social Change},
  volume={207},
  pages={123596},
  year={2024},
  doi={10.1016/j.techfore.2024.123596},
  url={https://doi.org/10.1016/j.techfore.2024.123596},
  publisher={Elsevier}
}

@article{macchi2024virtual,
  title={Virtual reality, face-to-face, and 2D video conferencing differently impact fatigue, creativity, flow, and decision-making in workplace dynamics},
  author={Macchi, Gregorio and De Pisapia, Nicola},
  journal={Scientific Reports},
  volume={14},
  number={1},
  pages={10260},
  year={2024},
  doi={10.1038/s41598-024-60942-6},
  url={https://doi.org/10.1038/s41598-024-60942-6},
  publisher={Nature Publishing Group}
}

@article{lin2023remote,
  title={Remote collaboration fuses fewer breakthrough ideas},
  author={Lin, Yiling and Frey, Carl Benedikt and Wu, Lingfei},
  journal={Nature},
  volume={623},
  number={7989},
  pages={987--991},
  year={2023},
  doi={10.1038/s41586-023-06767-1},
  url={https://doi.org/10.1038/s41586-023-06767-1},
  publisher={Nature Publishing Group}
}

@inproceedings{PuIdeaSynth,
  author = {Pu, Kevin and Feng, K. J. Kevin and Grossman, Tovi and Hope, Tom and Dalvi Mishra, Bhavana and Latzke, Matt and Bragg, Jonathan and Chang, Joseph Chee and Siangliulue, Pao},
  title = {IdeaSynth: Iterative Research Idea Development Through Evolving and Composing Idea Facets with Literature-Grounded Feedback},
  year = {2025},
  isbn = {9798400713941},
  publisher = {Association for Computing Machinery},
  address = {New York, NY, USA},
  url = {https://doi.org/10.1145/3706598.3714057},
  doi = {10.1145/3706598.3714057},
  booktitle = {Proceedings of the 2025 CHI Conference on Human Factors in Computing Systems},
  articleno = {145},
  numpages = {31},
  series = {CHI '25}
}

@inproceedings{TaoDesingWeaver,
  author = {Tao, Sirui and Liang, Ivan and Peng, Cindy and Wang, Zhiqing and Palani, Srishti and Dow, Steven P.},
  title = {DesignWeaver: Dimensional Scaffolding for Text-to-Image Product Design},
  year = {2025},
  isbn = {9798400713941},
  publisher = {Association for Computing Machinery},
  address = {New York, NY, USA},
  url = {https://doi.org/10.1145/3706598.3714211},
  doi = {10.1145/3706598.3714211},
  booktitle = {Proceedings of the 2025 CHI Conference on Human Factors in Computing Systems},
  articleno = {425},
  numpages = {26},
  series = {CHI '25}
}

@misc{miro2024ai,
  author       = {{Miro}},
  title        = {Miro AI},
  year         = {2024},
  howpublished = {\url{https://miro.com/ai/}},
  note         = {Accessed: 2025-04-24}
}

@article{Olson01092000,
  author = {Gary M. Olson and Judith S. Olson},
  title = {Distance Matters},
  journal = {Human--Computer Interaction},
  volume = {15},
  number = {2-3},
  pages = {139--178},
  year = {2000},
  publisher = {Taylor \& Francis},
  doi = {10.1207/S15327051HCI1523_4}
}

@techreport{barrero2023evolution,
  title={The Evolution of Working from Home},
  author={Barrero, Jose Maria and Bloom, Nicholas and Davis, Steven J.},
  year={2023},
  month={July},
  url={https://wfhresearch.com/wp-content/uploads/2023/07/SIEPR1.pdf},
  institution={Instituto Tecnol{\'o}gico Aut{\'o}nomo de M{\'e}xico, Stanford University, University of Chicago Booth School of Business and Hoover Institution}
}

@article{BrucksLevav2022,
  author    = {Melanie S. Brucks and Jonathan Levav},
  title     = {Virtual communication curbs creative idea generation},
  journal   = {Nature},
  year      = {2022},
  volume    = {605},
  number    = {7908},
  pages     = {108--112},
  doi       = {10.1038/s41586-022-04643-y},
  url       = {https://doi.org/10.1038/s41586-022-04643-y}
}

@inproceedings{Siangliulue,
  author = {Siangliulue, Pao and Arnold, Kenneth C. and Gajos, Krzysztof Z. and Dow, Steven P.},
  title = {Toward Collaborative Ideation at Scale: Leveraging Ideas from Others to Generate More Creative and Diverse Ideas},
  year = {2015},
  isbn = {9781450329224},
  publisher = {Association for Computing Machinery},
  address = {New York, NY, USA},
  url = {https://doi.org/10.1145/2675133.2675239},
  doi = {10.1145/2675133.2675239},
  booktitle = {Proceedings of the 18th ACM Conference on Computer Supported Cooperative Work \& Social Computing},
  pages = {937--945},
  numpages = {9},
  location = {Vancouver, BC, Canada},
  series = {CSCW '15}
}

@inproceedings{Leeexamples,
  author = {Lee, Brian and Srivastava, Savil and Kumar, Ranjitha and Brafman, Ronen and Klemmer, Scott R.},
  title = {Designing with Interactive Example Galleries},
  year = {2010},
  isbn = {9781605589299},
  publisher = {Association for Computing Machinery},
  address = {New York, NY, USA},
  url = {https://doi.org/10.1145/1753326.1753667},
  doi = {10.1145/1753326.1753667},
  booktitle = {Proceedings of the SIGCHI Conference on Human Factors in Computing Systems},
  pages = {2257--2266},
  numpages = {10},
  location = {Atlanta, Georgia, USA},
  series = {CHI '10}
}

@inproceedings{Dowprototyping,
  author = {Dow, Steven and Fortuna, Julie and Schwartz, Dan and Altringer, Beth and Schwartz, Daniel and Klemmer, Scott},
  title = {Prototyping Dynamics: Sharing Multiple Designs Improves Exploration, Group Rapport, and Results},
  year = {2011},
  isbn = {9781450302289},
  publisher = {Association for Computing Machinery},
  address = {New York, NY, USA},
  url = {https://doi.org/10.1145/1978942.1979359},
  doi = {10.1145/1978942.1979359},
  booktitle = {Proceedings of the SIGCHI Conference on Human Factors in Computing Systems},
  pages = {2807--2816},
  numpages = {10},
  location = {Vancouver, BC, Canada},
  series = {CHI '11}
}

@inproceedings{Herringinspired,
  author = {Herring, Scarlett R. and Chang, Chia-Chen and Krantzler, Jesse and Bailey, Brian P.},
  title = {Getting Inspired! Understanding How and Why Examples Are Used in Creative Design Practice},
  year = {2009},
  isbn = {9781605582467},
  publisher = {Association for Computing Machinery},
  address = {New York, NY, USA},
  url = {https://doi.org/10.1145/1518701.1518717},
  doi = {10.1145/1518701.1518717},
  booktitle = {Proceedings of the SIGCHI Conference on Human Factors in Computing Systems},
  pages = {87--96},
  numpages = {10},
  location = {Boston, MA, USA},
  series = {CHI '09}
}

@misc{nodaNodaMind,
  author = {{Noda}},
  title = {noda - mind mapping in virtual reality, solo or group --- noda.io},
  howpublished = {\url{https://noda.io/}},
  year = {2023},
  note = {[Accessed 09-09-2023]}
}

@inproceedings{gronbaek2024blended,
  title={Blended whiteboard: Physicality and reconfigurability in remote mixed reality collaboration},
  author={Gr{\o}nb{\ae}k, Jens Emil Sloth and S{\'a}nchez Esquivel, Juan and Leiva, Germ{\'a}n and Velloso, Eduardo and Gellersen, Hans and Pfeuffer, Ken},
  booktitle={Proceedings of the 2024 CHI Conference on Human Factors in Computing Systems},
  pages={1--16},
  year={2024},
  publisher={Association for Computing Machinery},
  address={New York, NY, USA},
  doi={10.1145/3613904.3642293},
  url={https://doi.org/10.1145/3613904.3642293},
  series={CHI '24}
}

@misc{arthur2023designthinking,
  author       = {{Arthur Technologies}},
  title        = {A Day in Design Thinking: Embracing Virtual Reality for Immersive Collaboration},
  year         = {2023},
  month        = {November},
  url          = {https://arthur.digital/blog/a-day-in-design-thinking-embracing-virtual-reality-for-immersive-collaboration},
  note         = {Accessed: 2025-05-03}
}

@misc{dimenteers2023immersivecollab,
  author       = {{Dimenteers GmbH}},
  title        = {Immersive Remote Collaboration},
  year         = {2023},
  url          = {https://www.dimenteers.com/en/immersive-remote-collaboration},
  note         = {Accessed: 2025-05-03}
}

@book{osborn1953applied,
  title = {Applied Imagination},
  author = {Osborn, Alex F.},
  year = {1953},
  publisher = {Charles Scribner's Sons},
  address = {New York}
}

@inproceedings{ShaerBrainwriting24,
  author = {Shaer, Orit and Cooper, Angelora and Mokryn, Osnat and Kun, Andrew L. and Ben Shoshan, Hagit},
  title = {AI-Augmented Brainwriting: Investigating the use of LLMs in group ideation},
  year = {2024},
  isbn = {9798400703300},
  publisher = {Association for Computing Machinery},
  address = {New York, NY, USA},
  url = {https://doi.org/10.1145/3613904.3642414},
  doi = {10.1145/3613904.3642414},
  booktitle = {Proceedings of the 2024 CHI Conference on Human Factors in Computing Systems},
  location = {Honolulu, HI, USA},
  series = {CHI '24}
}

@inproceedings{fromm2020vr_brainstorming,
  author       = {Fromm, J. and Mirbabaie, M. and Stieglitz, S.},
  title        = {The Effects of Virtual Reality Affordances and Constraints on Negative Group Effects during Brainstorming Sessions},
  booktitle    = {Proceedings of the 15th International Conference on Wirtschaftsinformatik},
  year         = {2020},
  address      = {Germany}
}

@article{kutak,
  author={Ku{\v{t}}{\'a}k, David and Dole{\v{z}}al, Milan and Kerous, Bojan and Eichler, Zdenek and Va{\v{s}}ek, Ji{\v{r}}{\'i} and Liarokapis, Fotis},
  title={An Interactive and Multimodal Virtual Mind Map for Future Workplace},
  journal={Frontiers in ICT},
  volume={6},
  year={2019},
  url={https://www.frontiersin.org/journals/ict/articles/10.3389/fict.2019.00014},
  doi={10.3389/fict.2019.00014},
  issn={2297-198X}
}

@article{bradley2021negativity,
  author       = {Bradley, B. and Henry, S. and Blake, B.},
  title        = {When Can Negativity Mean Success? Gender Composition, Negative Relationships and Team Performance},
  journal      = {Small Group Research},
  volume       = {52},
  pages        = {457--480},
  year         = {2021},
  doi          = {10.1177/1046496420959446}
}

@article{peter2021gender_diversity,
  author       = {Peter, L. and Michinov, N. and Besan{\c{c}}on, M. and Michinov, E. and Juhel, J. and Brown, G. and Jamet, E. and Cherbonnier, A. and {ProFAN Consortium}},
  title        = {Revisiting the Effects of Gender Diversity in Small Groups on Divergent Thinking: A Large-Scale Study Using Synchronous Electronic Brainstorming},
  journal      = {Frontiers in Psychology},
  volume       = {12},
  pages        = {723235},
  year         = {2021},
  doi          = {10.3389/fpsyg.2021.723235},
  url          = {https://doi.org/10.3389/fpsyg.2021.723235}
}

@misc{owid_female_labor_supply,
  author       = {Esteban Ortiz-Ospina and Sandra Tzvetkova and Max Roser},
  title        = {Female Labor Supply},
  year         = {2023},
  howpublished = {\url{https://ourworldindata.org/female-labor-supply}},
  note         = {Accessed: 2025-01-04}
}

@techreport{vorderer2004mec_spq,
  author       = {Peter Vorderer and Werner Wirth and Feliz R. Gouveia and Frank Biocca and Timo Saari and Lutz J{\"a}ncke and Saskia B{\"o}cking and Holger Schramm and Andre Gysbers and Tilo Hartmann and others},
  title        = {MEC Spatial Presence Questionnaire (MEC-SPQ): Short Documentation and Instructions for Application},
  year         = {2004},
  institution  = {Report to the European Community, Project Presence: MEC (IST-2001-37661)}
}

@inproceedings{IdeAvatar2020,
  author = {Ide, Masahiro and Oshima, Shoji and Mori, Shingo and Yoshimi, Masato and Ichino, Junko and Tano, Shunichi},
  title = {Effects of Avatar's Symbolic Gesture in Virtual Reality Brainstorming},
  year = {2021},
  isbn = {9781450389754},
  publisher = {Association for Computing Machinery},
  address = {New York, NY, USA},
  url = {https://doi.org/10.1145/3441000.3441081},
  doi = {10.1145/3441000.3441081},
  booktitle = {Proceedings of the 32nd Australian Conference on Human-Computer Interaction},
  pages = {170--177},
  numpages = {8},
  location = {Sydney, NSW, Australia},
  series = {OzCHI '20}
}

@article{kirsh2010thinking,
  author = {Kirsh, David},
  title = {Thinking with external representations},
  journal = {AI \& Society},
  volume = {25},
  number = {4},
  pages = {441--454},
  year = {2010},
  doi = {10.1007/s00146-010-0272-8},
  url = {https://doi.org/10.1007/s00146-010-0272-8}
}

@article{HILL2025104107,
  title = {Gender differences in virtual collaboration effectiveness in hybrid teams},
  journal = {Information \& Management},
  volume = {62},
  number = {3},
  pages = {104107},
  year = {2025},
  issn = {0378-7206},
  doi = {10.1016/j.im.2025.104107},
  url = {https://www.sciencedirect.com/science/article/pii/S0378720625000102},
  author = {N. Sharon Hill and Maria Ximena Hincapie}
}

@article{Zhang2025AIVoice,
  author  = {Zhang, Qiaoning and Yang, X. Jessie and Robert, Jr., Lionel P.},
  title   = {Artificial intelligence voice gender, gender role congruity, and trust in automated vehicles},
  journal = {Scientific Reports},
  volume  = {15},
  number  = {1},
  pages   = {16364},
  year    = {2025},
  month   = {May},
  doi     = {10.1038/s41598-025-00884-9},
  url     = {https://doi.org/10.1038/s41598-025-00884-9},
  urldate = {2025-05-21}
}

@inproceedings{SuminCHI25,
  author = {Heo, Sumin and Chen, Erika R and Khuu, Jasmine},
  title = {Exploring Gender Biases in LLM-based Voice Chatbots for Job Interviews},
  year = {2025},
  isbn = {9798400713958},
  publisher = {Association for Computing Machinery},
  address = {New York, NY, USA},
  url = {https://doi.org/10.1145/3706599.3719281},
  doi = {10.1145/3706599.3719281},
  booktitle = {Proceedings of the Extended Abstracts of the CHI Conference on Human Factors in Computing Systems},
  articleno = {899},
  numpages = {8},
  series = {CHI EA '25}
}

@inproceedings{HwangCHI24,
  author = {Hwang, Angel Hsing-Chi and Won, Andrea Stevenson},
  title = {The Sound of Support: Gendered Voice Agent as Support to Minority Teammates in Gender-Imbalanced Team},
  year = {2024},
  isbn = {9798400703300},
  publisher = {Association for Computing Machinery},
  address = {New York, NY, USA},
  url = {https://doi.org/10.1145/3613904.3642202},
  doi = {10.1145/3613904.3642202},
  booktitle = {Proceedings of the 2024 CHI Conference on Human Factors in Computing Systems},
  articleno = {877},
  numpages = {22},
  location = {Honolulu, HI, USA},
  series = {CHI '24}
}

@article{Ollesch2021,
  author    = {Ollesch, Lena and Heimbuch, Steffen and Bodemer, Daniel},
  title     = {Improving learning and writing outcomes: {I}nfluence of cognitive and behavioral group awareness tools in wikis},
  journal   = {International Journal of Computer-Supported Collaborative Learning},
  volume    = {16},
  pages     = {225--259},
  year      = {2021},
  doi       = {10.1007/s11412-021-09346-6},
  url       = {https://doi.org/10.1007/s11412-021-09346-6}
}

@inproceedings{Rocha,
  author = {Rocha, Filipa and Sim\~{a}o, Hugo and Nogueira, Jo\~{a}o and Neto, Isabel and Guerreiro, Tiago and Nicolau, Hugo},
  title = {Awareness in Collaborative Mixed-Visual Ability Tangible Programming Activities},
  year = {2025},
  isbn = {9798400713941},
  publisher = {Association for Computing Machinery},
  address = {New York, NY, USA},
  url = {https://doi.org/10.1145/3706598.3713610},
  doi = {10.1145/3706598.3713610},
  booktitle = {Proceedings of the 2025 CHI Conference on Human Factors in Computing Systems},
  articleno = {1146},
  numpages = {15},
  series = {CHI '25}
}

@misc{openai2024gpt4omini,
  author       = {{OpenAI}},
  title        = {{GPT-4o mini: Advancing Cost-Efficient Intelligence}},
  year         = {2024},
  howpublished = {\url{https://openai.com/index/gpt-4o-mini-advancing-cost-efficient-intelligence/}},
  note         = {Accessed: 2025-01-04}
}

@techreport{buckman2025measuring,
  author      = {Buckman, Shelby R. and Barrero, Jose Maria and Bloom, Nicholas and Davis, Steven J.},
  title       = {Measuring Work from Home},
  institution = {National Bureau of Economic Research},
  year        = {2025},
  month       = feb,
  type        = {Working Paper},
  number      = {33508},
  doi         = {10.3386/w33508},
  url         = {https://www.nber.org/papers/w33508}
}

@techreport{aksoy2025remote,
  author      = {Aksoy, Cevat Giray and Bloom, Nicholas and Davis, Steven J. and Marino, Victoria and Ozguzel, Cem},
  title       = {Remote Work, Employee Mix, and Performance},
  institution = {National Bureau of Economic Research},
  year        = {2025},
  type        = {Working Paper},
  number      = {33851},
  url         = {https://www.nber.org/papers/w33851}
}

@inproceedings{liu2024structured,
  author = {Liu, Michael Xieyang and Liu, Frederick and Fiannaca, Alexander J. and Koo, Terry and Dixon, Lucas and Terry, Michael and Cai, Carrie J.},
  title = {"We Need Structured Output": Towards User-centered Constraints on Large Language Model Output},
  year = {2024},
  isbn = {9798400703317},
  publisher = {Association for Computing Machinery},
  address = {New York, NY, USA},
  url = {https://doi.org/10.1145/3613905.3650756},
  doi = {10.1145/3613905.3650756},
  booktitle = {Extended Abstracts of the CHI Conference on Human Factors in Computing Systems},
  articleno = {10},
  numpages = {9},
  location = {Honolulu, HI, USA},
  series = {CHI EA '24}
}

@inproceedings{weitz2024designpriciples,
  author = {Weisz, Justin D. and He, Jessica and Muller, Michael and Hoefer, Gabriela and Miles, Rachel and Geyer, Werner},
  title = {Design Principles for Generative AI Applications},
  year = {2024},
  isbn = {9798400703300},
  publisher = {Association for Computing Machinery},
  address = {New York, NY, USA},
  url = {https://doi.org/10.1145/3613904.3642466},
  doi = {10.1145/3613904.3642466},
  booktitle = {Proceedings of the 2024 CHI Conference on Human Factors in Computing Systems},
  articleno = {378},
  numpages = {22},
  location = {Honolulu, HI, USA},
  series = {CHI '24}
}

@inproceedings{Gutwin04,
  author = {Gutwin, Carl and Penner, Reagan and Schneider, Kevin},
  title = {Group awareness in distributed software development},
  year = {2004},
  isbn = {1581138105},
  publisher = {Association for Computing Machinery},
  address = {New York, NY, USA},
  url = {https://doi.org/10.1145/1031607.1031621},
  doi = {10.1145/1031607.1031621},
  booktitle = {Proceedings of the 2004 ACM Conference on Computer Supported Cooperative Work},
  pages = {72--81},
  numpages = {10},
  location = {Chicago, Illinois, USA},
  series = {CSCW '04}
}

@article{Yang23,
  author = {Yang, Stephanie and Poonawala, Amreen Amin and Jiang, Tian-Shun Allan and Schneider, Bertrand},
  title = {Can Synchronous Code Editing and Awareness Tools Support Remote Tutoring? Effects on Learning and Teaching},
  year = {2023},
  issue_date = {October 2023},
  publisher = {Association for Computing Machinery},
  address = {New York, NY, USA},
  volume = {7},
  number = {CSCW2},
  url = {https://doi.org/10.1145/3610177},
  doi = {10.1145/3610177},
  journal = {Proc. ACM Hum.-Comput. Interact.},
  month = oct,
  articleno = {328},
  numpages = {30}
}

@article{Marchant2021ZoomCalls,
  author  = {Marchant, Natalie},
  title   = {Why women don't speak up on Zoom calls - and why that's a problem},
  journal = {World Economic Forum},
  year    = {2021},
  month   = {January},
  url     = {https://www.weforum.org/stories/2021/01/women-gender-equality-workplace-meetings/},
  urldate = {2025-05-21}
}

@inproceedings{DelcourtCHI22,
  author = {Delcourt, Catherine Grevet and Charmaraman, Linda and Durrani, Sidrah and Gu, Quan and Xiao, Le Fan},
  title = {Innovating Novel Online Social Spaces with Diverse Middle School Girls: Ideation and Collaboration in a Synchronous Virtual Design Workshop},
  year = {2022},
  isbn = {9781450391573},
  publisher = {Association for Computing Machinery},
  address = {New York, NY, USA},
  url = {https://doi.org/10.1145/3491102.3517576},
  doi = {10.1145/3491102.3517576},
  booktitle = {Proceedings of the 2022 CHI Conference on Human Factors in Computing Systems},
  articleno = {248},
  numpages = {16},
  location = {New Orleans, LA, USA},
  series = {CHI '22}
}

@article{Gupta2020OnlineMeetings,
  author  = {Gupta, Alisha Haridasani},
  title   = {It's Not Just You: In Online Meetings, Many Women Can't Get a Word In},
  journal = {The New York Times},
  year    = {2020},
  month   = {April},
  url     = {https://www.nytimes.com/2020/04/14/us/zoom-meetings-gender.html},
  urldate = {2025-05-21}
}

@article{Proudfoot2015GenderBias,
  author  = {Proudfoot, David and Kay, Aaron C. and Koval, Carrie Z.},
  title   = {A Gender Bias in the Attribution of Creativity: Archival and Experimental Evidence for the Perceived Association Between Masculinity and Creative Thinking},
  journal = {Psychological Science},
  volume  = {26},
  number  = {11},
  pages   = {1751--1761},
  year    = {2015},
  doi     = {10.1177/0956797615598739},
  url     = {https://doi.org/10.1177/0956797615598739}
}

@article{breum2024persuasive,
  author    = {Breum, S. M. and Egdal, D. V. and Gram Mortensen, V. and M{\o}ller, A. G. and Aiello, L. M.},
  title     = {The Persuasive Power of Large Language Models},
  journal   = {Proceedings of the International AAAI Conference on Web and Social Media},
  volume    = {18},
  number    = {1},
  pages     = {152--163},
  year      = {2024},
  doi       = {10.1609/icwsm.v18i1.31304},
  url       = {https://doi.org/10.1609/icwsm.v18i1.31304}
}

@article{gong2022virtual_brainstorming,
  author       = {Gong, Z. and Lee, L. H. and Soomro, S. A. and Nanjappan, V. and Georgiev, G. V.},
  title        = {A systematic review of virtual brainstorming from the perspective of creativity: affordances, framework, and outlook},
  journal      = {Digital Creativity},
  volume       = {33},
  number       = {2},
  pages        = {96--127},
  year         = {2022},
  doi          = {10.1080/14626268.2022.2064879},
  url          = {https://doi.org/10.1080/14626268.2022.2064879}
}

@inproceedings{CoralSpatial,
  author = {Olaosebikan, Monsurat and Aranda Barrios, Claudia and Kolawole, Blessing and Cowen, Lenore and Shaer, Orit},
  title = {Identifying Cognitive and Creative Support Needs for Remote Scientific Collaboration using VR: Practices, Affordances, and Design Implications},
  year = {2022},
  isbn = {9781450393270},
  publisher = {Association for Computing Machinery},
  address = {New York, NY, USA},
  url = {https://doi.org/10.1145/3527927.3532797},
  doi = {10.1145/3527927.3532797},
  booktitle = {Proceedings of the 14th Conference on Creativity and Cognition},
  pages = {97--110},
  numpages = {14},
  location = {Venice, Italy},
  series = {C\&C '22}
}

@article{Wilson2013,
  title={Using Brainwriting For Rapid Idea Generation},
  url={https://www.smashingmagazine.com/2013/12/using-brainwriting-for-rapid-idea-generation/},
  journal={Smashing Magazine},
  author={Wilson, Chauncey},
  year={2013},
  month={Dec}
}

@book{braun2012thematic,
  title={Thematic analysis.},
  author={Braun, Virginia and Clarke, Victoria},
  year={2012},
  publisher={American Psychological Association}
}

@book{Bono1999,
  address={New York},
  title={Six thinking hats},
  publisher={Back Bay Books},
  author={De Bono, Edward},
  year={1999}
}

@book{buzan1996mindmap,
  author       = {Buzan, Tony and Buzan, Barry},
  title        = {The Mind Map Book: How to Use Radiant Thinking to Maximize Your Brain's Untapped Potential},
  year         = {1996},
  edition      = {Reprint},
  publisher    = {Plume},
  address      = {New York},
  pages        = {320},
  isbn         = {978-0452273221}
}

@inproceedings{ishizaka2024sparkit,
  author       = {Ishizaka, Masaki and Taya, Akira and Tobe, Yoshiko},
  title        = {{SPARKIT: A Mind Map-Based MAS for Idea Generation Support}},
  booktitle    = {Engineering Multi-Agent Systems. EMAS 2024},
  editor       = {Briola, Davide and Cardoso, Rafael C. and Logan, Brian},
  series       = {Lecture Notes in Computer Science},
  volume       = {15152},
  year         = {2024},
  publisher    = {Springer, Cham},
  doi          = {10.1007/978-3-031-71152-7_1},
  url          = {https://doi.org/10.1007/978-3-031-71152-7_1}
}

@article{radensky2024scideator,
  author       = {Radensky, Marissa and Shahid, Simra and Fok, Raymond and Siangliulue, Pao and Hope, Tom and Weld, Daniel S.},
  title        = {Scideator: Human-LLM Scientific Idea Generation Grounded in Research-Paper Facet Recombination},
  year         = {2024},
  journal      = {arXiv preprint},
  volume       = {arXiv:2409.14634},
  url          = {https://arxiv.org/abs/2409.14634}
}

@inproceedings{HeCanvasCHIWork24,
  author = {He, Jessica and Houde, Stephanie and Gonzalez, Gabriel E. and Silva Moran, Dar\'{\i}o Andr\'{e}s and Ross, Steven I. and Muller, Michael and Weisz, Justin D.},
  title = {AI and the Future of Collaborative Work: Group Ideation with an LLM in a Virtual Canvas},
  year = {2024},
  isbn = {9798400710179},
  publisher = {Association for Computing Machinery},
  address = {New York, NY, USA},
  url = {https://doi.org/10.1145/3663384.3663398},
  doi = {10.1145/3663384.3663398},
  booktitle = {Proceedings of the 3rd Annual Meeting of the Symposium on Human-Computer Interaction for Work},
  articleno = {9},
  numpages = {14},
  location = {Newcastle upon Tyne, United Kingdom},
  series = {CHIWORK '24}
}

@misc{xmind,
  author       = {{XMind}},
  title        = {XMind: Mind Mapping and Brainstorming Software},
  year         = {2025},
  note         = {Accessed: 2025-01-03},
  url          = {https://xmind.app/},
  publisher    = {XMind Ltd.}
}

@article{eppler2006conceptmaps,
  author       = {Eppler, Martin J.},
  title        = {A Comparison between Concept Maps, Mind Maps, Conceptual Diagrams, and Visual Metaphors as Complementary Tools for Knowledge Construction and Sharing},
  journal      = {Information Visualization},
  volume       = {5},
  number       = {3},
  pages        = {202--210},
  year         = {2006},
  doi          = {10.1057/palgrave.ivs.9500131},
  url          = {https://doi.org/10.1057/palgrave.ivs.9500131}
}

@techreport{novak2008conceptmaps,
  author       = {Novak, Joseph D. and Ca{\~n}as, Alberto J.},
  title        = {The Theory Underlying Concept Maps and How to Construct and Use Them},
  year         = {2008},
  institution  = {Florida Institute for Human and Machine Cognition},
  type         = {Technical Report},
  number       = {IHMC CmapTools 2006-01 Rev 01-2008},
  address      = {Pensacola},
  url          = {http://cmap.ihmc.us/Publications/ResearchPapers/TheoryUnderlyingConceptMaps.pdf}
}

@incollection{ju2007mouse,
  title={The Mouse, the Demo, and the Big Idea},
  author={Ju, Wendy},
  booktitle={HCI Remixed: Reflections on Works That Have Influenced the HCI Community},
  editor={Erickson, Thomas and McDonald, David W.},
  publisher={MIT Press},
  address={Cambridge, MA},
  year={2008},
  url={https://direct.mit.edu/books/edited-volume/3814/chapter-abstract/125137/}
}

@inproceedings{engelbart1968research,
  title={A research center for augmenting human intellect},
  author={Engelbart, Douglas C and English, William K},
  booktitle={Proceedings of the December 9-11, 1968, fall joint computer conference, part I},
  pages={395--410},
  year={1968}
}

@article{mokryn2025interpretable,
  title={Interpretable transformation and analysis of timelines through learning via surprisability},
  author={Mokryn, Osnat and Lazebnik, Teddy and Ben-Shoshan, Hagit},
  journal={Chaos: An Interdisciplinary Journal of Nonlinear Science},
  volume={35},
  number={7},
  pages={073134},
  year={2025},
  doi={10.1063/5.0269365},
  url={https://doi.org/10.1063/5.0269365},
  publisher={AIP Publishing}
}

@inproceedings{olson1995mix,
  title={What mix of video and audio is useful for small groups doing remote real-time design work?},
  author={Olson, Judith S and Olson, Gary M and Meader, David K},
  booktitle={Proceedings of the SIGCHI conference on Human factors in computing systems},
  pages={362--368},
  year={1995}
}

@article{muller2011leaving,
  title={Leaving the beaten tracks in creative work--A design theory for systems that support convergent and divergent thinking},
  author={M{\"u}ller-Wienbergen, Felix and M{\"u}ller, Oliver and Seidel, Stefan and Becker, J{\"o}rg},
  journal={Journal of the Association for Information Systems},
  volume={12},
  number={11},
  pages={2},
  year={2011},
  url={https://aisel.aisnet.org/jais/vol12/iss11/2/}
}

@inproceedings{ansah2022need,
  author = {Ansah, Alberta A. and Xing, Yilun and Kamaraj, Amudha Varshini and Tosca, Diana and Boyle, Linda and Iqbal, Shamsi and Kun, Andrew L. and Lee, John D. and Pahud, Michel and Shaer, Orit},
  title = {''I need to respond to this'' -- Contributions to group creativity in remote meetings with distractions},
  year = {2022},
  isbn = {9781450396554},
  publisher = {Association for Computing Machinery},
  address = {New York, NY, USA},
  url = {https://doi.org/10.1145/3533406.3533411},
  doi = {10.1145/3533406.3533411},
  booktitle = {Proceedings of the 1st Annual Meeting of the Symposium on Human-Computer Interaction for Work},
  articleno = {7},
  numpages = {12},
  location = {Durham, NH, USA},
  series = {CHIWORK '22}
}

@article{mednick1968remote,
  title={The Remote Associates Test},
  author={Mednick, Sarnoff A.},
  journal={The Journal of Creative Behavior},
  volume={2},
  number={3},
  pages={213--214},
  year={1968},
  doi={10.1002/j.2162-6057.1968.tb00104.x},
  url={https://doi.org/10.1002/j.2162-6057.1968.tb00104.x}
}

@book{guilford1978alternate,
  title={Alternate Uses: Manual of Instructions and Interpretations},
  author={Guilford, Joy Paul and Christensen, Paul R. and Merrifield, Philip R. and Wilson, Robert C.},
  year={1978},
  publisher={Sheridan Psychological Services},
  address={Orange, CA}
}

@article{friedman2003attentional,
  title={Attentional priming effects on creativity},
  author={Friedman, Ronald S and Fishbach, Ayelet and F{\"o}rster, Jens and Werth, Lioba},
  journal={Creativity Research Journal},
  volume={15},
  number={2-3},
  pages={277--286},
  year={2003},
  publisher={Taylor \& Francis}
}

@article{grice1975logic,
  title={Logic and conversation},
  author={Grice, Herbert Paul},
  journal={Syntax and Semantics},
  volume={3},
  pages={43--58},
  year={1975}
}

@techreport{dell2025cybernetic,
  title={The cybernetic teammate: A field experiment on generative AI reshaping teamwork and expertise},
  author={Dell'Acqua, Fabrizio and Ayoubi, Charles and Lifshitz, Hila and Sadun, Raffaella and Mollick, Ethan and Mollick, Lilach and Han, Yi and Goldman, Jeff and Nair, Hari and Taub, Stewart and others},
  year={2025},
  institution={National Bureau of Economic Research},
  type={Working Paper},
  number={33641},
  url={https://www.nber.org/papers/w33641}
}

@article{lubart2005partners,
  author  = {Lubart, Todd},
  title   = {How can computers be partners in the creative process: Classification and commentary on the Special Issue},
  journal = {International Journal of Human-Computer Studies},
  volume  = {63},
  number  = {4-5},
  pages   = {365--369},
  year    = {2005},
  doi     = {10.1016/j.ijhcs.2005.04.002},
  url     = {https://doi.org/10.1016/j.ijhcs.2005.04.002}
}

@inproceedings{guzdial2019friend,
  author = {Guzdial, Matthew and Liao, Nicholas and Chen, Jonathan and Chen, Shao-Yu and Shah, Shukan and Shah, Vishwa and Reno, Joshua and Smith, Gillian and Riedl, Mark O.},
  title = {Friend, Collaborator, Student, Manager: How Design of an AI-Driven Game Level Editor Affects Creators},
  year = {2019},
  publisher = {Association for Computing Machinery},
  address = {New York, NY, USA},
  url = {https://doi.org/10.1145/3290605.3300854},
  doi = {10.1145/3290605.3300854},
  booktitle = {Proceedings of the 2019 CHI Conference on Human Factors in Computing Systems},
  pages = {1--13},
  series = {CHI '19}
}

@inproceedings{guo2025pen,
  author = {Guo, Alicia and Sathyanarayanan, Shreya and Wang, Leijie and Heer, Jeffrey and Zhang, Amy X.},
  title = {From Pen to Prompt: How Creative Writers Integrate AI into their Writing Practice},
  year = {2025},
  publisher = {Association for Computing Machinery},
  address = {New York, NY, USA},
  url = {https://doi.org/10.1145/3698061.3726910},
  doi = {10.1145/3698061.3726910},
  booktitle = {Proceedings of the 2025 Conference on Creativity and Cognition},
  pages = {527--545},
  series = {C\&C '25}
}

@article{mccomb2023focus,
  author  = {McComb, Christopher and Boatwright, Peter and Cagan, Jonathan},
  title   = {Focus and Modality: Defining a Roadmap to Future AI-Human Teaming in Design},
  journal = {Proceedings of the Design Society},
  volume  = {3},
  pages   = {1905--1914},
  year    = {2023},
  doi     = {10.1017/pds.2023.191},
  url     = {https://doi.org/10.1017/pds.2023.191},
  publisher = {Cambridge University Press}
}

@inproceedings{kantosalo2016modes,
  author    = {Kantosalo, Anna and Toivonen, Hannu},
  title     = {Modes for Creative Human-Computer Collaboration: Alternating and Task-Divided Co-Creativity},
  booktitle = {Proceedings of the Seventh International Conference on Computational Creativity (ICCC 2016)},
  editor    = {Pachet, Fran{\c{c}}ois and Cardoso, Amilcar and Corruble, Vincent and Ghedini, Fiammetta},
  pages     = {77--84},
  year      = {2016},
  publisher = {Sony CSL},
  address   = {Paris},
  isbn      = {978-274-669-155-1},
  url       = {http://www.computationalcreativity.net/iccc2016/proceedings-2016/}
}

@inproceedings{koch2019mayai,
  author = {Koch, Janin and Lucero, Andr{\'e}s and Hegemann, Lena and Oulasvirta, Antti},
  title = {May AI? Design Ideation with Cooperative Contextual Bandits},
  year = {2019},
  publisher = {Association for Computing Machinery},
  address = {New York, NY, USA},
  url = {https://doi.org/10.1145/3290605.3300863},
  doi = {10.1145/3290605.3300863},
  booktitle = {Proceedings of the 2019 CHI Conference on Human Factors in Computing Systems},
  pages = {1--12},
  series = {CHI '19}
}

@article{zhang2025agency,
  author  = {Zhang, Shuning and Wang, Hui and Yi, Xin},
  title   = {Exploring Collaboration Patterns and Strategies in Human-AI Co-creation through the Lens of Agency: A Scoping Review of the Top-tier HCI Literature},
  journal = {Proceedings of the ACM on Human-Computer Interaction},
  volume  = {9},
  number  = {7},
  numpages = {43},
  year    = {2025},
  doi     = {10.1145/3757594},
  url     = {https://doi.org/10.1145/3757594},
  publisher = {Association for Computing Machinery}
}

@inproceedings{horvitz1999mixed,
  author = {Horvitz, Eric},
  title = {Principles of mixed-initiative user interfaces},
  year = {1999},
  publisher = {Association for Computing Machinery},
  address = {New York, NY, USA},
  url = {https://doi.org/10.1145/302979.303030},
  doi = {10.1145/302979.303030},
  booktitle = {Proceedings of the SIGCHI Conference on Human Factors in Computing Systems},
  pages = {159--166},
  series = {CHI '99}
}

@inproceedings{amershi2019guidelines,
  author = {Amershi, Saleema and Weld, Dan and Vorvoreanu, Mihaela and Fourney, Adam and Nushi, Besmira and Collisson, Penny and Suh, Jina and Iqbal, Shamsi and Bennett, Paul N. and Inkpen, Kori and Teevan, Jaime and Kikin-Gil, Ruth and Horvitz, Eric},
  title = {Guidelines for Human-AI Interaction},
  year = {2019},
  publisher = {Association for Computing Machinery},
  address = {New York, NY, USA},
  url = {https://doi.org/10.1145/3290605.3300233},
  doi = {10.1145/3290605.3300233},
  booktitle = {Proceedings of the 2019 CHI Conference on Human Factors in Computing Systems},
  pages = {1--13},
  series = {CHI '19}
}

@inproceedings{gaver2003ambiguity,
  author = {Gaver, William W. and Beaver, Jacob and Benford, Steve},
  title = {Ambiguity as a resource for design},
  year = {2003},
  publisher = {Association for Computing Machinery},
  address = {New York, NY, USA},
  url = {https://doi.org/10.1145/642611.642653},
  doi = {10.1145/642611.642653},
  booktitle = {Proceedings of the SIGCHI Conference on Human Factors in Computing Systems},
  pages = {233--240},
  series = {CHI '03}
}

@inproceedings{sengers2006staying,
  author = {Sengers, Phoebe and Gaver, William},
  title = {Staying open to interpretation: engaging multiple meanings in design and evaluation},
  year = {2006},
  publisher = {Association for Computing Machinery},
  address = {New York, NY, USA},
  url = {https://doi.org/10.1145/1142405.1142422},
  doi = {10.1145/1142405.1142422},
  booktitle = {Proceedings of the 6th Conference on Designing Interactive Systems},
  pages = {99--108},
  series = {DIS '06}
}

\onecolumn

\begin{appendices}
\section*{Appendix}

\section{Supplementary Tables}
\label{appx:supptbls}
\renewcommand{\thetable}{A.\arabic{table}}
\setcounter{table}{0}



\newpage

\begin{table}[t]
\centering
\caption[MEC-SPQ presence results by platform]{\rev{Presence (MEC-SPQ) \cite{vorderer2004mec_spq} responses by platform, in administration order. Mdn = median; IQR = interquartile range. Mann-Whitney $U$ is computed with the Miro group first; $p$ values are two-tailed. Per-item sample sizes reflect the omissions documented in Methods.}}
\label{tab:presence-appendix}
\rev{%
\begin{tabular}{@{}p{0.40\linewidth}cc cc cc cc@{}}
\toprule
& \multicolumn{2}{c}{$N$} & \multicolumn{2}{c}{Miro} & \multicolumn{2}{c}{Noda} & & \\
\cmidrule(lr){2-3} \cmidrule(lr){4-5} \cmidrule(lr){6-7}
Item & Miro & Noda & Mdn & IQR & Mdn & IQR & $U$ & $p$ \\
\midrule
\emph{In the computer generated world I had a sense of ``being there''.}
  & 20 & 21 & 4 & 2--4 & 6 & 5--6 & 56.5 & $<$.001*** \\
\addlinespace
\emph{How real did the virtual world seem to you?} (first occurrence)
  & 19 & 21 & 4 & 4--5 & 4 & 3--5 & 235.0 & .325 \\
\addlinespace
\emph{I did not feel present in the virtual space.}
  & 19 & 21 & 4 & 3--4 & 5 & 3--6 & 140.0 & .102 \\
\addlinespace
\emph{How aware were you of the real world surrounding while navigating in the virtual world? (i.e. sounds, room temperature, other people, etc.)?}
  & 19 & 21 & 4 & 2.5--4 & 5 & 4--6 & 113.0 & .017* \\
\addlinespace
\emph{How much did your experience in the virtual environment seem consistent with your real world experience?}
  & 19 & 21 & 4 & 3--4 & 3 & 2--4 & 233.5 & .323 \\
\addlinespace
\emph{How real did the virtual world seem to you?} (second occurrence)
  & 19 & 21 & 4 & 3.5--4 & 2 & 1--3 & 273.5 & .039* \\
\addlinespace
\emph{Somehow I felt that the virtual world surrounded me.}
  & 20 & 21 & 4 & 2--4 & 6 & 5--6 & 50.0 & $<$.001*** \\
\addlinespace
\emph{I felt like I was just perceiving pictures.}
  & 20 & 21 & 4 & 4--5.25 & 3 & 2--4 & 339.5 & $<$.001*** \\
\addlinespace
\emph{I was not aware of my real environment.}
  & 20 & 21 & 4 & 2--4 & 4 & 3--6 & 155.0 & .142 \\
\addlinespace
\emph{I had a sense of acting in the virtual space, rather than operating something from outside.}
  & 20 & 20 & 4 & 4--4.25 & 6 & 4--6 & 106.5 & .010** \\
\addlinespace
\emph{I felt present in the virtual space.}
  & 20 & 21 & 4 & 4--4 & 5 & 4--6 & 100.0 & .003** \\
\addlinespace
\emph{I was completely captivated by the virtual world.}
  & 20 & 21 & 4 & 3--4 & 5 & 3--5 & 158.0 & .168 \\
\addlinespace
\emph{The virtual world seemed more realistic than the real world.}
  & 20 & 21 & 4 & 2--4 & 1 & 1--2 & 326.5 & .002** \\
\addlinespace
\emph{I still paid attention to the real environment.}
  & 20 & 21 & 4 & 4--6.25 & 4 & 2--5 & 263.0 & .159 \\
\bottomrule
\end{tabular}%
}
 
\smallskip
\rev{\footnotesize Note. *** $p < .001$, ** $p < .01$, * $p < .05$. Item wording is verbatim; response options are listed in Table~B.8. The two ``How real did the virtual world seem to you?'' items are distinct questionnaire items with identical stems and different response anchors (first occurrence: Fully disagree to Fully agree; second occurrence: about as real as an imagined world to indistinguishable from the real world); both are reported. Higher values on ``I did not feel present,'' ``aware of the real world,'' ``just perceiving pictures,'' and ``I still paid attention to the real environment'' indicate lower presence.}
\end{table}

\begin{table}[t]
\centering
\caption[NASA-TLX results by platform]{\rev{NASA-TLX responses by platform. Items were administered on a 1--7 scale. Mdn = median; IQR = interquartile range. Mann-Whitney $U$ is computed with the planar Miro group first; $p$ values are two-tailed.}}
\label{tab:tlx-appendix}
\rev{%
\begin{tabular}{@{}p{0.36\linewidth}cccccc@{}}
\toprule
& \multicolumn{2}{c}{Miro} & \multicolumn{2}{c}{Noda} & & \\
\cmidrule(lr){2-3} \cmidrule(lr){4-5}
TLX item & Mdn & IQR & Mdn & IQR & $U$ & $p$ \\
\midrule
Mental demand: \emph{How mentally demanding was the task?}
  & 4.00 & 3.10--4.90 & 4.30 & 3.40--5.20 & 180.0 & .312 \\
\addlinespace
Physical demand: \emph{How physically demanding was the task?}
  & 1.00 & 1.00--1.90 & 1.90 & 1.30--3.70 & 126.0 & .015* \\
\addlinespace
Pace: \emph{How hurried or rushed was the pace of the task?}
  & 4.30 & 3.70--4.90 & 4.30 & 3.10--4.90 & 241.5 & .603 \\
\addlinespace
Success: \emph{How successful were you in accomplishing what you were asked to do?}
  & 5.20 & 4.60--5.50 & 5.20 & 4.30--6.10 & 212.0 & .840 \\
\addlinespace
Effort: \emph{How hard did you have to work to accomplish your level of performance?}
  & 4.30 & 3.40--4.90 & 4.60 & 3.10--5.20 & 206.5 & .733 \\
\addlinespace
Frustration: \emph{How insecure, discouraged, irritated, stressed, and annoyed were you?}
  & 3.10 & 1.60--4.30 & 2.80 & 2.20--4.30 & 229.0 & .840 \\
\bottomrule
\end{tabular}%
}
 
\smallskip
\rev{\footnotesize Note. $N = 21$ per platform; no missing responses. * $p < .05$. Physical demand is the only item differing significantly between platforms.}
\end{table}

\begin{table}[t]
\centering
\caption[Collaboration perceptions by platform]{\rev{Perceptions of collaboration by platform. All items were administered on five-point scales (anchors in Table~B.8); responses were coded 1--5. Mdn = median; IQR = interquartile range. Mann-Whitney $U$ is computed with the planar Miro group first; $p$ values are two-tailed. $N = 21$ per platform; no missing responses.}}
\label{tab:collab-appendix}
\rev{%
\begin{tabular}{@{}p{0.44\linewidth}cc cc cc@{}}
\toprule
& \multicolumn{2}{c}{Miro} & \multicolumn{2}{c}{Noda} & & \\
\cmidrule(lr){2-3} \cmidrule(lr){4-5}
Item & Mdn & IQR & Mdn & IQR & $U$ & $p$ \\
\midrule
\emph{To what extent did you feel that team decisions were made collaboratively, taking into account different perspectives?}
  & 4 & 3--4 & 4 & 4--5 & 160.5 & .108 \\
\addlinespace
\emph{Did you feel comfortable providing feedback to your team members and receiving feedback from them?}
  & 4 & 4--5 & 5 & 4--5 & 148.0 & .047* \\
\addlinespace
\emph{How well do you think team members understood and appreciated each other's strengths and weaknesses?}
  & 3 & 3--4 & 4 & 3--4 & 182.0 & .322 \\
\addlinespace
\emph{How much of the time did you feel that everyone's ideas were encouraged and considered during team decision-making?}
  & 4 & 4--5 & 4 & 4--5 & 235.0 & .694 \\
\addlinespace
\emph{On a scale of 1-5, how satisfied are you with the overall level of collaboration within the team?}
  & 4 & 4--5 & 5 & 4--5 & 152.0 & .063 \\
\bottomrule
\end{tabular}%
}
 
\smallskip
\rev{\footnotesize Note. * $p < .05$. Comfort with feedback is the only item differing significantly between platforms.}
\end{table}


\begin{table}[t]
\centering
\caption[Idea quality and ideation process by platform]{\rev{Participant perceptions of the quality of ideas and the ideation process by platform. All items were administered on a five-point agreement scale (strongly disagree to strongly agree), coded 1--5. Mdn = median; IQR = interquartile range. Mann-Whitney $U$ is computed with the planar Miro group first; $p$ values are two-tailed. $N = 21$ per platform; no missing responses.}}
\label{tab:quality-process-appendix}
\rev{%
\begin{tabular}{@{}p{0.46\linewidth}cc cc cc@{}}
\toprule
& \multicolumn{2}{c}{Miro} & \multicolumn{2}{c}{Noda} & & \\
\cmidrule(lr){2-3} \cmidrule(lr){4-5}
Item & Mdn & IQR & Mdn & IQR & $U$ & $p$ \\
\midrule
\multicolumn{7}{@{}l}{\emph{Own ideas}} \\
The ideas I contributed were relevant to the problem.
  & 4 & 4--5 & 5 & 4--5 & 174.5 & .190 \\
The ideas I contributed were innovative.
  & 4 & 3--4 & 4 & 3--4 & 203.5 & .657 \\
The ideas I contributed were insightful.
  & 3 & 3--4 & 4 & 3--4 & 180.5 & .295 \\
\addlinespace
\multicolumn{7}{@{}l}{\emph{AI-generated ideas}} \\
The ideas generated by AI were relevant to the problem.
  & 4 & 4--5 & 4 & 4--5 & 258.5 & .303 \\
The ideas generated by AI were innovative.
  & 3 & 3--4 & 2 & 2--3 & 298.5 & .046* \\
The ideas generated by AI were insightful.
  & 4 & 3--4 & 2 & 2--3 & 334.0 & .004** \\
\addlinespace
\multicolumn{7}{@{}l}{\emph{Group-authored ideas (labeled ``the team'' in the questionnaire)}} \\
Overall, the ideas created by the team were relevant to the problem.
  & 5 & 4--5 & 5 & 5--5 & 147.0 & .010* \\
Overall, the ideas created by the team were innovative.
  & 4 & 3--5 & 4 & 4--5 & 175.0 & .212 \\
Overall, the ideas created by the team were insightful.
  & 4 & 4--5 & 4 & 4--5 & 205.5 & .687 \\
\addlinespace
\multicolumn{7}{@{}l}{\emph{Ideation process}} \\
Using AI in our idea generation sessions made us consider new ideas.
  & 5 & 4--5 & 4 & 4--5 & 275.0 & .133 \\
Using AI in our idea generation session improved the ideation session results in terms of quantity of ideas.
  & 5 & 4--5 & 4 & 4--5 & 284.5 & .080 \\
Using AI in our idea generation session improved the ideation session results in terms of quality of ideas.
  & 4 & 3--5 & 4 & 3--4 & 274.5 & .159 \\
\bottomrule
\end{tabular}%
}
 
\smallskip
\rev{\footnotesize Note. * $p < .05$, ** $p < .01$. The full item wording, including the dimension glosses shown to participants, is given in Table~B.8 (Q22\_1 to Q22\_12). The nine idea-quality items are the item-level counterparts of the cross-platform comparisons in Table~7; the parametric and rank-based tests agree on every significance decision except AI-generated innovation (see Methods).}
\end{table}


\newpage

\label{appx:supptbls}

\renewcommand{\thetable}{B.\arabic{table}}
\setcounter{table}{0}
\begin{table}[t]
\centering
\caption[Codebook, Q19]{\rev{Codebook for Q19: What challenges did you have interacting as a team?}}
\label{tab:codebook-q19}
\rev{%
\footnotesize
\begin{tabular}{@{}p{0.13\linewidth}p{0.20\linewidth}p{0.10\linewidth}p{0.49\linewidth}@{}}
\toprule
Theme & Code & Participant & Quote \\
\midrule
No challenges & None & P\_3 (Miro) & There was pretty much no challenges. We all helped each other and listened to each other's ideas. \\
\midrule
Unfamiliarity & Unfamiliarity with teammates & P\_45 (Noda) & It was harder to exchange ideas with the person I didn't know prior to the experiment because I didn't know their name \\
\addlinespace
 & Application navigation & P\_28 (Noda) & It was a bit awkward moving around the space at first, and all of our nodes were in different areas, so I couldn't really see what other people were thinking without moving around. \\
 &  & P\_1 (Miro) & Just navigating the site, learning different short cuts \\
\addlinespace
 & Lack of subject knowledge & P\_33 (Miro) & It was a bit slow at first because we do not know each other, and we do not know a lot of the specifics of the problem we were tasked to solve \\
\midrule
Disconnection with Teammates & Perceived virtual distance & P\_29 (Noda) & Not being able to physically move to see each other's ideas; being really far away in the beginning, etc. \\
\addlinespace
 & Lack of face-to-face communication & P\_30 (Noda) & The space awareness made it a bit complicated to connect with the team; I like to look at the person speaking in the eyes to use my full attention to the ideas of the presenter, but when we were discussing the ideas in the virtual world, it as very hard to follow who is speaking and it took me twice as much effort to process all the information only through the speech \\
 &  & P\_4 (Miro) & Being able to speak to each other without interrupting since we can't see each other's face. \\
\midrule
Technical Issues & Technical issues with apps & P\_49 (Noda) & glitches in making large mind maps \\
 &  & P\_6 (Miro) & The Miro platform was kind of irritating to use at the same time as others, since you couldn't edit a node while someone else was editing, and I kept having to zoom in and out to see everything. \\
\addlinespace
 & Technical issues with hardware & P\_18 (Noda) & Functioning with the joystick \\
\addlinespace
 & Issues with volume/hearing & P\_47 (Noda) & Hearing them from across the room and trying to navigate the mind maps getting in the way virtually. \\
\bottomrule
\end{tabular}%
}
\end{table}
 
\begin{table}[t]
\centering
\caption[Codebook, Q20]{\rev{Codebook for Q20: In what ways did using AI features contribute to or hinder the idea generation process?}}
\label{tab:codebook-q20}
\rev{%
\footnotesize
\begin{tabular}{@{}p{0.13\linewidth}p{0.20\linewidth}p{0.10\linewidth}p{0.49\linewidth}@{}}
\toprule
Theme & Code & Participant & Quote \\
\midrule
Positive & Quick & P\_13 (Miro) & We only had 3 minutes to generate ideas and in that time each only had 2-5 ideas, whereas using AI we each found 10+ ideas in seconds. \\
\addlinespace
 & Many ideas & P\_16 (Noda) & They contributed by generating many ideas. \\
 &  & P\_14 (Miro) & Using AI greatly expanded our idea pool, although sometimes it gave redundant or un-innovative answers. \\
\addlinespace
 & Provided a starting point, simplified & P\_25 (Miro) & It helped give a starting point for the individuals and let us think more creatively about ideas we would not have come up with by ourselves. \\
 &  & P\_43 (Noda) & the AI features didn't necessarily come up with concrete ideas but they did provide me with good building blocks, like it suggested cost saving and that made me think of doing a clothing share [...] \\
\addlinespace
 & Insightful or high quality & P\_20 (Miro) & I thought AI contributed to the idea generation process because it added project ideas I hadn't considered, such as a sustainable fashion show. \\
 &  & P\_41 (Noda) & The AI was able to help me think outside of the box and helped me realize different ways to achieve the same goal. \\
\midrule
Negative & Hindered & P\_51 (Noda) & It was only giving us ideas of my idea without any real content which hindered the idea generation process. \\
 &  & P\_45 (Noda) & I feel like I would've typed out my ideas much faster \\
\addlinespace
 & Lack of relevancy or low quality & P\_1 (Miro) & AI generated a lot of ideas quickly, but some of them were not relevant to Wellesley specifically \\
 &  & P\_29 (Noda) & I think the AI feature is very superficial and obvious, even with more specific prompts. It is good for an initial brainstorming, but as you move on in a project, you become the expert in what makes sense for your project, and AI's suggestions no longer seemed relevant/insightful. \\
\addlinespace
 & Overwhelming & P\_27 (Miro) & The AI features gave a lot of new ideas, but it was overwhelming in a way because it would give too many ideas or repeat the same ideas in different categories. \\
 &  & P\_28 (Noda) & [...] I don't think all of them were super original or meaningful, but there were so many that some of them had to be good. \\
\addlinespace
 & Replaced human ingenuity & P\_20 (Miro) & [...] It didn't hinder the idea generation process as we came up with our own ideas first. However, I found that after using AI to generate ideas in the second part of the challenge, I was having a harder time coming up with my own ideas. \\
\bottomrule
\end{tabular}%
}
\end{table}
 
\begin{table}[t]
\centering
\caption[Codebook, Q21]{\rev{Codebook for Q21: What positive experiences did you have interacting as a team?}}
\label{tab:codebook-q21}
\rev{%
\footnotesize
\begin{tabular}{@{}p{0.13\linewidth}p{0.20\linewidth}p{0.10\linewidth}p{0.49\linewidth}@{}}
\toprule
Theme & Code & Participant & Quote \\
\midrule
Dialogue & Agreement between teammates & P\_20 (Miro) & I thought it was great to get to brainstorm solutions to a problem I hadn't considered in depth before with a group of people, and I thought we came up with some great ideas together. I had a positive experience because I felt that all our perspectives were represented in the final product. \\
\addlinespace
 & Compatible/respectful dialogue & P\_28 (Noda) & I think we did a good job of communicating, so asking questions and offering feedback \\
 &  & P\_15 (Miro) & We didn't talk over each other and listened to everyone's ideas. \\
\midrule
Task & Physically seeing each other's ideas & P\_38 (Noda) & It was fun to talk and see the physical display of ideas in front of each person, it feels like we all contributed. \\
\addlinespace
 & New insight/perspective & P\_42 (Noda) & I enjoyed brainstorming and learning about ideas I had not thought about. My team had prompted the AI in different ways which led to drastically different results which was interesting \\
\addlinespace
 & Ability to focus on task well & P\_6 (Miro) & We were able to talk through our ideas and build on them with each other, expressing agreement or disagreement and using that to come up with a final result. \\
 &  & P\_30 (Noda) & The team-building or the ice-breaking part went way faster than in the real life, I think it somehow affects that we don't spend time evaluating peoples appearances or body language which sometimes makes people feel colder to each other at the very beginning. but in teh virtual world we were fully focused on the task, without any second thought on who I am working with \\
\bottomrule
\end{tabular}%
}
\end{table}
 
\begin{table}[t]
\centering
\caption[Codebook, Q25]{\rev{Codebook for Q25: In what ways did using AI features contribute to or hinder collaboration with your team members?}}
\label{tab:codebook-q25}
\rev{%
\footnotesize
\begin{tabular}{@{}p{0.13\linewidth}p{0.20\linewidth}p{0.10\linewidth}p{0.49\linewidth}@{}}
\toprule
Theme & Code & Participant & Quote \\
\midrule
Positive & Generated more ideas or expanded on previous & P\_43 (Noda) & it helped each of us come up with more ideas to share between the group. \\
 &  & P\_2 (Miro) & AI helped with expanding on our ideas and adding more details. \\
\addlinespace
 & Helped break the ice & P\_16 (Noda) & The AI features helped us find common ground in our ideas. \\
\addlinespace
 & Starting point for collaboration & P\_40 (Noda) & It seemed we mainly used AI features beforehand that we shared with each other, but did not use while collaborating. So, I would say that AI only helped our individual selves share our ideas with each other, but it did not contribute to our full group ideas. \\
 &  & P\_33 (Miro) & The AI allowed us to easily pick from a larger set of possibilities without having to spend much time thinking beforehand. However, most of our discussion revolved around specifics which the AI did not know about. \\
\midrule
Hindered & Required excessive pruning & P\_28 (Noda) & I think that, although the AI gave a lot of ideas and suggestions, it became overwhelming how many there were, and it became a little challenging to sort through everything. \\
 &  & P\_15 (Miro) & AI would give a lot of ideas, which made it difficult to filter through. When we would all click the AI tool, a whole bunch of branches would appear making it hard to read. \\
\addlinespace
 & Technically overwhelming & P\_13 (Miro) & It hindered collaboration because we all were struggling with using the features and it made mistakes a few times, so we were very consumed in the individual ways we were applying it. \\
\midrule
Neither & Neither & P\_30 (Noda) & Didn't really notice the difference, in the idea discussion we did pretty much the same both with and without AI \\
 &  & P\_27 (Miro) & I don't think using the AI features contributed to or hindered team collaboration. It really only affected he idea generation process. \\
\addlinespace
 & Reduced need to collaborate & P\_1 (Miro) & I think because it was so easy, there wasn't a huge reason to collaborate when generating ideas. \\
 &  & P\_29 (Noda) & [...] using AI to brainstorm alone then share my findings actually produced less-insightful findings than I had previously when I was brainstorming without AI. \\
\bottomrule
\end{tabular}%
}
\end{table}
 
\begin{table}[t]
\centering
\caption[Codebook, Q27]{\rev{Codebook for Q27: Given your answer to the above question (Q26), why do you think your team had that preference?}}
\label{tab:codebook-q27}
\rev{%
\footnotesize
\begin{tabular}{@{}p{0.13\linewidth}p{0.20\linewidth}p{0.10\linewidth}p{0.49\linewidth}@{}}
\toprule
Theme & Code & Participant & Quote \\
\midrule
No preference & None & P\_47 (Noda) & Good ideas were able to come from both my team mates and the AI. \\
\midrule
Preferred AI & Generated higher number of ideas & P\_15 (Miro) & AI had more ideas instantly. We had a short amount of time to think of ideas, which made it difficult to fully process. \\
 &  & P\_45 (Noda) & The AI generated way more answers but their answers were also way more general. So it was important to balance out its ideas with our understanding of Wellesley. \\
\addlinespace
 & Provided a starting point & P\_39 (Noda) & human generated ideas seemed more specific and particularly relevant to wellesley, but AI helped create a jumping off point for them \\
\addlinespace
 & More detailed/insightful & P\_13 (Miro) & The AI generated ideas went much more in depth, whereas ours were more bullet points or non descriptive ideas. \\
\addlinespace
 & Easier to use AI & P\_18 (Noda) & I feel like most of our ideas were similar to the AI generated ideas, therefore ultimately we preferred our own. \\
\midrule
Preferred Human & More applicable/relevant & P\_52 (Noda) & Because AI lacked the context of our school culture and our experiences here, so its ideas were not fully applicable or relevent \\
 &  & P\_4 (Miro) & We are more closely related to the prompt so it makes sense for us to speak on it, whereas the ai doesn't necessarily know what it's like to be a student at Wellesley College. \\
\addlinespace
 & More original/creative & P\_30 (Noda) & AI ideas made the diversity but were very generic, therefore it was way harder to include that into the final proposal of the project, while human-generated ideas already had some depth and development, plus were more applied to the context we are in. \\
\addlinespace
 & Easier to use human ideas & P\_40 (Noda) & I think our team had that preference because we would rather work out the problem through communication with each other. It would take time to pause our conversation and use AI to create a concrete plan. It seemed we also preferred the creativity of our minds that wasn't necessarily provided by the AI. We also know our campus more personally than AI, so we believed better idea would come from our minds. \\
\bottomrule
\end{tabular}%
}
\end{table}
 
\begin{table}[t]
\centering
\caption[Codebook, Q29]{\rev{Codebook for Q29: Think about the platform you used to complete the study (either Miro or Noda). To what extent was this platform helpful? Why or why not?}}
\label{tab:codebook-q29}
\rev{%
\footnotesize
\begin{tabular}{@{}p{0.13\linewidth}p{0.20\linewidth}p{0.10\linewidth}p{0.49\linewidth}@{}}
\toprule
Theme & Code & Participant & Quote \\
\midrule
Helpful & Applicable to future & P\_3 (Miro) & I feel like I would definitely use this website for my personal planning taks in the future. \\
\addlinespace
 & Good organization/visualization & P\_41 (Noda) & The platform, Noda, was very helpful in brainstorming, connecting ideas, and keeping track of them. The only problem I had was that I was running out of space and it started to become cluttered. \\
\addlinespace
 & Inspired collaboration/creativity & P\_14 (Miro) & The platform was very helpful in generating ideas. The mind map allowed us to branch our ideas off accordingly and the AI feature helped us generate new ideas when we were stuck. \\
 &  & P\_16 (Noda) & Noda, allowed us to visualize our thoughts, connect the, and expand on them with ease. \\
\addlinespace
 & Starting point & P\_5 (Miro) & I think Miro is a good start, but in terms of planning and giving more actions, we need more effort. not just relying on this platform \\
 &  & P\_15 (Miro) & Miro was useful because it allowed us to organize our thoughts in a way that is easier to understand. The AI tool was also helpful in this case, as I was having some trouble getting started. \\
\midrule
Challenging & Unfamiliarity & P\_4 (Miro) & It was good organization wise because it has an easy way to look at everything in a web map. However, it felt hard to move around and zooming in and out was a little frustrating. \\
\addlinespace
 & Limited functionality/accessibility & P\_17 (Noda) & Noda was not as helpful because you have to type the ideas using a one finger at a time which takes a lot longer than typing. The drawing of the lines to connect the ideas is helpful but I think it could be done on an ipad. \\
 &  & P\_21 (Miro) & I thought Miro was somewhat helpful, and I liked the customizability of it. I did feel like understanding the controls and features hindered me a little, and we could have achieved the same results faster on a whiteboard. \\
 &  & P\_29 (Noda) & I think the Noda platform is extremely helpful to brainstorm or meet as a group to discuss a project. It allows you to feel like you're getting together physically without needing to actually meet. The down side might be just the costs of the VR glasses and platform, and the physical effects it might cause (strained eyes, etc). \\
\addlinespace
 & Visually confusing or unhelpful & P\_47 (Noda) & Pretty helpful, I think the UI coudlve been set up in a visually-simpler way. \\
 &  & P\_13 (Miro) & Miro was helpful but a little bit confusing to implement the AI tool and not very visually appealing. \\
\bottomrule
\end{tabular}%
}
\end{table}

\begin{table}[t]
\centering
\caption[Codebook, video analysis]{\rev{Codebook for the video analysis: group collaboration strategies for the study tasks.}}
\label{tab:codebook-video}
\rev{%
\footnotesize
\begin{tabular}{@{}p{0.16\linewidth}p{0.16\linewidth}p{0.60\linewidth}@{}}
\toprule
Theme & Code & Example \\
\midrule
Teamwork Strategy & Individual & M7 - Subtask 1: all participants started working on the task in separate areas of the Miro board, no collaboration in any way. \\
\addlinespace
 & Equal Collaboration & M4 - Subtask 3: Participants created one central shared mind map/nodes for listing the 3 ideas. \\
\addlinespace
 & Driver/Navigator & N6 - Subtask 3: P2 verbally suggests and shares out ideas while P1 and P3 create and label the nodes. \\
\midrule
Physical state (of VR avatar or cursor) & Facing & N3 - Subtask 1: Participants made individual small maps, remained in their spawn position, but are all facing the center of the room. \\
\addlinespace
 & Not Facing & N2 - Subtask 1: Participants are in a triangle formation but facing outwards. \\
\addlinespace
 & Long Distance & M5 - Subtask 1: Participants work in opposite corners of the Miro Board and their maps do not touch. \\
\addlinespace
 & Short Distance & N6 - Subtask 2: While P3 is further away, P1 and P2 have mind maps that begin to overlap as they develop. \\
\addlinespace
 & Lots of Movement & M3 - Subtask 3: Participants copy/paste the content of previously existing nodes into a new node under a sticky note that denotes the cost of the idea. Lots of movement with cursors and with the creation of new maps or re-organization of existing maps. \\
\addlinespace
 & Little to No Movement & N4: After choosing the 3 ideas, participants changed the color of corresponding nodes to mark it in their own maps, they did not congregate or create a new map. \\
\bottomrule
\end{tabular}%
}
 
\smallskip
\rev{\footnotesize Note. M = Miro group; N = Noda group. The Facing and Not Facing codes apply only to the immersive Noda platform, where avatars have a facing direction. The Equal Collaboration and Driver/Navigator codes are the basis of the equal and unequal collaboration classifications reported in the Results; the spatial codes were operationalized per platform as described in the Video Analysis section.}
\end{table}
 
\newpage
 
\begin{table*}[p]
\begingroup
\footnotesize
\centering
\caption{Survey questions and response options as administered (Part 1 of 2: workload, platform, presence, and collaboration items). Curly-bracketed lists give the full response options; for endpoint-labeled scales, the anchors shown to participants are listed. Items Q1--Q6 (NASA-TLX) were recorded on a 1--7 scale.}
\label{tab:questionnaire}
\begin{tabular}{@{}p{0.08\linewidth}p{0.47\linewidth}p{0.41\linewidth}@{}}
\toprule
Key & Question & Response options \\
\midrule
Q1 & How mentally demanding was the task? & Scale: \{Very Low, Medium, Very High\} \\
Q2 & How physically demanding was the task? & Scale: \{Very Low, Medium, Very High\} \\
Q3 & How hurried or rushed was the pace of the task? & Scale: \{Very Low, Medium, Very High\} \\
Q4 & How successful were you in accomplishing what you were asked to do? & Scale: \{Very Low, Medium, Very High\} \\
Q5 & How hard did you have to work to accomplish your level of performance? & Scale: \{Very Low, Medium, Very High\} \\
Q6 & How insecure, discouraged, irritated, stressed, and annoyed were you? & Scale: \{Very Low, Medium, Very High\} \\
\addlinespace
Q7 & Which platform did you use to complete the study? & \{Noda (immersive VR space)\}; \{Miro (browser-based whiteboard space)\} \\
\addlinespace
\multicolumn{3}{@{}p{0.96\linewidth}}{\emph{Instructions: The following questions have to do with the platform on which you completed the task. Please indicate the extent to which each statement applies to your experience.}} \\
\addlinespace
Q8\_1 & Somehow I felt that the virtual world surrounded me. & Likert scale: \{Fully Disagree, Fully Agree\} \\
Q8\_2 & I felt like I was just perceiving pictures. & Likert scale: \{Fully Disagree, Fully Agree\} \\
Q8\_3 & I was not aware of my real environment. & Likert scale: \{Fully Disagree, Fully Agree\} \\
Q8\_4 & I had a sense of acting in the virtual space, rather than operating something from outside. & Likert scale: \{Fully Disagree, Fully Agree\} \\
Q8\_5 & I felt present in the virtual space. & Likert scale: \{Fully Disagree, Fully Agree\} \\
Q8\_6 & I was completely captivated by the virtual world. & Likert scale: \{Fully Disagree, Fully Agree\} \\
Q8\_7 & The virtual world seemed more realistic than the real world. & Likert scale: \{Fully Disagree, Fully Agree\} \\
Q8\_8 & I still paid attention to the real environment. & Likert scale: \{Fully Disagree, Fully Agree\} \\
Q9 & How aware were you of the real world surrounding while navigating in the virtual world? (i.e. sounds, room temperature, other people, etc.)? & Likert scale: \{Fully Disagree, Fully Agree\} \\
Q10 & I did not feel present in the virtual space. & Likert scale: \{Fully Disagree, Fully Agree\} \\
Q11 & In the computer generated world I had a sense of ``being there''. & Likert scale: \{Fully Disagree, Fully Agree\} \\
Q12 & How real did the virtual world seem to you? & Likert scale: \{Fully Disagree, Fully Agree\} \\
\addlinespace
\multicolumn{3}{@{}p{0.96\linewidth}}{\emph{Instructions: The following questions have to do with your experience of collaboration during the task.}} \\
\addlinespace
Q14 & To what extent did you feel that team decisions were made collaboratively, taking into account different perspectives? & \{Not well at all\}; \{Slightly well\}; \{Moderately well\}; \{Very well\}; \{Extremely well\} \\
Q15 & Did you feel comfortable providing feedback to your team members and receiving feedback from them? & \{Extremely uncomfortable\}; \{Somewhat uncomfortable\}; \{Neither comfortable nor uncomfortable\}; \{Somewhat comfortable\}; \{Extremely comfortable\} \\
Q16 & How well do you think team members understood and appreciated each other's strengths and weaknesses? & \{Not well at all\}; \{Slightly well\}; \{Moderately well\}; \{Very well\}; \{Extremely well\} \\
Q17 & How much of the time did you feel that everyone's ideas were encouraged and considered during team decision-making? & \{Never\}; \{Sometimes\}; \{About half the time\}; \{Most of the time\}; \{Always\} \\
Q18 & On a scale of 1-5, how satisfied are you with the overall level of collaboration within the team? & \{Extremely dissatisfied\}; \{Somewhat dissatisfied\}; \{Neither satisfied nor dissatisfied\}; \{Somewhat satisfied\}; \{Extremely satisfied\} \\
\bottomrule
\end{tabular}
\endgroup
\end{table*}
 
\begin{table*}[p]
\begingroup
\footnotesize
\centering
\addtocounter{table}{-1}
\caption{Survey questions and response options as administered (Part 2 of 2: open-text items, idea-quality and ideation-process items, and additional presence and preference items). Response options are given as in Part 1.}
\label{tab:questionnaire2}
\begin{tabular}{@{}p{0.08\linewidth}p{0.47\linewidth}p{0.41\linewidth}@{}}
\toprule
Key & Question & Response options \\
\midrule
Q19 & What challenges did you have interacting as a team? & Open text response \\
Q20 & In what ways did using AI features contribute to or hinder the idea generation process? & Open text response \\
Q21 & What positive experiences did you have interacting as a team? & Open text response \\
\addlinespace
Q22\_1 & The ideas I contributed were relevant to the problem. & \{Strongly disagree\}; \{Somewhat disagree\}; \{Neither agree nor disagree\}; \{Somewhat agree\}; \{Strongly agree\} \\
Q22\_2 & The ideas I contributed were innovative--- reflected original and creative ways to think of the problem. & \{Strongly disagree\}; \{Somewhat disagree\}; \{Neither agree nor disagree\}; \{Somewhat agree\}; \{Strongly agree\} \\
Q22\_3 & The ideas I contributed were insightful--- reflected deep and perceptive ways to think of the problem. & \{Strongly disagree\}; \{Somewhat disagree\}; \{Neither agree nor disagree\}; \{Somewhat agree\}; \{Strongly agree\} \\
Q22\_4 & The ideas generated by AI were relevant to the problem. & \{Strongly disagree\}; \{Somewhat disagree\}; \{Neither agree nor disagree\}; \{Somewhat agree\}; \{Strongly agree\} \\
Q22\_5 & The ideas generated by AI were innovative--- reflected original and creative ways to think of the problem. & \{Strongly disagree\}; \{Somewhat disagree\}; \{Neither agree nor disagree\}; \{Somewhat agree\}; \{Strongly agree\} \\
Q22\_6 & The ideas generated by AI were insightful--- reflected deep and perceptive ways to think of the problem. & \{Strongly disagree\}; \{Somewhat disagree\}; \{Neither agree nor disagree\}; \{Somewhat agree\}; \{Strongly agree\} \\
Q22\_7 & Overall, the ideas created by the team were relevant to the problem. & \{Strongly disagree\}; \{Somewhat disagree\}; \{Neither agree nor disagree\}; \{Somewhat agree\}; \{Strongly agree\} \\
Q22\_8 & Overall, the ideas created by the team were innovative--- reflected original and creative ways to think of the problem. & \{Strongly disagree\}; \{Somewhat disagree\}; \{Neither agree nor disagree\}; \{Somewhat agree\}; \{Strongly agree\} \\
Q22\_9 & Overall, the ideas created by the team were insightful - reflected deep and perceptive ways to think of the problem. & \{Strongly disagree\}; \{Somewhat disagree\}; \{Neither agree nor disagree\}; \{Somewhat agree\}; \{Strongly agree\} \\
Q22\_10 & Using AI in our idea generation sessions made us consider new ideas. & \{Strongly disagree\}; \{Somewhat disagree\}; \{Neither agree nor disagree\}; \{Somewhat agree\}; \{Strongly agree\} \\
Q22\_11 & Using AI in our idea generation session improved the ideation session results in terms of quantity of ideas. & \{Strongly disagree\}; \{Somewhat disagree\}; \{Neither agree nor disagree\}; \{Somewhat agree\}; \{Strongly agree\} \\
Q22\_12 & Using AI in our idea generation session improved the ideation session results in terms of quality of ideas. & \{Strongly disagree\}; \{Somewhat disagree\}; \{Neither agree nor disagree\}; \{Somewhat agree\}; \{Strongly agree\} \\
\addlinespace
Q23 & How much did your experience in the virtual environment seem consistent with your real world experience? & Likert scale: \{Not consistent, Moderately consistent, Very consistent\} \\
Q24 & How real did the virtual world seem to you? & Likert scale: \{About as real as an imagined world, Indistinguishable from the real world\} \\
\addlinespace
Q25 & In what ways did using AI features contribute to or hinder collaboration with your team members? & Open text response \\
Q26 & To what extent do you feel that your team preferred human-generated ideas compared to AI-generated ideas? & \{Entirely preferred AI-generated ideas\}; \{Mostly preferred AI-generated ideas\}; \{Preferred AI-generated and human-generated ideas equally\}; \{Mostly preferred human-generated ideas\}; \{Entirely preferred human-generated ideas\} \\
Q27 & Given your answer to the above question (Q26), why do you think your team had that preference? & Open text response \\
Q28 & How did you like the task? e.g. Was it engaging, too simple, etc.? & Open text response \\
Q29 & Think about the platform you used to complete the study (either Miro or Noda). To what extent was this platform helpful? Why or why not? & Open text response \\
Q30 & Would you use or recommend this platform to others? Why? & Open text response \\
Q31 & Would you use or recommend the AI feature of this platform to others? & Open text response \\
\bottomrule
\end{tabular}
\endgroup
\end{table*}
 








\newpage

\newpage
\label{appendix:ideas}

\renewcommand{\thetable}{C.\arabic{table}}
\setcounter{table}{0}





\begin{table*}[p]
\begingroup
\footnotesize
\centering
\caption{Finalist ideas selected by each group in Subtask 3, with cost tier and coded source.}
\label{tab:finalists}
\begin{tabular}{@{}p{0.06\linewidth}p{0.14\linewidth}p{0.76\linewidth}@{}}
\toprule
Group & Source & Idea \\
\midrule
M1 & AI & (High) Renewable Energy Projects: Investing in solar panels or wind turbines to generate clean energy for campus buildings. \\
   & Human & (Medium) Expand on the bike share program (more accessible) \\
   & AI & (Low) Community Gardens: Creating spaces for students to grow their own food, fostering a sense of community and promoting sustainable agriculture \\
\midrule
M2 & AI & (High) Establish a student-led sustainability committee specifically focused on the dining hall, allowing students to actively participate in decision-making processes and initiatives related to sustainability in the dining hall. \\
   & AI & (Medium) Initiate a `Green Dorm Challenge' where dorms compete to reduce energy and water consumption, with progress tracked and shared through a digital dashboard. \\
   & Human & (Low) Card Swipe incentives: Have a particular amount of Flex dollars that can be earned for swiping at least a certain number of times in a semester to make sure dining hall workers can track and be prepared for students at meal times \\
\midrule
M3 & AI & (High) Install solar panels on campus buildings to generate renewable energy and reduce reliance on fossil fuels. \\
   & AI & (Medium) Creating a community garden where residents can come home together to grow their own fruits, vegetables, and herbs, promoting sustainable agriculture practices. \\
   & AI & (Low) Hosting a community clean-up event where participants can come together to pick up litter and beautify public spaces. \\
\midrule
M4 & Group-authored & (High) Solar panels or another form of renewable energy \\
   & Group-authored & (Medium) Campus waste - sus bins revamp, composting, trainings for awareness \\
   & Group-authored & (Low) Fund for student projects that will increase sustainability \\
\midrule
M5 & Group-authored & (High) Solar panel installation \\
   & AI & (Medium) Water conservation Program inspired park\ldots\ target will be locals and kids from the children's center \\
   & AI & (Low) Implementing a waste reduction program \\
\midrule
M6 & Group-authored & (High) Sustainable Energy \\
   & Human & (Medium) Green Startups \\
   & Group-authored & (Low) Sustainable Dorms \\
\midrule
M7 & Human & (High) Dorm renovation \\
   & Group-authored & (Medium) Fix the compost \\
   & AI & (Low) Sustainable transportation \\
\midrule
N1 & Group-authored & (High) Energy initiatives \\
   & AI & (Medium) Waste Reduction initiatives \\
   & Human & (Low) More sustainable transportation \\
\midrule
N2 & Human & (High) Solar Panels \\
   & Group-authored & (Medium) Financial Support for A Student Led Project \\
   & Group-authored & (Low) Research Project \\
\midrule
N3 & Group-authored & (High) Water Conservation in dorms and buildings \\
   & Human & (Medium) Making sure fauna suited to local habitat \\
   & Human & (Low) Food waste reduction in dining halls \\
\midrule
N4 & Human & (High) Solar Panels \\
   & --- & (Medium) N/A \\
   & Human & (Low) Sharing Clothes, recycling materials \\
\midrule
N5 & Group-authored & (High) Energy efficient BOW shuttle \\
   & Human & (Medium) Donating Food \\
   & Human & (Low) Swipe in Technology and Culture \\
\midrule
N6 & Group-authored & (High) Radiator Renovations \\
   & AI & (Medium) Educational seminars with industry professionals \\
   & Human & (Low) Community garden/composting and cleanup \\
\midrule
N7 & Group-authored & (High) More convenient transportation within the area and to the airport \\
   & AI & (Medium) Eliminate single use plastics by using comportable alternatives and hiring staff \\
   & Group-authored & (Low) Hiring of new staff to make recycling transportation more efficient \\
\bottomrule
\multicolumn{3}{@{}p{0.96\linewidth}}{\emph{Note.} M = Miro group; N = Noda group. Each group selected one high-cost (\$60,000), one medium-cost (\$30,000), and one low-cost (\$10,000) idea. Source was coded from the final workspace and the session video as Human (a Subtask 1 participant-entered node), AI (an unmodified AI-generated node), or Group-authored (a node the group composed during Subtask 2). Ideas are reproduced as the group wrote them in the workspace, including abbreviations and typographical errors. Group N4 did not record a medium-cost selection; the row is retained as ``N/A'' and is excluded from the counts in Table ~10, giving 20 Noda finalists and 21 Miro finalists.}\\
\end{tabular}
\endgroup
\end{table*}


\newpage
\label{appendix:codebook}

\end{appendices}

\end{document}